\documentclass[prd,amsmath,amssymb,floatfix,nofootinbib,superscriptaddress]{revtex4}
\pdfoutput=1
\usepackage{graphicx}
\usepackage{bm}
\usepackage{amsmath,amssymb}
\usepackage{epsfig}
\usepackage{color}
\usepackage{natbib, ifthen}
\usepackage{enumerate}
\usepackage{subfig}

\usepackage[pdftex]{hyperref}

\def\eprinttmp@#1arXiv:#2 [#3]#4@{
\ifthenelse{\equal{#3}{x}}{\href{http://arxiv.org/abs/#1}{#1}}{\href{http://arxiv.org/abs/#2}{arXiv:#2} [#3]}}

\providecommand{\eprint}[1]{\eprinttmp@#1arXiv: [x]@}
\newcommand{\dsurl}[1]{\href{#1}{ADS}}
\providecommand{\bibinfo}[2]{\ifthenelse{\equal{#1}{isbn}}{
\href{http://cosmologist.info/ISBN/#2}{#2}}{#2}}

\providecommand{\e}[1]{\ensuremath{\times 10^{#1}}}

\newcommand{\highaccuracydefault}{{\ttfamily high\_accuracy\_default}}
\newcommand{\lSampleBoost}{{\ttfamily lSampleBoost}}
\newcommand{\AccuracyBoost}{{\ttfamily  AccuracyBoost}}
\newcommand{\lAccuracyBoost}{{\ttfamily  lAccuracyBoost}}
\newcommand{\chisqeff}{ \chi_{\text{eff}}^{2}}
\newcommand{\kf}{\beta}
\newcommand{\nonflat}[1]{#1}
\newcommand{\lmin}{l_{\rm min}}

\newcommand{\adcq}[1]{\textcolor{magenta}{}}

\newcommand{\Mpc}{\text{Mpc}}

\newcommand{\lmax}{l_{\text{max}}}

\providecommand{\CAMB}{\textsc{camb}}
\providecommand{\COSMOMC}{\textsc{CosmoMC}}
\providecommand{\CMBFAST}{\textsc{cmbfast}}
\providecommand{\COSMICS}{\textsc{Cosmics}}
\providecommand{\CLASS}{\textsc{class}}
\providecommand{\CMBEASY}{\textsc{cmbeasy}}
\providecommand{\LCDM}{{$\rm{\Lambda CDM}$}}

\newcommand{\begm}{\begin{pmatrix}}
\newcommand{\enm}{\end{pmatrix}}

\newcommand\ba{\begin{eqnarray}}
\newcommand\ea{\end{eqnarray}}
\newcommand\bea{\begin{eqnarray}}
\newcommand\eea{\end{eqnarray}}

\newcommand\be{\begin{equation}}
\newcommand\ee{\end{equation}}

\providecommand{\Tr}{\text{Tr}}
\newcommand{\la}{\langle}
\newcommand{\ra}{\rangle}

\newcommand{\ud}{{\rm d}}

\newcommand{\mC}{\bm{C}}

\newcommand{\boldvec}[1]{{{\mathbf{#1}}}}

\newcommand{\vn}{\boldvec{n}}

\newcommand{\cll}{\mathcal{L}}

\newcommand{\clo}{\mathcal{O}}

\newcommand{\clz}{\mathcal{Z}}

\begin{document}

%%%%%%%%%%%%%%%%%%%%%%%%%%%%%%%%%%%%%%%%%%%%%%%%%%%%%%%%%%%%%%%%%%
%                       Title matter                             %
%%%%%%%%%%%%%%%%%%%%%%%%%%%%%%%%%%%%%%%%%%%%%%%%%%%%%%%%%%%%%%%%%%
% title and affiliations

\title{CMB power spectrum parameter degeneracies in the era of precision cosmology}

\author{Cullan Howlett}
\affiliation{Department of Physics \& Astronomy, University of Sussex, Brighton BN1 9QH, UK}

\author{Antony Lewis}
\homepage{http://cosmologist.info}
\affiliation{Department of Physics \& Astronomy, University of Sussex, Brighton BN1 9QH, UK}

\author{Alex Hall}
\affiliation{Institute of Astronomy and Kavli Institute for Cosmology, Madingley
 Road, Cambridge, CB3 0HA, UK}

\author{Anthony Challinor}
\affiliation{Institute of Astronomy and Kavli Institute for Cosmology, Madingley
 Road, Cambridge, CB3 0HA, UK}
 \affiliation{DAMTP, Centre
for Mathematical Sciences, Wilberforce Road, Cambridge CB3 0WA, UK}

\begin{abstract}
Cosmological parameter constraints from the CMB power spectra alone suffer several well-known degeneracies.
These degeneracies can be broken by numerical artefacts and also a variety of physical effects that become quantitatively important with high-accuracy data e.g. from the Planck satellite.  We study degeneracies in models with flat and non-flat spatial sections, non-trivial dark energy and massive neutrinos, and investigate the importance of various physical degeneracy-breaking effects. We test the {\CAMB} power spectrum code for numerical accuracy, and demonstrate that the numerical calculations are accurate enough for degeneracies to be broken mainly by true physical effects (the integrated Sachs-Wolfe effect, CMB lensing and geometrical and other effects through recombination) rather than numerical artefacts. We quantify the impact of CMB lensing on the power spectra, which inevitably provides degeneracy-breaking information even without using information in the non-Gaussianity. Finally we check the numerical accuracy of sample-based parameter constraints using {\CAMB} and {\COSMOMC}. In an appendix we document recent changes to {\CAMB}'s numerical treatment of massive neutrino perturbations, which are tested along with other recent improvements by our degeneracy exploration results.
\end{abstract}

\date{\today}

\maketitle

\pagenumbering{arabic}

\section{Introduction}
Observations of the CMB can provide accurate constraints on cosmological models given various fairly weak assumptions. Numerical results for comparison with observations are usually calculated using a linear line-of-sight Boltzmann code, with additional modelling of non-linear effects such as CMB lensing and the Sunyaev-Zel'dovich effects. Analysis of the WMAP data~\cite{Komatsu:2010fb} uses {\CAMB}~\cite{Lewis:1999bs}, a code evolved from an early version of {\CMBFAST}~\cite{Seljak:1996is}, itself developed from the Boltzmann hierarchy code \COSMICS~\cite{Ma:1995ey}. As with the other recent codes \CLASS~\cite{Lesgourgues:2011re} and \CMBEASY~\cite{Doran:2003ua}, {\CAMB} and the latest version of {\CMBFAST} aim to calculate the temperature and polarization power spectra at sub-percent precision in a matter of seconds, and use of such codes is now routine. Forthcoming data, especially from Planck, will greatly increase the precision of available data on small-scales, and it remains important to check numerical robustness and understand the physical limitations on what the CMB data can in principle constrain, for example due to the well-known geometrical degeneracy.

Previous work has shown that {\CAMB} is consistent with independent numerical codes and can reliably be used for precision cosmology when used with appropriate accuracy settings (and assuming that the physical model including recombination history is correct)~\cite{Hamann:2009yy,Lesgourgues:2011rg}. Older work has also studied the stability of {\CMBFAST} by comparison against a full Boltzmann hierarchy code~\cite{Seljak:2003th}. However these studies have focused on spatially-flat or otherwise restricted models. Recent new approximations and coding development aimed towards Planck data analysis have led to faster high accuracy calculations, and accuracy settings have been adjusted to allow efficient calculation of power spectra at the $\alt 0.1\%$ accuracy level ~\cite{CyrRacine:2010bk,Blas:2011rf,camb_notes}. As the code has developed the approximations and accuracy settings have been extensively tested as part of routine development, and we shall not labour the reader by giving an extensive presentation here. In this paper we focus on providing a powerful check on numerical accuracy for non-flat (and flat) models by exploring parameter degeneracies, where small numerical errors could potentially lead to spurious degeneracy breaking.

This paper has three objectives: (i) testing the numerical accuracy of {\CAMB} and determining the precision that is required; (ii) examining to what extent degeneracies are broken by physical processes when precision CMB power spectrum data is available; and (iii) demonstrating the extent of degeneracies that can be expected from Planck when using CMB power spectrum data alone, independently of parameter priors. We use the numerical code {\CAMB}\footnote{\url{http://camb.info}} exclusively since at the time of writing \CLASS~\cite{Lesgourgues:2011re} does not support non-flat models, and {\CMBFAST}~\cite{Seljak:1996is} and \CMBEASY~\cite{Doran:2003sy} are no longer actively maintained. The July 2011 version of {\CAMB} was originally used for much of this work, and we made various adjustments as indicated by the tests of this paper to maintain accuracy; however for consistency of presentation the final plots are shown from the October 2011 version after we made various accuracy parameter tweaks. In Sec.~\ref{newinterpolation} we discuss a further minor modification made following the results of this paper that is now implemented in the January 2012 {\CAMB} version; the new version only improves the accuracy of the results shown in this paper.

As is well-known, CMB lensing (for reviews see Refs.~\cite{Lewis:2006fu,Hanson:2009kr}) can break the geometrical degeneracy since the lensing deflections are sourced all along the line of sight, and hence are sensitive to both the geometry and growth of structure after recombination~\cite{Stompor:1998zj,Smith:2006nk,Sherwin:2011gv}. Lensing does not bias parameter constraints if it is modelled consistently, and the effect on the error bars of the power spectrum estimators is also small~\cite{Smith:2006nk} (unless considering low-noise $B$-mode observations~\cite{Smith:2006vq}). Reference~\cite{Stompor:1998zj} has looked at the effect of lensing on the power spectrum on parameter degeneracies for flat and open models, and many authors have also considered the additional information available by using the lensing three- and four-point functions (typically by performing lensing reconstruction). Here we focus on updating the analysis of lensing on the power spectra to the sub-percent precision era and current understanding of the cosmological model, allowing for flat, open, and closed models. Note that the unlensed power spectra are not observable directly, so this lensing information is inevitably present in any consistent cosmological CMB power spectrum analysis. The effect of lensing on the power spectra has recently been used to constrain dark energy using only CMB data from ACT~\cite{Sherwin:2011gv} and SPT~\cite{vanEngelen:2012va}, but future data constraints will require validation to significantly higher precision.
The more complicated question of how correctly to use non-Gaussian lensing information in combination with the information in the lensed power spectra is left for future work.

We shall assume the recombination history is accurately calculated, as studied in detail by many authors~\cite{Seager:1999km,Wong:2007ym,Switzer:2007sn,RubinoMartin:2009ry,Chluba:2010ca,AliHaimoud:2010dx,Shaw:2011ez},  so that it is not a source of bias or uncertainty. We approximate reionization as being fairly sharp, using {\CAMB}'s standard parametrization where hydrogen reionization and the first reionization of helium happen together~\cite{Lewis:2008wr}.

This paper is organized as follows. We start in Sec.~\ref{sec:geo} by considering the geometrical degeneracy in non-flat \LCDM\ models, considering both numerical and physical effects that determine the extent of the degeneracy. In Sec.~\ref{sec:DE} we then restrict to flat models but allow more general dark energy. In Sec.~\ref{sec:approxdeg} we analyse the approximate degeneracy that persists even in a standard flat model with a cosmological constant. In Secs~\ref{sec:flatnu} and~\ref{sec:nonflatnu} we then include massive neutrinos, for the case of both flat and non-flat models (Appendix~\ref{app:massive_nu} describes recent improvements in \CAMB's massive neutrino modelling which are tested by these sections). Finally in Sec.~\ref{sec:MCMC} we look at parameter constraints expected from the Planck satellite using a standard MCMC analysis, demonstrating the extent of the degeneracy expected, and quantifying the impact of residual numerical errors on parameter constraints. There, we also describe a new interpolation scheme that removes the leading numerical artefact shown in the previous sections; we demonstrate that this limits numerical biases in parameter constraints to being $\alt 5\%$ of the random error while maintaining acceptable numerical speed.

\section{Geometrical degeneracy with a cosmological constant}
\label{sec:geo}

\subsection{Calculating the degenerate models}

In this section we test the numerical accuracy of {\CAMB} using the well-known geometrical degeneracies. We also look at physical effects that break the degeneracy including CMB lensing, a geometrical averaging effect through recombination, and the late-time integrated-Sachs-Wolfe (ISW) effect.

We first recap the reason for the geometrical degeneracy.
A parameter degeneracy effectively describes our inability to distinguish certain cosmological parameter combinations, in this context through using CMB anisotropies alone. The primary CMB anisotropies are generated around recombination and what we observe is a projection of conditions on the last-scattering surface. If we keep the physical densities in baryons, cold dark matter and the number of (massless) neutrinos fixed, the pre-recombination physics is unchanged. Since the mapping of physical scales at last-scattering to observed angular scales depends only on the angular-diameter distance to last-scattering, there are generally degenerate combinations of ``late-time'' parameters (such as the curvature parameter $\Omega_K$ and expansion rate today $H_0$) that yield very nearly the same power spectra of primary anisotropies~\cite{Zaldarriaga:1997ch,Bond:1997wr,Efstathiou:1998xx}. We will consider various combinations of parameters that can give nearly identical unlensed CMB power spectra, starting with the degeneracy in non-flat \LCDM\ models.

%Specifically, when  we measure the CMB we only measure some temperature and direction on the surface of last scattering.
%How physical scales at last-scattering map into observed angular scales depends only on the angular diameter distance, which depends on combination of cosmological parameters (such as the curvature parameter $\Omega_K$ and expansion rate today $H_0$). Without other information it is impossible to constrain the individual parameters, since many different combinations can give the same angular diameter distance: different combinations produce nearly identical CMB anisotropies and hence power spectra~\cite{Zaldarriaga:1997ch,Bond:1997wr,Efstathiou:1998xx}.
%We will consider various combinations of parameters that can give nearly identical CMB power spectra, starting with the degeneracy in non-flat \LCDM\ models.
%One such pair of parameters is called the geometrical degeneracy and is between \(h\) and \(\Omega_{\Lambda}\). This is the first of several degeneracies that we will be investigating.

\begin{table}[h!]
 \begin{center}
  \begin{tabular}{cc}
   \hline
   Parameter           & Value         \\ \hline
   \(\Omega_{\Lambda}\)& 0.733         \\
   \(\Omega_{b}h^{2}\) & 0.0226         \\
   \(\Omega_{c}h^{2}\) & 0.112         \\
   \(\Omega_{K}\)      & 0.0           \\
   \(\Omega_{\nu}\)    & 0.0           \\
   \(h\)               & 0.71          \\
%AL I changed A_s to give it at the pivot scale used by rather than WMAP (0.002)
 %  \(A_{s}\)           & 2.43\e{-9}    \\
   \(A_{s}\)           & 2.1364\e{-9}    \\
   \(n_{s}\)           & 0.96          \\
   \(\tau\)            & 0.088         \\
   \hline
  \end{tabular}
 \end{center}
 \caption{Parameter values for our fiducial model. Note that $A_s$ is defined at a pivot scale of $k_0=0.05 \Mpc^{-1}$ and running of the spectral index $n_s$ is assumed to be zero.}
 \label{ParamsTable}
\end{table}

The first step is to determine which sets of parameters give these nearly identical CMB power spectra. We define a fiducial model, and then explore other combinations of parameters that are (nearly) degenerate.
We use as fiducial parameters the best-fit seven-year WMAP parameters~\cite{Komatsu:2010fb} given in Table~\ref{ParamsTable}. We run {\CAMB} at boosted high accuracy settings to ensure that the fiducial model itself has minimal numerical error: we set the \highaccuracydefault\ parameter to `true', and set the additional three accuracy parameters {\lSampleBoost}, \lAccuracyBoost\ and \AccuracyBoost\ to values of 2, thus ensuring all numerical calculations are performed with extremely high precision\footnote{\lSampleBoost\ changes the sampling in $l$ over which the $C_l$ are interpolated. \lAccuracyBoost\ changes the number of multipoles maintained when integrating the Boltzmann hierarchies. \AccuracyBoost\ changes wavenumber sampling and integration step sizes (and various other parameters). The {\highaccuracydefault} switch increases the density and range of wavenumber samples, increases the number of multipoles retained in the Boltzmann hierarchies that are evolved, switches from the tight coupling approximation slightly earlier, uses a larger range of
unlensed template $C_l$ when calculating the lensed $C_l$, and, prior to January
2012, increases the density of $l$ samples that are interpolated (see
Sec.~\ref{newinterpolation}). Unlike blindly increasing the accuracy parameters, which changes many parameters simultaneously by the same amount, the changes adopted by the  {\highaccuracydefault} switch give the minimal internal parameter tweaks required to give the target accuracy $\alt 0.1\%$  at $l\agt 500$ and hence do not
dramatically increase the running time.
For further discussion see Refs.~\cite{Hamann:2009yy,Lesgourgues:2011rg}, though changes have been made since those papers.}.
%In our case all these parameters were given values of 2.
Henceforth it should also be noted that the term `high accuracy' corresponds to the \highaccuracydefault\ parameter having a value of `true' (designed to be appropriate for Planck analysis), whilst `low accuracy' corresponds to that parameter having value `false' (appropriate for WMAP analysis).

Once the fiducial model has been specified it is possible to determine which combinations of parameters create the geometrical degeneracy by direct comparison of the power spectra returned by {\CAMB}. We want to quantify the difference between the spectra in a way that is relevant for observations: for example cosmic variance means that much lower accuracy is required on large-scales than on small-scales.  What we observe is the CMB temperature \(T(\hat{\vn})\) along the line of sight $\hat{\vn}$, which can be decomposed into spherical harmonics as
\begin{equation}
  T(\hat{\vn}) = \sum_{lm} T_{lm} Y_{lm}(\hat{\vn}).
\end{equation}
The power spectrum for statistically-isotropic fluctuations is defined by
\begin{equation}
 \la T_{lm}T_{l'm'}^*\ra = \delta_{ll'}\delta_{mm'}C_{l},
\end{equation}
which in a perfect experiment can be estimated by
\begin{equation}
 \hat{C_{l}} = \frac{1}{2l+1} \sum_{m}|T_{lm}|^{2}.
\end{equation}
From these estimators the probability of the true power spectrum $C_l$ (with a flat prior) is given by the likelihood $\cll$, where for convenience we define $\chisqeff\equiv -2\log \cll$. For Gaussian perturbations
\begin{equation}
\chisqeff= \sum^{\lmax}_{\lmin} (2l+1)\left(\frac{\hat{C_{l}}}{C_{l}} + \ln\left(\frac{C_{l}}{\hat{C_{l}}}\right) - 1\right).
\label{eq:chisq}
\end{equation}
In any particular realization of the sky, the estimators $\hat{C}_l$ have cosmic variance about the true power spectrum, and hence scatter in a realization-dependent way. To explore the degeneracies in a realization-independent way we replace $\hat{C}_l$ with the power spectrum in the fiducial model, $C_{l}^{\rm fid}$.  We can then quantify how close a degenerate model is to our fiducial model by using the effective chi-squared value $\chisqeff$. Note that we have normalized this equation such that any model that is exactly equal to the fiducial model will give a value of $\chisqeff= 0$. Throughout, we use a value of $\lmax = 2000$ with $\lmin=2$ or $\lmin=100$ as indicated below (depending on whether we are focusing on high-$l$ numerical errors or a more realistic analysis). The cut at $\lmax=2000$ is somewhat arbitrary, but reflects that lower precision is required on small-scales due to rapidly growing uncertainties from beams, point sources, secondary anisotropies, and other foregrounds. In practice the numerical performance of CAMB above $\lmax=2000$ is not dramatically worse, and should be sufficient if data on smaller scales can be used.

\begin{figure}[h!]
\begin{center}
\includegraphics[width = 8.5cm]{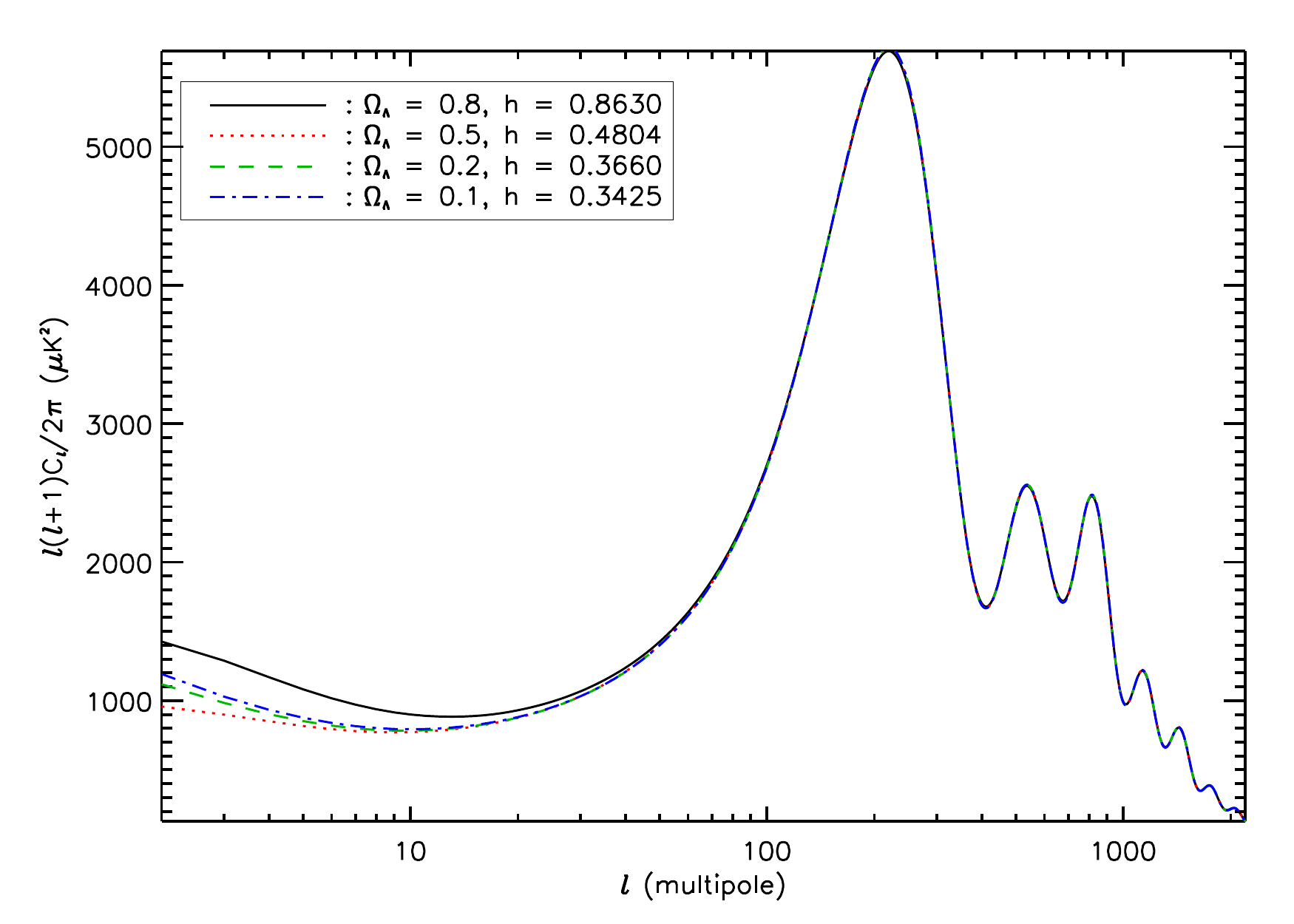}
\includegraphics[width = 8.5cm]{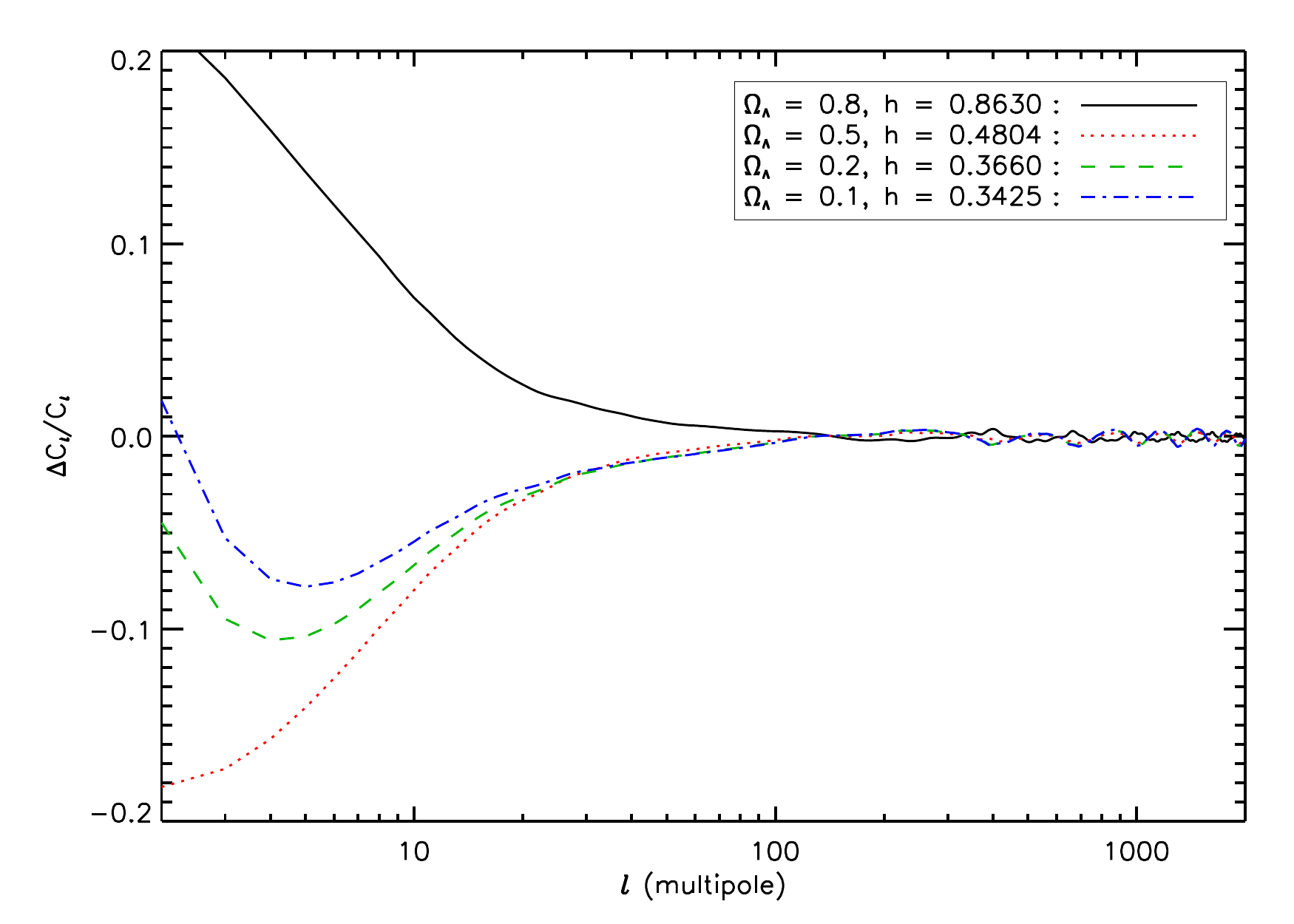}
\end{center}
 \caption{CMB power spectrum obtained using {\CAMB} for nearly degenerate geometries in non-flat $\Lambda$CDM models with no lensing (left) and the fractional differences from the fiducial-model spectrum (right).
%This shows the geometric degeneracies with nearly identical CMB anisotropies but very different geometries. The fiducial model used throughout has values \(\Omega_{b}h^{2}\) = 0.0226, \(\Omega_{c}h^{2}\) = 0.112, \(\Omega_{k}\) = 0.0 and \(h\) = 0.71.
Both \(\Omega_{b}h^{2}\) and \(\Omega_{c}h^{2}\) were fixed to their fiducial values in all cases to preserve the pre-recombination physics.
Low accuracy and values of 1 for {\lSampleBoost}, \AccuracyBoost\ and \lAccuracyBoost\ were used for the calculations.
 \label{CLNonFlat}
 }
\end{figure}

By minimizing $\chisqeff$, we can  find sets of parameters that give power spectra that are very close to the fiducial model. Example unlensed power spectra are shown in non-flat $\Lambda$CDM models in Fig.~\ref{CLNonFlat}. In this case, the
geometric degeneracy is within the two-dimensional space of $\Omega_\Lambda$ and
$h$.

\subsection{Degeneracy breaking effects}

\subsubsection{Numerical accuracy}

As shown in Fig.~\ref{CLNonFlat} the power spectra are not entirely degenerate. On large-scales this is due to the late-time ISW effect; on small scales the power spectra are much closer. However the small-scale spectra will not be quite identical due to both numerical artefacts and small physical effects. For the purpose of having reliable parameter constraints we would of course like the numerical effects to be small, so the degeneracy is broken only by the physical effects.  Here, we quantify the difference between these degenerate power spectra as a check that the numerical accuracy of {\CAMB} is sufficient for forthcoming data, and investigate in more detail degeneracy breaking effects such as lensing, geometrical effects through last scattering, and also the late-time ISW effect.

We expect any numerical errors in the power spectrum as a result of {\CAMB}'s calculations to be comparable to, if not less than, the quoted values: 0.3\% for low accuracy calculations (\highaccuracydefault=F) and 0.1\% (at $l\agt 500$) for high accuracy calculations (\highaccuracydefault=T). Figure~\ref{CLNonFlat} (right-hand panel) shows the fractional differences between the power spectra in four degenerate geometries and the fiducial model. The spectra are closely similar at high $l$ where the ISW contributions to the spectra all become small.

\begin{figure}[h!]
 \vspace{-160pt}
 \includegraphics[width = 18cm]{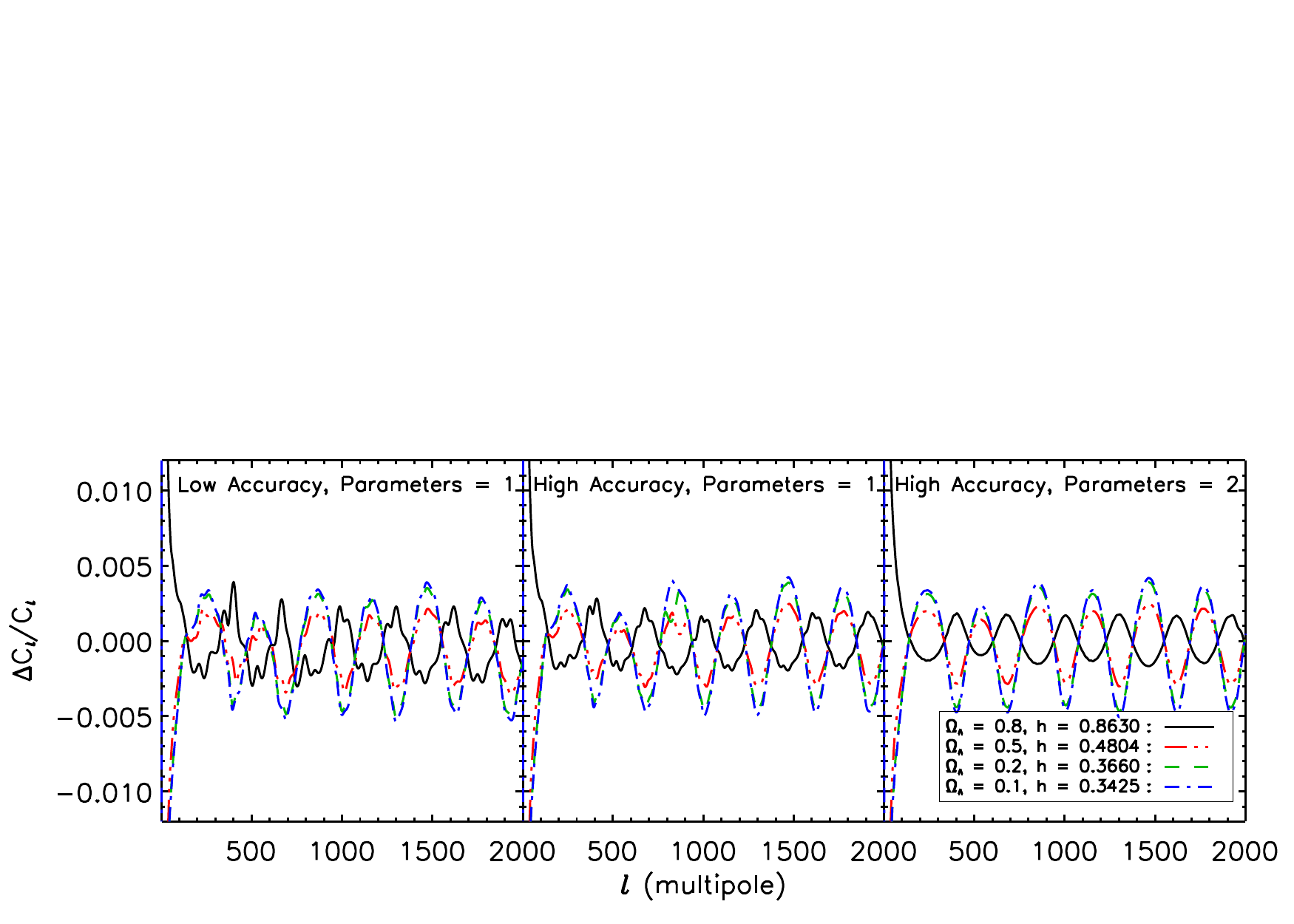}
 \caption{Difference between the unlensed power spectra of a range of non-flat degenerate models and the fiducial model.  Models are computed at low accuracy with default accuracy parameters (left), high accuracy with default accuracy parameters (middle) and  high accuracy with parameters boosted to 2 (right).
The right-hand figure shows the physical differences in the spectra, with very little residual numerical error.
 \label{NonFLatHighAccuracyBoosted}
 }
\end{figure}

Figure~\ref{NonFLatHighAccuracyBoosted} shows differences between the high-$l$ power spectra for a
%the effect of increasing the accuracy parameters for a
selection of nearly-degenerate models.
 We can see a larger periodic oscillation in the difference plots that roughly corresponds to the location of the peaks and troughs in the power spectra. These are dependent on the degenerate model used (and increase with $|\Omega_K|$), which indicates they are due to physical effects (see discussion in the next subsection). However, we also see much smaller more random oscillations within this larger periodicity. These are due to numerical errors and are reduced slightly by increasing from low accuracy calculations to high accuracy calculations; we shall quantify whether this improvement is good enough in more detail later. Boosting the accuracy removes most of the residual numerical error, and we are left with smoothly varying differences due to physical effects.

\begin{figure}[h!]
 \vspace{-60pt}
 \includegraphics[width = 14cm]{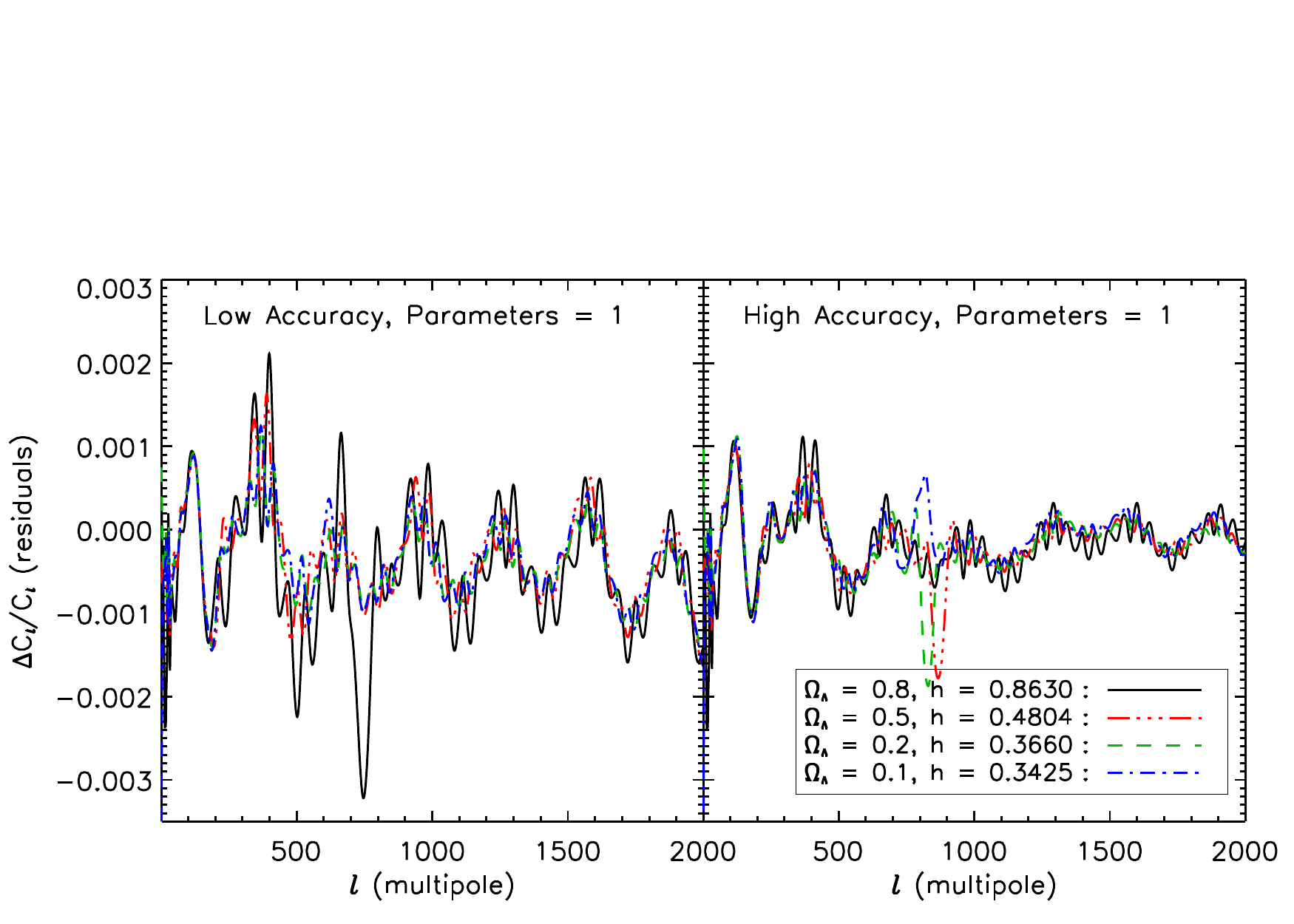}
 \caption{Numerical errors in the computation of a range of non-flat degenerate models. For a given model, the numerical errors are estimated by subtracting the spectrum from one calculated at high accuracy with accuracy parameters boosted to 2. Errors are plotted for a low (left) and high (right) accuracy calculation. In all cases, the other accuracy parameters are at their default values.
%A plot showing the difference between the unlensed CMB power spectra compared to the fiducial model (upper figures), and the  fractional error for low and high accuracy calculations (with {\lSampleBoost}= \lAccuracyBoost\ = \AccuracyBoost\ = 1) compared to when the spectra are calculated using a very accurate calculation (high accuracy and all parameters set to 2).
 \label{NonFlatDoubleDIfference}
 }
\end{figure}

%Figure~\ref{NonFLatHighAccuracyBoosted} shows the effect of increasing the accuracy parameters for a selection of nearly-degenerate models, showing the physical differences that remain when numerical precision is increased.
We can nearly isolate the numerical errors in the computation of a given model by subtracting a very accurate computation (high accuracy and all parameters boosted to 2) of the same model.
Our results are plotted in Fig.~\ref{NonFlatDoubleDIfference}.
We see that the numerical error for low accuracy calculations is of the order of 0.2\% (rather better than the quoted accuracy, indicating that the calculation is slower than it needs to be). There is a small increase in numerical precision when high accuracy calculations are used, decreasing the error to within the quoted value of 0.1\% at $l\agt 500$ where cosmic variance becomes small.

\begin{figure}[h!]
 \subfloat[\(\Omega_{\Lambda}\) = 0.80, \(\Omega_{K}\) = 0.01925]{\includegraphics[width = 0.33\textwidth, type = pdf,ext=.pdf,read=.pdf]{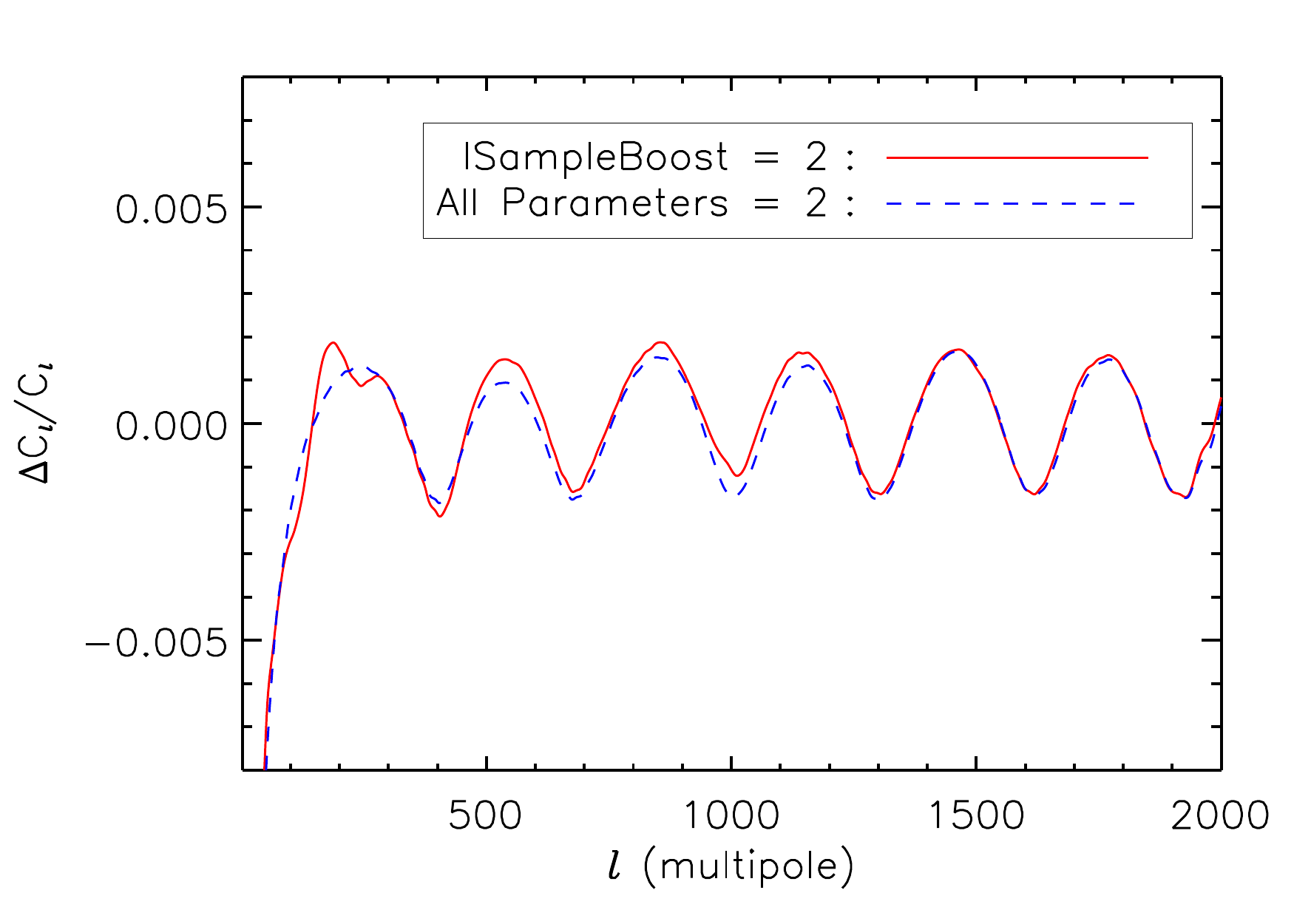}}
 \subfloat[\(\Omega_{\Lambda}\) = 0.73, \(\Omega_{K}\) = -0.00092]{\includegraphics[width = 0.33\textwidth, type = pdf,ext=.pdf,read=.pdf]{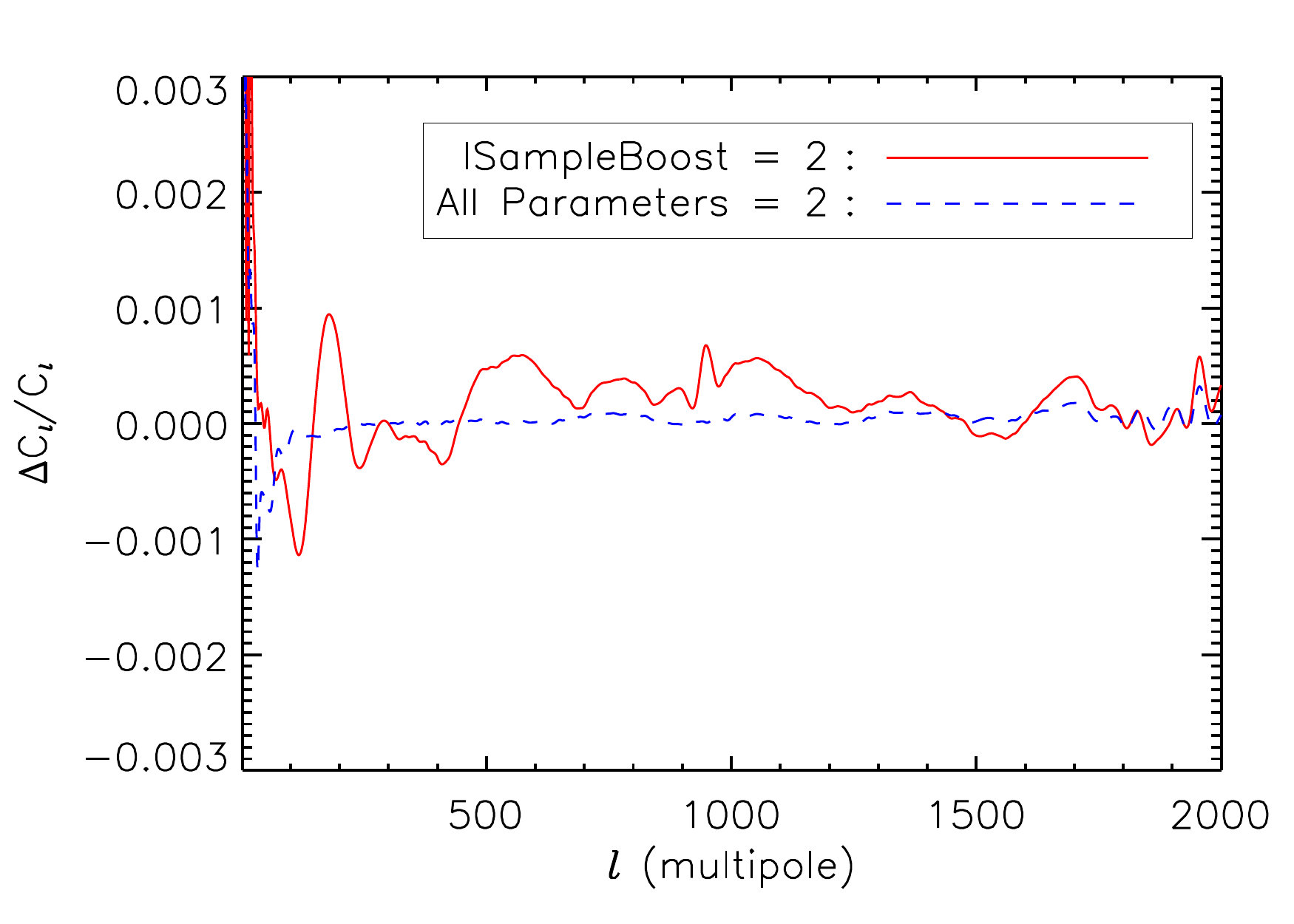}}
 \subfloat[\(\Omega_{\Lambda}\) = 0.40, \(\Omega_{K}\) = -0.12273]{\includegraphics[width = 0.33\textwidth, type = pdf,ext=.pdf,read=.pdf]{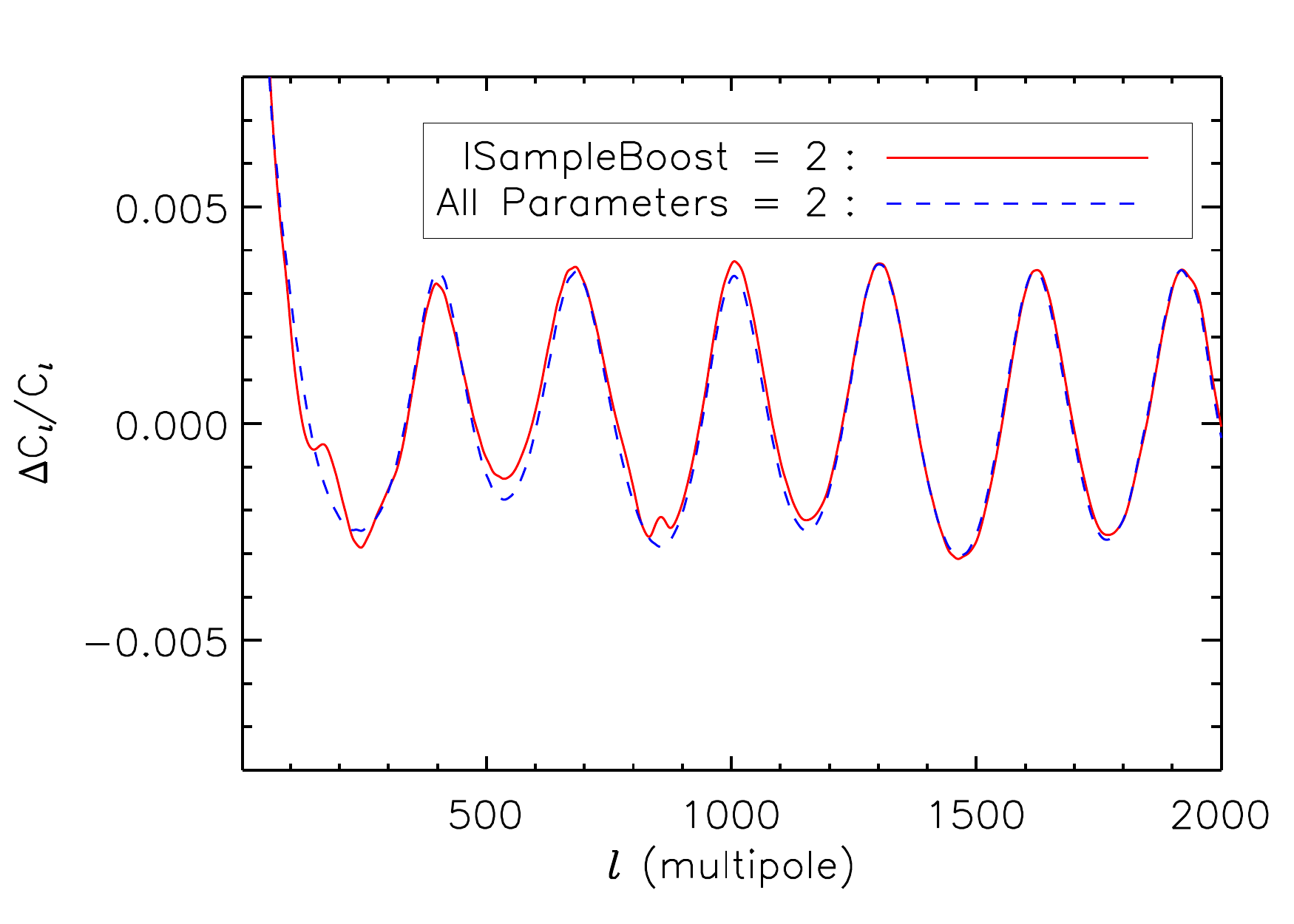}}
 \caption{
 Fractional differences between nearly degenerate geometries and the fiducial model, comparing boosts of 2 in all the accuracy parameters (solid red) to boosts of 2 in only \lSampleBoost\ (dashed blue). In all cases, high accuracy is used for the calculations.
 \label{lSampleBoosted}
 }
\end{figure}

If numerical precision is the only objective, increasing the accuracy settings will give accurate smooth results. However, such settings can increase the running time required very significantly, so rather than simply increasing these parameters it is best to optimize them such that a balance between speed and precision is found (see Ref.~\cite{Hamann:2009yy} for previous work on this). It is easy to identify the main cause of the high-frequency numerical wiggles: this is just due to small interpolation errors from the rather sparse $l$ sampling used when calculating the power spectra ($\Delta l\sim 50$). This effect is suggested by the clear correlations between models of the numerical errors isolated in Fig.~\ref{NonFlatDoubleDIfference}.
Figure~\ref{lSampleBoosted} shows that increasing \lSampleBoost\ to 2 can effectively remove these wiggles, as expected.
However using twice the density of $l$ samples takes nearly twice as long (in non-flat models) to calculate spectra.
Amongst other things, setting \highaccuracydefault\ lowers the $l$ sampling at high $l$ to $\Delta l=42$ from $\Delta l=50$, which significantly lowers the residuals; however this still leaves some interpolation artefacts. A further boost of \lSampleBoost\ $\sim 1.2$--$1.5$ is enough to remove most of these if required (so that $\Delta l \rightarrow \Delta l /\text{\lSampleBoost}$) while maintaining good efficiency. A better alternative to increasing blindly the $l$ sampling is to improve the interpolation method as discussed later in Sec.~\ref{newinterpolation}. On large scales the numerical errors are larger, $\sim 0.3\%$, but still well below the physical differences from different late-time ISW contributions between nearly-degenerate models, and always smaller than cosmic variance.

\begin{figure}[h!]
 \includegraphics[width = 9.5cm]{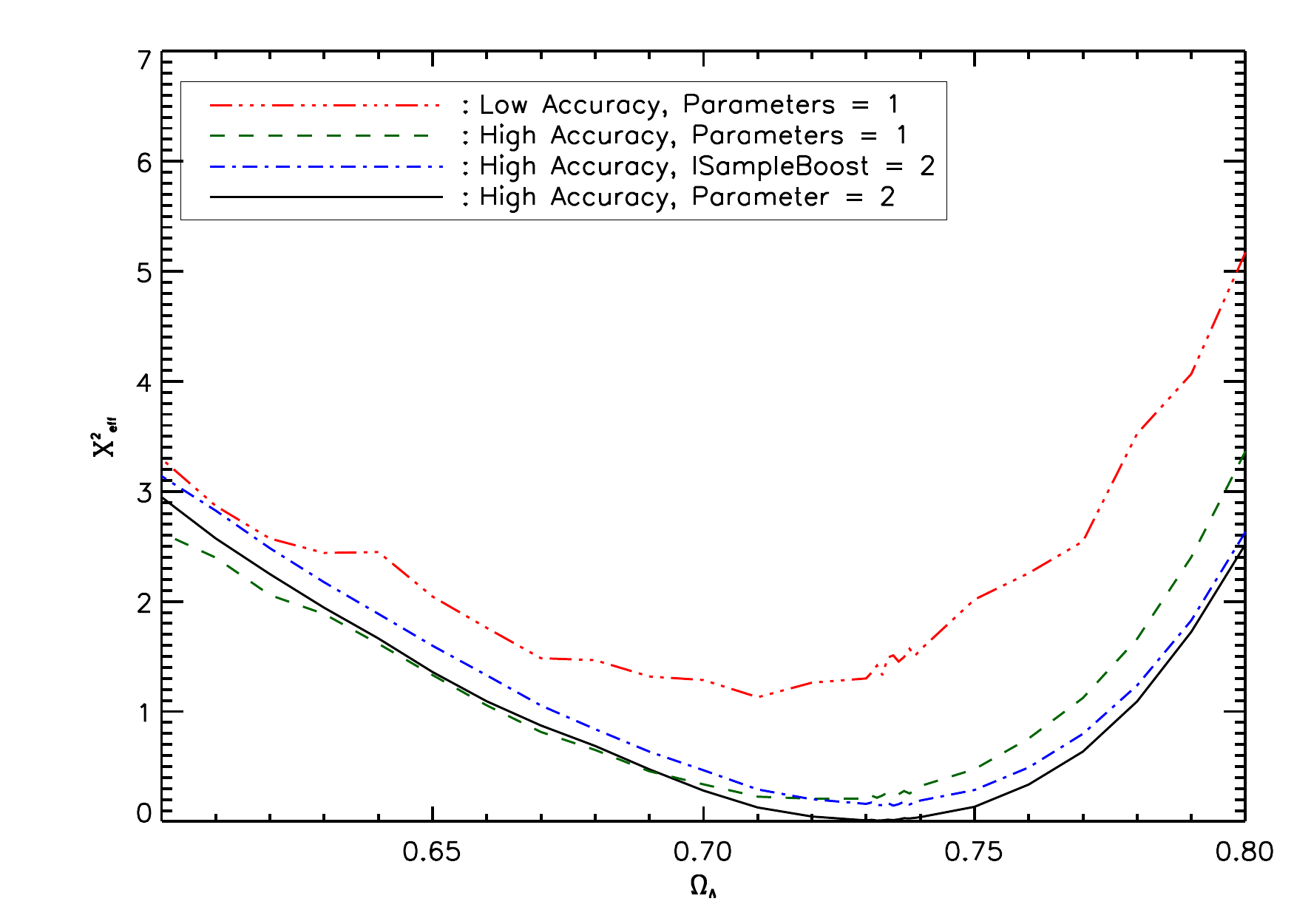}
 \caption{Minimum $\chisqeff$ for a range of degenerate geometries close to the fiducial model. Solid (black) is high accuracy and each of {\lSampleBoost}, \lAccuracyBoost\ and \AccuracyBoost\ boosted to 2; dashed (green) is high accuracy with default accuracy parameter values; dot-dashed (blue) is high accuracy with only {\lSampleBoost} boosted to 2; and triple-dot-dashed (red) is low accuracy with no accuracy boosts. In all cases we use only $l \ge 100$ in the calculation of~$\chisqeff$.
\label{NonFlatChiSq}
}
\end{figure}

We now test to what extent numerical errors break the geometric degeneracy compared to physical effects by comparing, for a range of values of $\Omega_\Lambda$, the minimum values of $\chisqeff$ (with the fiducial model for $\hat{C}_l$ and including only multipoles $l\ge100$) with respect to $h$.
%themselves in terms of the effective chi-squared values returned for each model compared to the fiducial model.
Figure~\ref{NonFlatChiSq} shows the minimum $\chisqeff$ as a function of $\Omega_\Lambda$ close to the fiducial value; the models are calculated for a range of accuracy settings. The solid black line indicates the closest we can get to our fiducial model in terms of numerical accuracy, and the shape of the curve reflects almost entirely the physical effects that weakly break the degeneracy; these are quantitatively more important the further away the degenerate geometry is from the fiducial model.

Numerical errors change this result, shifting the curve to a different offset and shape. The low accuracy settings are significantly biased (by around $0.5\sigma$) but use of default high accuracy settings recovers a much more accurate likelihood curve. However the likelihood curve is still slightly shifted compared to the accurate calculation. If we increase the value of \lSampleBoost\ to 2, effectively removing all the small-scale interpolation wiggles, we find the curve returns to nearly the correct shape, with only a slight offset due to residual numerical effects. Offsets do not affect inferences about parameter constraints and are relatively harmless. However shifts in the curves give rise to errors in the posterior mean and maximum likelihood point, and give rise to a parameter biases. We assess this in more detail below (Fig.~\ref{NonFlatChisqlsamp}) in the more observationally-relevant case that both noise, low-$l$ modes and lensing are included.

\subsubsection{Physical effects and Planck data}

Figure~\ref{NonFLatHighAccuracyBoosted} shows that the geometric degeneracy for non-flat models is broken at the $10^{-3}$ level (for $\Omega_K h^2 \sim 0.02$) by physical (geometrical) effects even in the absence of lensing. The effect of curvature on the evolution until recombination is very small, $\clo(10^{-4})$, so the physical anisotropy sources are essentially the same in the various degenerate models. However there is a larger effect due to the finite thickness of the last-scattering surface (c.f. Ref.~\cite{Zaldarriaga:1997va}). The change in the (comoving) angular-diameter distance $d_A$ over a fixed thickness of last-scattering centred on a given $d_A(z_*)$ (fixed by the peak of the visibility function) in curved models differs from the flat case; it is smaller in closed models and larger in open. In detail, in a closed model $d_A = \sin(\sqrt{K} \chi)/\sqrt{K}$ at radial distance $\chi$, so that $d d_A / d\chi =
\cos(\sqrt{K}\chi) = (1-K d_A^2)^{1/2}$ for $\sqrt{K}\chi \leq \pi /2$. In an open
model, the equivalent result is $d d_A / d\chi = (1+|K|d_A^2)^{1/2}$.
If we look at an angular scale corresponding to an acoustic peak --- probing a perturbation scale that is a maximum or minimum at the peak of the visibility --- then we will also generally see slightly larger scales at earlier times near the start of recombination, which are not yet at an extremum. However, in a closed universe the change in transverse comoving scale going through last-scattering is smaller: the perturbations probed earlier would be closer to an extremum.
In an open universe there is therefore a slight suppression of the power in the acoustic peaks, and in closed universe a slight boost, as shown in Fig.~\ref{NonFLatHighAccuracyBoosted}.
Assuming the thickness of recombination is a significant fraction of $\eta_*$, the conformal time at the peak of the visibility function, this effect is ${\cal O}(\eta_* d_A K)\sim 10^{-3}$--$10^{-2}$ for the parameter range of interest. This is large compared to other effects of curvature on the pre-recombination dynamics because we are viewing last scattering from a large distance ( $d_A \gg \eta_*$), and the effect of curvature on the fractional change in $d_A$ through a fixed thickness increases with radial distance.

In principle perfect small-scale unlensed data could therefore determine the curvature, even without additional information. However
in the immediate future, the best data available on intermediate scales will come from the Planck satellite. To assess more carefully the accuracy required for Planck we need to look in more detail at what is actually observed, and the relevant errors. For example the CMB is inevitably lensed, so we only actually observe the lensed power spectra, and it is the accuracy of parameter constraints including the lensing effects that really matter. We also introduce a very simple noise in the form relevant for Planck.

In terms of the power spectrum, lensing serves to smooth out the peaks corresponding to the CMB anisotropies~\cite{Stompor:1998zj}. For example, random gravitational lenses between the observer and the surface of last scattering cause the temperature anisotropies that we see effectively to be `smeared' out, causing us to loose definition in these regions and damping the sharpness of the features. This gives a smoothing of the power spectrum at the peak locations. This causes a breaking of the geometric degeneracy, as the amount of lensing is dependent on the geometry and growth of structure between us and last scattering~\cite{Stompor:1998zj,Smith:2006nk,Sherwin:2011gv,vanEngelen:2012va}.

\begin{figure}[h!]
 \includegraphics[width = 10cm]{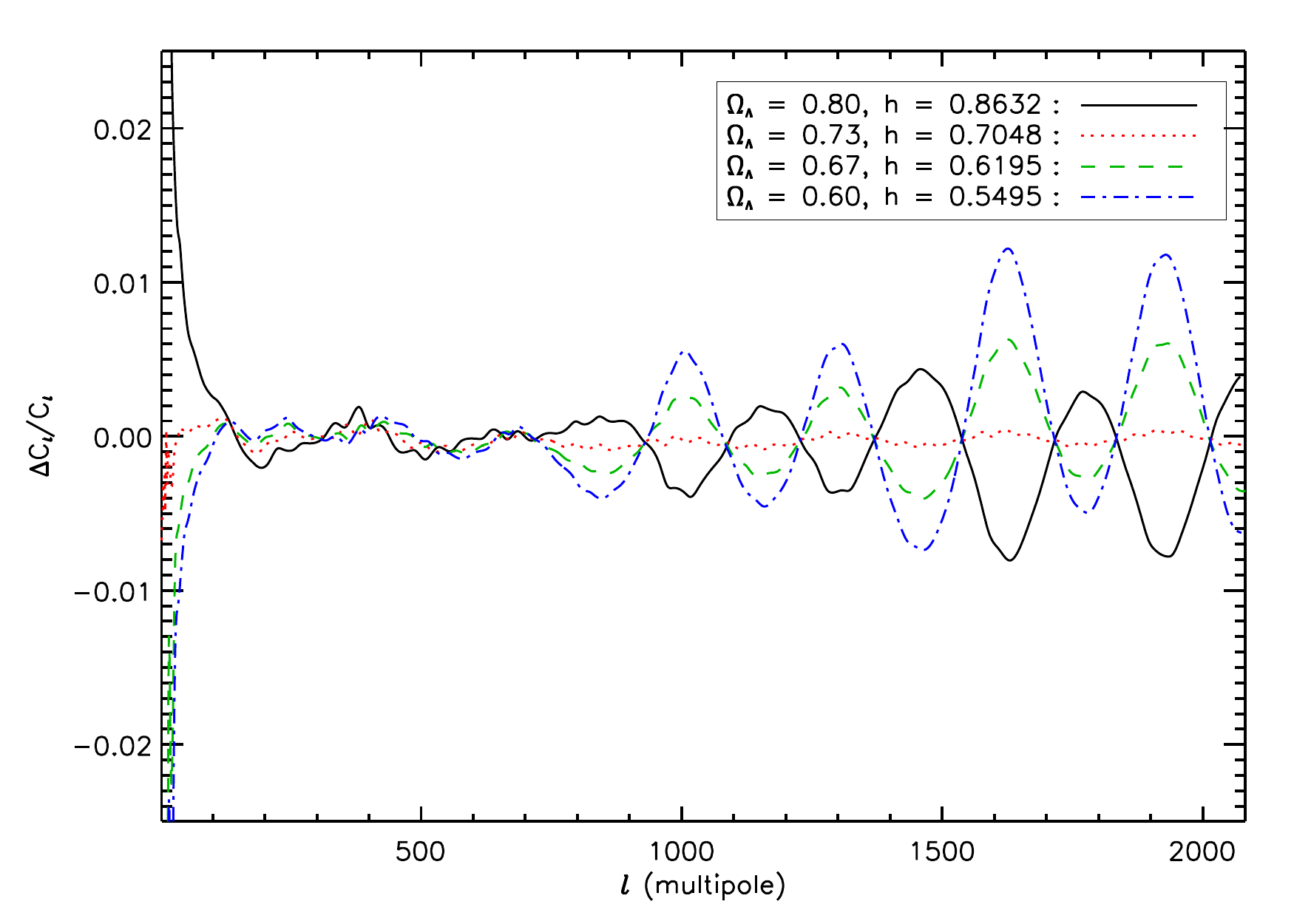}
 \caption{Fractional change in the \emph{lensed} CMB power spectra at different nearly-degenerate non-flat models. Note that the fractional differences between power spectra are much larger with lensing included, so here we plot a significantly narrower range of nearly-degenerate models than in the unlensed case of Fig.~\ref{NonFLatHighAccuracyBoosted}. Here results are calculated at high accuracy with \lSampleBoost\ = 1.17 to remove most of the small interpolation wiggles.
 \label{CLDIffWIthLensing}
 }
\end{figure}

When we include lensing the differences between the previously nearly-degenerate geometries increases significantly, as shown in Fig.~\ref{CLDIffWIthLensing}. Whilst the periodicity we observed earlier is still present, the amplitude of these peak differences increases nearly ten-fold on small scales as the different geometries cause different amounts of lensing, and hence different amounts of smoothing in the power spectra. As we will see later this is also apparent quantitatively, where we see a large increase in the minimum effective chi-squared values for each degenerate model in the presence of lensing, allowing us to place a much better constraint on $\Omega_\Lambda$ (or $\Omega_K$).

Including noise in the procedure, on the other hand, has the opposite effect to lensing, in that it strengthens the degeneracy by decreasing the sensitivity to small changes in the power spectra at high $l$. As a simple test we include noise on the temperature, \(N_{T}\), and polarization, \(N_{E}\), spectra with
\begin{equation}
 N_{T} = \frac{N_{E}}{4} = N_{0}e^{l(l+1)/\sigma^{2}},
\end{equation}
where \(N_{0} = 0.5 \e{-4}\, \mu \mathrm{K}^{2}\) is the white-noise level (corresponding to $24\,\mu\mathrm{K}\,\mathrm{arcmin}$) and $\sigma = 1.7\e{-3}\,\mathrm{rad}$ for a beam of $7\,$arcmin full-width at
half-maximum.
%and \(\sigma^{2}\) is given in terms of the beam full-width half-maximum, \(\text{fwhm} = \sfrac{7}{60}\), in degrees, by
%\begin{equation}
% \sigma^{2} = \left(\frac{2\pi}{180}\frac{\text{fwhm}}{\sqrt{8\log(2)}}\right)^{2}.
%\end{equation}

Our choice of noise level is at the lower end of what is likely to be achievable by combining multiple sky scans and frequencies from Planck, and is therefore conservative (in that if numerical accuracy is sufficient in our tests, it will almost certainly be sufficient in reality). We then simply add this noise to the polarization and temperature \(C_{l}\)s, \(C^{EE}_{l}\) and \(C^{TT}_{l}\) respectively, returned by {\CAMB}. Since the polarization signal is much smaller than the temperature it has significant noise, but nonetheless can have some degeneracy breaking power (as well as constraining the optical depth). For current purposes we can neglect any $B$-mode polarization, and the $\chisqeff$ including polarization (and noise) is then given by

\begin{equation}
 \chi_{\text{eff}}^{2} = \sum^{\lmax}_{\lmin} (2l+1)\left(\Tr(\hat{\mC_{l}}\mC_{l}^{-1}) + \ln\left(\frac{|\mC_{l}|}{|\hat{\mC_{l}}|}\right) - 2\right)
 \quad \text{where} \quad \mC_{l} = \begin{bmatrix}
                                          C^{TT}_{l} & C^{TE}_{l} \\
                                          C^{TE}_{l} & C^{EE}_{l} \\
                                         \end{bmatrix} .
\label{MatrixChisq}
\end{equation}
Again note that this has been normalized so that any degenerate model exactly equal to the fiducial model $\hat{\mC}_l=\mC^{\rm fid}_l$ will return a value of \(\chi_{\text{eff}}^{2} = 0\).

\begin{figure}[h!]
 \includegraphics[width = 13cm]{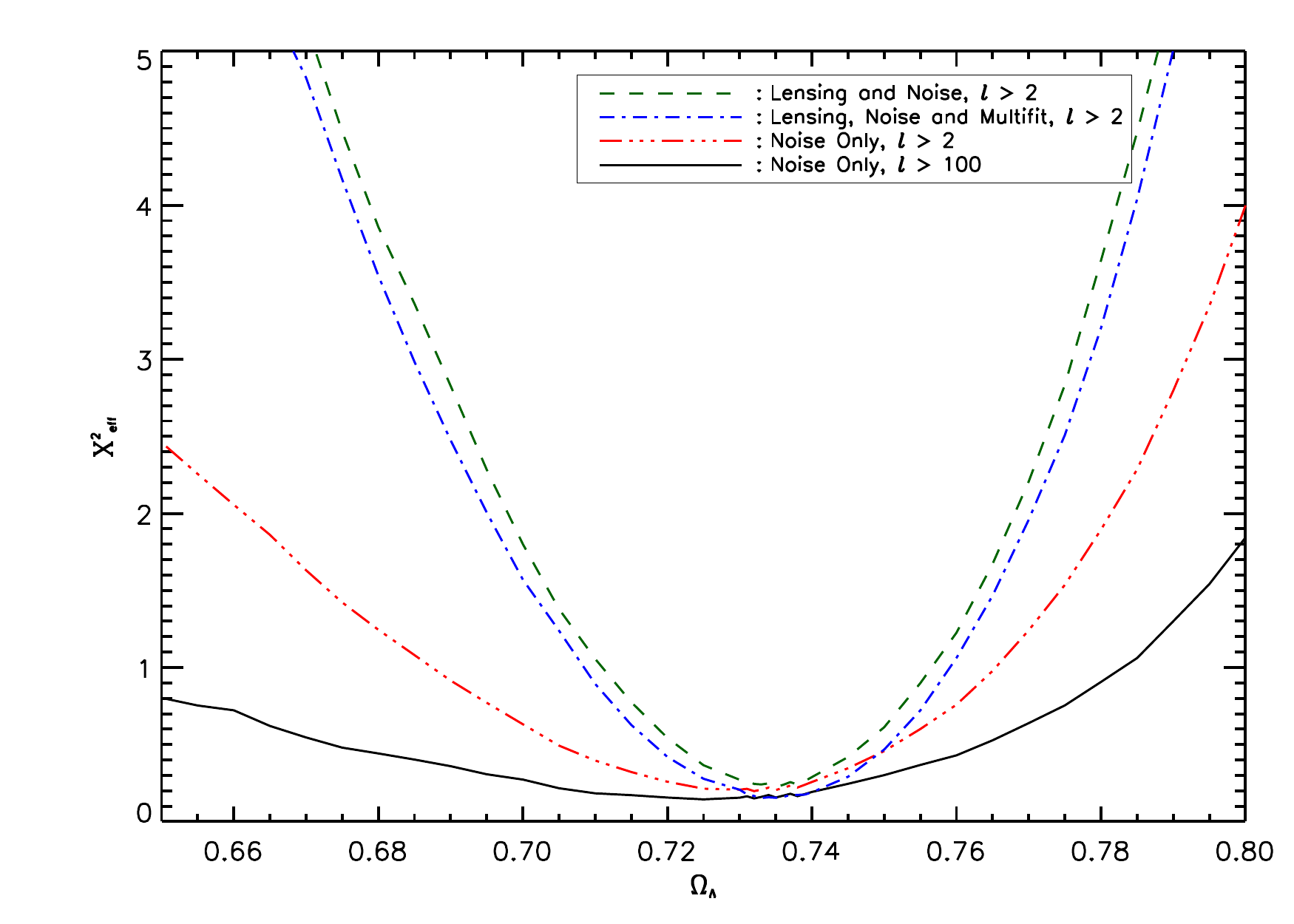}
 \caption{Minimum effective chi-squared values as a function of $\Omega_\Lambda$. We consider unlensed spectra with noise, using $l>2$ (red triple-dot-dashed) or only $l>100$ (black solid), and lensed spectra with noise using $l>2$.
With lensing included, we consider both a simple minimisation with respect to
$h$ (green dashed) and a standard six-parameter fit (Multifit) minimising with
respect to \(h\), \(\Omega_{b}h^{2}\), \(\Omega_{c}h^{2}\), the spectral index, \(n_{s}\), the optical depth, \(\tau\), and \(A_{s}e^{-2\tau}\) (blue dot-dashed).
%Effect on the lensed spectra for the non-flat geometric degeneracy when we include noise and when we do a standard 6-parameter fit (Multifit), i.e. setting a value for \(\Omega_{\Lambda}\) then minimizing with respect to \(H_{0}\), \(\Omega_{b}h^{2}\), \(\Omega_{c}h^{2}\), the spectral index, \(n_{s}\), the optical depth, \(\tau\), and \(A_{s}e^{-2\tau}\), (we minimize over this function rather than just the scalar amplitude \(A_{s}\) as it allows us to constrain both \(A_{s}\) and \(\tau\) better).
 \label{NonFlatLensNoiseChisq}
 }
\end{figure}
\begin{figure}[h!]
 \includegraphics[width = 10cm]{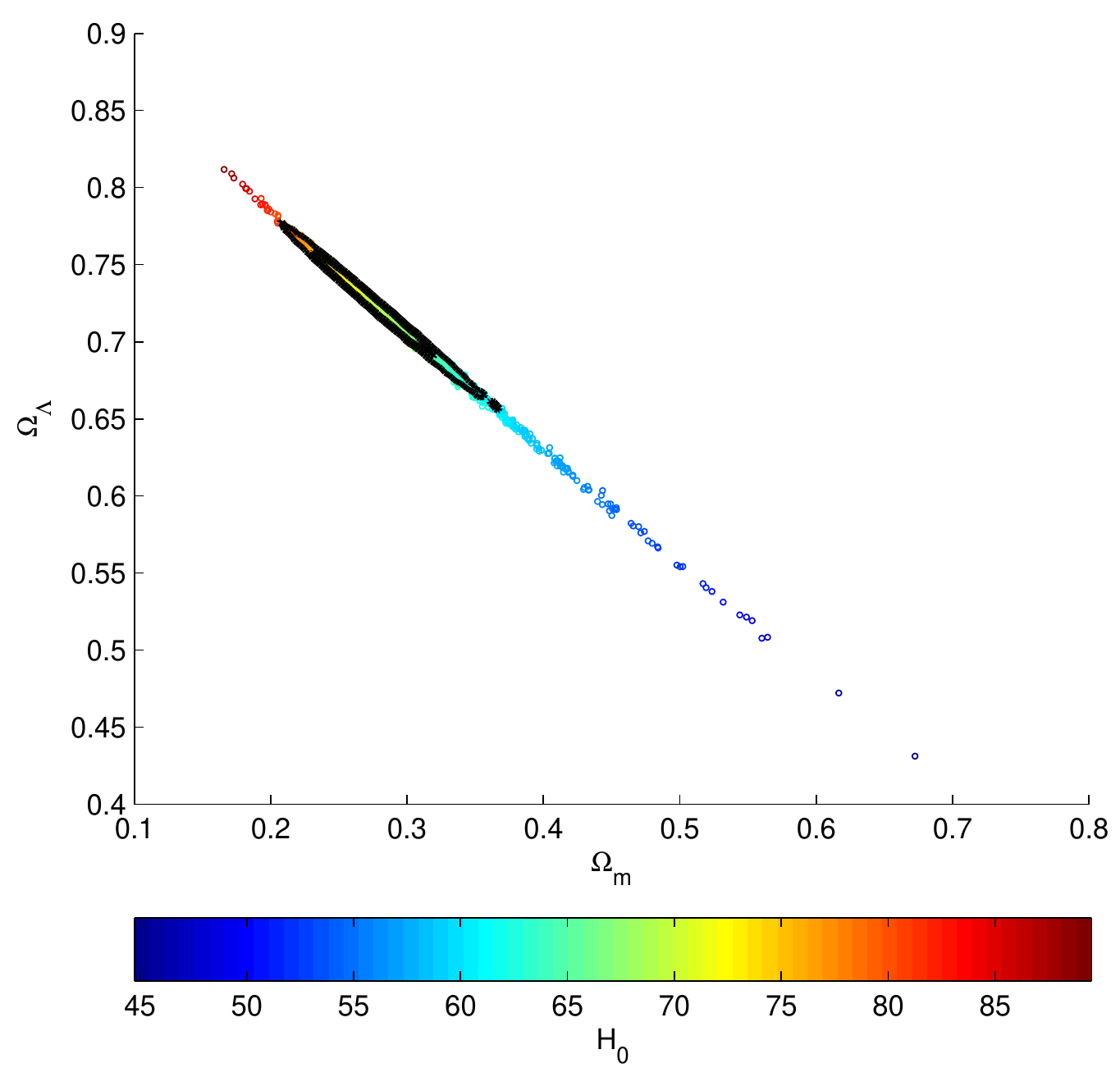}
 \caption{Idealized forecast parameter constraints from Planck power spectra only, varying $A_s, n_s, \Omega_b h^2, \Omega_c h^2, \tau, \theta,$ and $\Omega_K$ with flat priors, and  $H_0$, $\Omega_m$ and $\Omega_\Lambda$ being derived parameters. Points show samples from the expected posterior if the unlensed power spectra were observed, the black contours the better constraint obtainable in reality accounting for power spectrum lensing. Adding lensing reconstruction information could further shrink the extent of the degeneracy by a factor of roughly two.
 \label{NonFlatMCMC}
  }
\end{figure}

Once noise is included, constraints become significantly weaker, but this is compensated by lensing degeneracy breaking.
%, as illustrated in Fig.~\ref{ChiSqWithNoise}.
Maximum likelihood curves are shown in Fig.~\ref{NonFlatLensNoiseChisq},
where we now also vary other cosmological parameters, as small changes in these may be able to compensate partly degeneracy-breaking due to lensing and other physical effects: in addition to $h$ we minimize over \(\Omega_{b}h^{2}, \Omega_{c}h^{2}, A_{s}, n_{s}\) and \(\tau\) for each value of $\Omega_\Lambda$.
As expected, even in the presence of noise the effect of lensing is enough to break the degeneracy and constrain $\Omega_\Lambda$ well. When we use a six-parameter fit the breaking of the degeneracy is lessened somewhat, though not substantially, indicating that the lensing effect cannot easily be mimicked by other changes of parameters (at least if dark energy is assumed to be a cosmological constant).

\begin{figure}[h!]
 \includegraphics[width = 9.5cm]{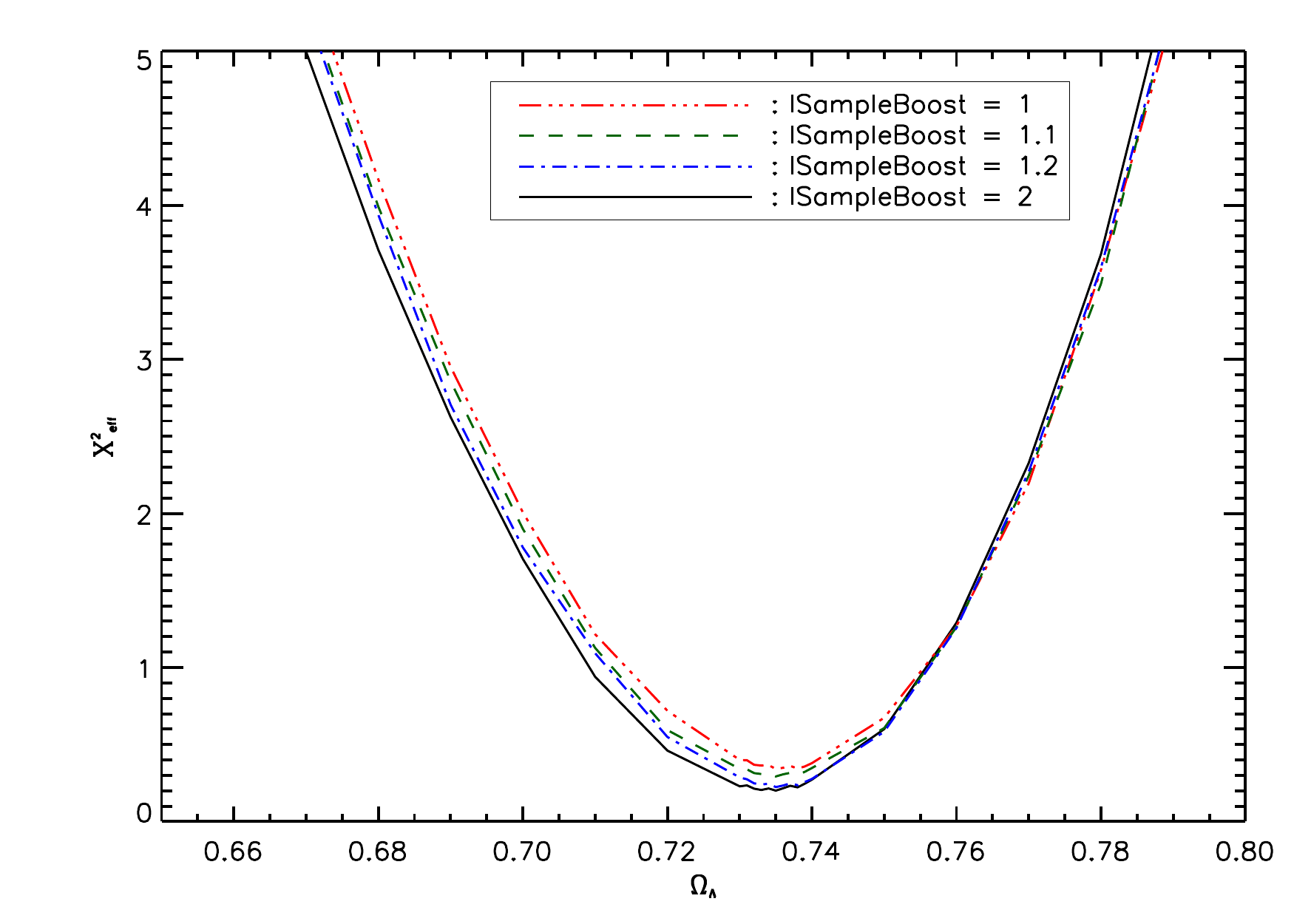}
 \caption{Minimum effective chi-squared values for non-flat geometries, in the presence of both lensing and noise, using high accuracy and varying the value of \lSampleBoost. Note that here we use all $l>2$ to make this figure comparable to Fig.~\ref{NonFlatLensNoiseChisq}.
 \label{NonFlatChisqlsamp}
 }
\end{figure}

We also see from Fig.~\ref{NonFlatLensNoiseChisq} that the late-time ISW signal in the low-$l$ anisotropies can break the degeneracy significantly, though the lensing information makes the constraint about  a factor of two tighter. This is shown further in Fig.~\ref{NonFlatMCMC} for a full MCMC parameter analysis using {\COSMOMC}~\cite{Lewis:2002ah} from the same likelihood but with four times larger noise (as might be nearer to reality if only one frequency is used).

We finally assess the impact of the numerical errors we found previously, in particular the small shifts in the inferred value of $\Omega_\Lambda$ when different values of \lSampleBoost\ were used. Figure~\ref{NonFlatChisqlsamp} shows the likelihood curves for various values of \lSampleBoost\  with both noise and lensing included. High accuracy with default settings is slightly shifted compared to a more accurate calculation, though the shift is not large compared to the overall degeneracy width. For a more accurate calculation \lSampleBoost\ $\agt 1.2$ could be used.

\section{Geometrical degeneracy in a flat universe with a more general dark energy}
\label{sec:DE}

\begin{figure}[h!]
 \includegraphics[width = 8.5cm]{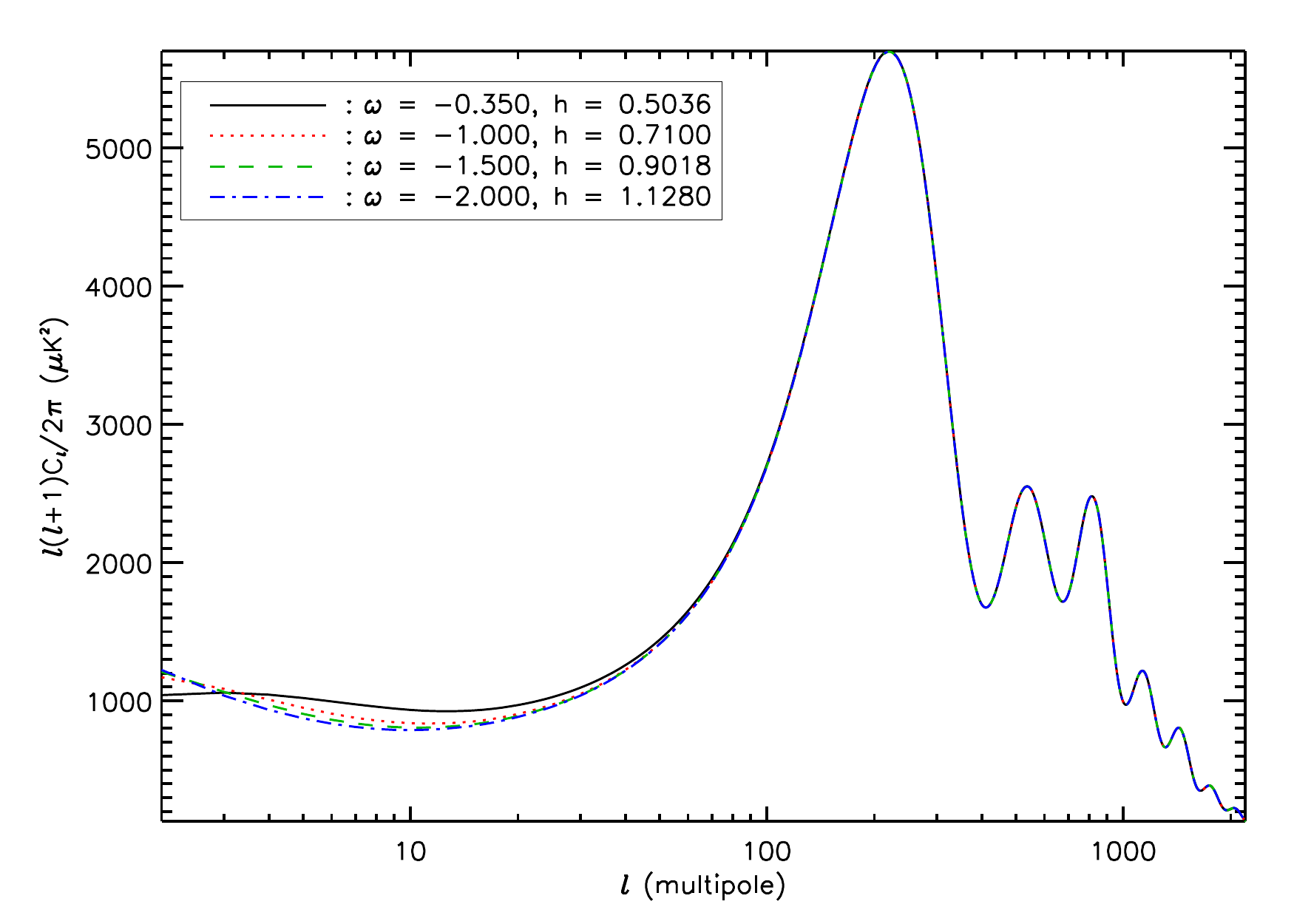}
 \includegraphics[width = 8.5cm]{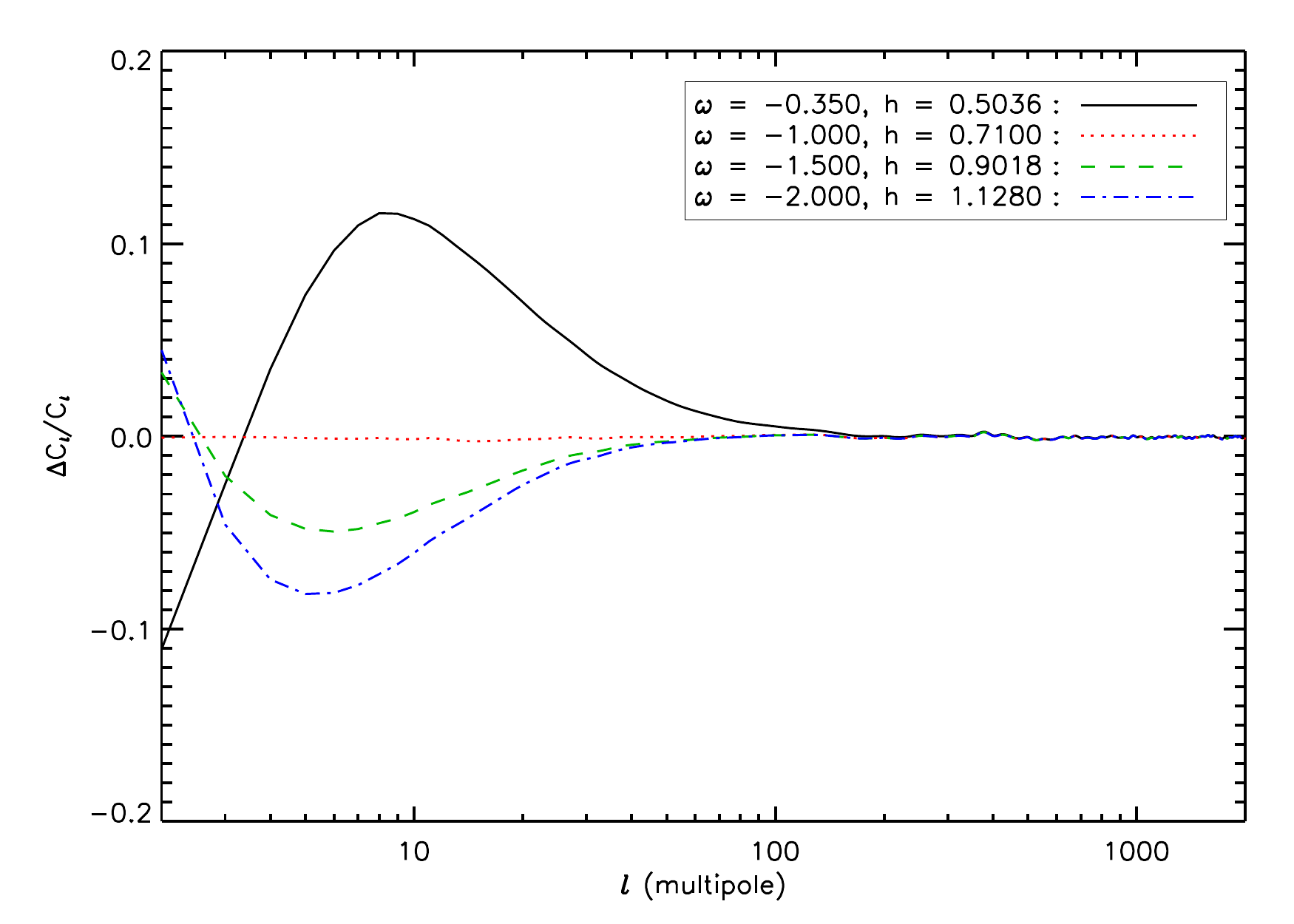}
 \caption{Geometric degeneracy in flat models with $w \neq -1$. Unlensed CMB power spectra for several different nearly degenerate flat models are plotted (left), along with their fractional differences with respect to the fiducial model (right). The fiducial model is the same as used in Sec.~\ref{sec:geo} (and
so has $w=-1$).
 \label{FlatDarkEnergyCL}
 }
\end{figure}

In this section we investigate the geometrical degeneracy in a flat universe when there is dark energy parameterized by a constant equation of state parameter $w\equiv P/\rho$ (we restrict to \(w < -1/3\)). We take a fiducial value $w=-1$ corresponding to a cosmological constant.
Figure~\ref{FlatDarkEnergyCL} shows four different nearly degenerate models obtained by varying $w$ and $h$.
Very different values for $w$ can still produce nearly identical CMB anisotropies: the degeneracy is only significantly broken at low \(l\) by the late-time ISW effect.
The flat-model dark energy degeneracy is more exact in the unlensed CMB than the non-flat case, and hence provides a stringent test of numerical accuracy of {\CAMB} as the geometry is varied (though a much less sensitive test of errors in the pre-recombination physics).
We focus on the high-$l$ regime as an accuracy test, where differences in the spectrum are well below a percent.

\subsection{Degeneracy breaking effects}

\subsubsection{Numerical Accuracy}

%put in with previous
%\begin{figure}[h!]
% \includegraphics[width = 10cm]{CAMB_wlam_Cls_subtract_lowacc_allparams1.pdf}
% \caption{A plot showing the fractional difference between the power spectra of several flat-model dark energy degenerate models and the fiducial model. Low accuracy was used here, but differences on large-scales are dominated by the significantly different ISW contributions to the power spectrum.
% \label{FlatDarkEnergyClDiff}
% }
%\end{figure}

%Fig.~\ref{FlatDarkEnergyClDiff} shows the differences between the power spectra of a degenerate model and that of the fiducial model. At high \(l\) (greater than 100) we see only small differences less than 1\%, dominated by numerical inaccuracies. At low \(l\) (less than 100) the ISW effect changes the $C_l$ at the \(\sim\)10\%-level. We focus on the high-$l$ regime as an accuracy test.

\begin{figure}[h!]
 \vspace{-160pt}
 \includegraphics[width = 18cm]{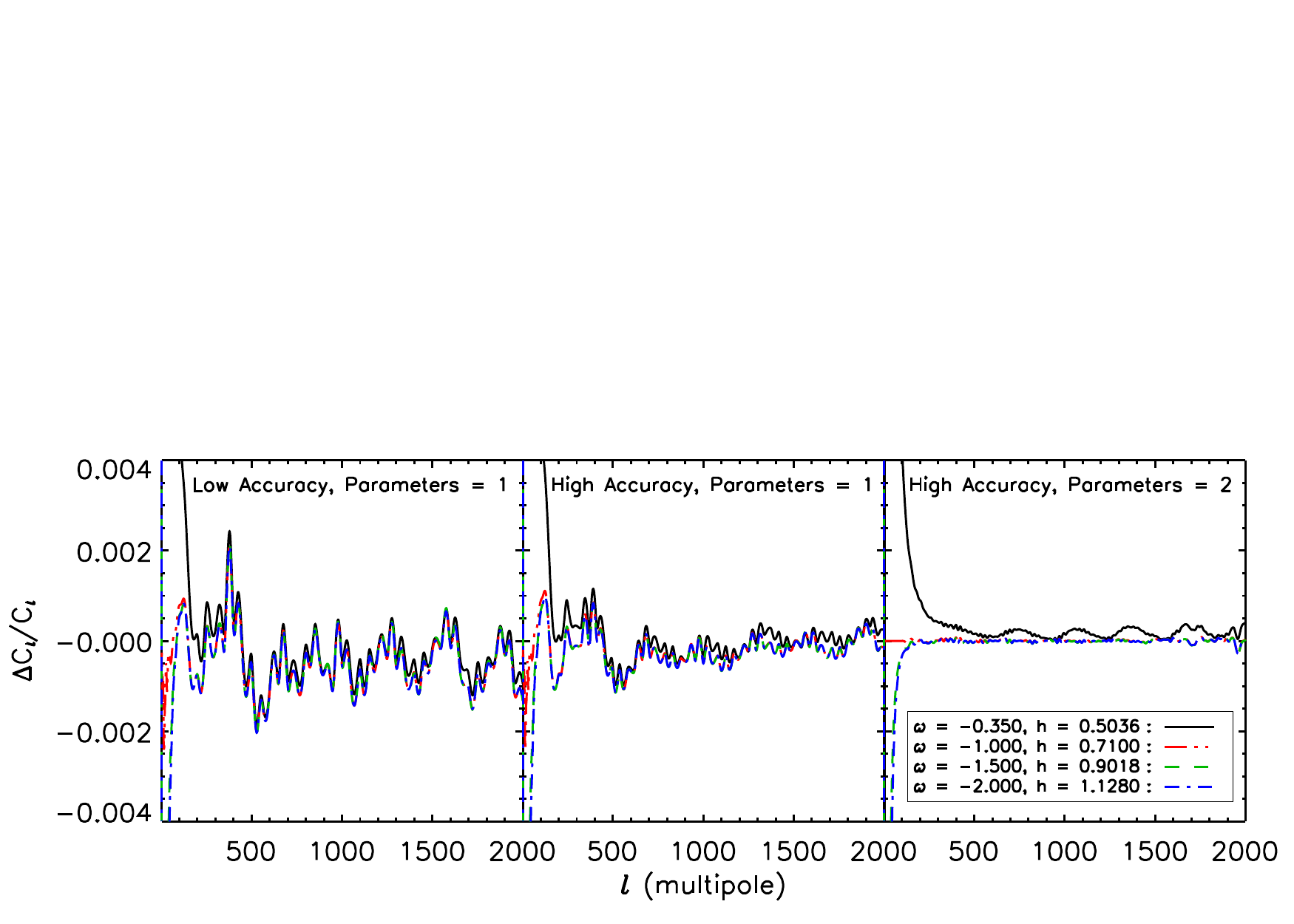}
 \caption{Fractional differences between unlensed power spectra and the fiducial model for flat dark energy models. The nearly-degenerate models are computed at low accuracy with all accuracy parameters set to 1 (left), and at high accuracy with all accuracy parameters set to 1 (middle) and 2 (right). With sufficient numerical accuracy, the high-$l$ unlensed power spectra are essentially identical for different flat dark energy models unless $w$ is very close to $-1/3$.
 \label{FlatDarkEnergyBoosts}
 }
\end{figure}

Figure~\ref{FlatDarkEnergyBoosts} shows the change in the unlensed power spectrum for different degenerate models at various accuracy settings. As expected we see some small numerical wiggles at standard settings which subsequently disappear at boosted accuracy. As expected there is hardly any breaking of the degeneracy due to physical effects at \(l > 100\). The degeneracy is much more accurate than in the non-flat case (with $w=-1$) because the physical effect of the dark energy is negligible in the early universe, and the distance to last-scattering and geometry is now fixed, so there is no change in the geometrical effect from averaging through last-scattering.  %the dark energy density modelled by a cosmological constant has no dependence on the scale factor \(a\), but the curvature and matter densities are proportional to \(\frac{1}{a^{2}}\) and \(\frac{1}{a^{3}}\) respectively. Hence the ratio of dark energy to curvature at the time of recombination, where the scale factor is \(a \sim \frac{1}{1000}\), is $\clo(\e{-6})$, so dark energy effects are negligible in comparison. For a non-zero curvature, the ratio of curvature to matter at the time of recombination is on the order of 1\e{-3}, which, while small, still has some effect, causing the small physical breaking of the degeneracy for open and closed models that we observed in the previous section, even in the absence of lensing and ISW.

At low accuracy, the numerical errors are, on the whole, less than the quoted accuracy of 0.3\%. There is a small `drift' at low accuracy; however, even including this drift, the errors are still generally significantly below 0.3\% at high $l$. The small `drift' disappears with high accuracy calculations.

\begin{figure}[h!]
 \subfloat[Original]{\includegraphics[width = 0.5\textwidth]{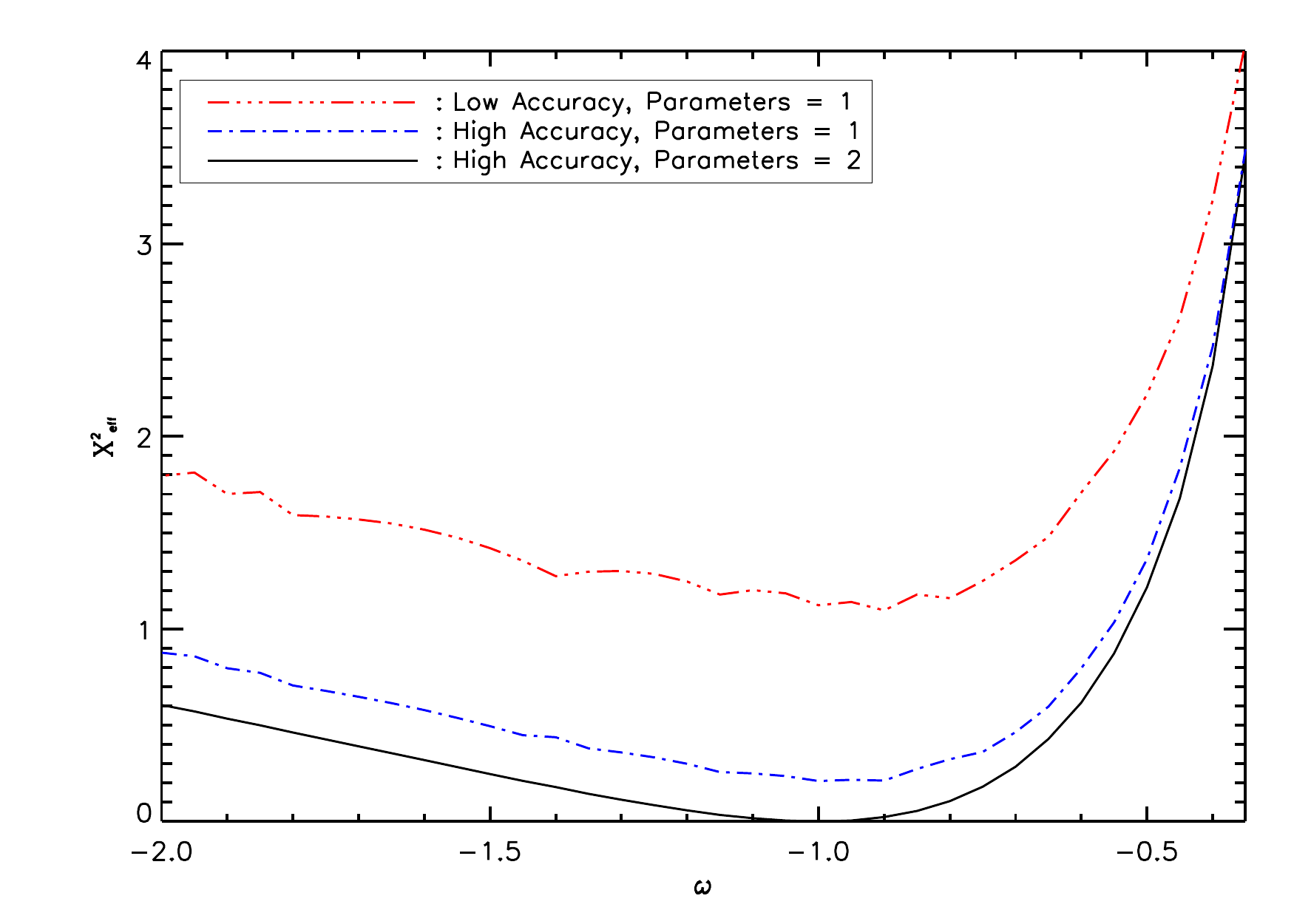}}
 \subfloat[Normalized]{\includegraphics[width = 0.5\textwidth]{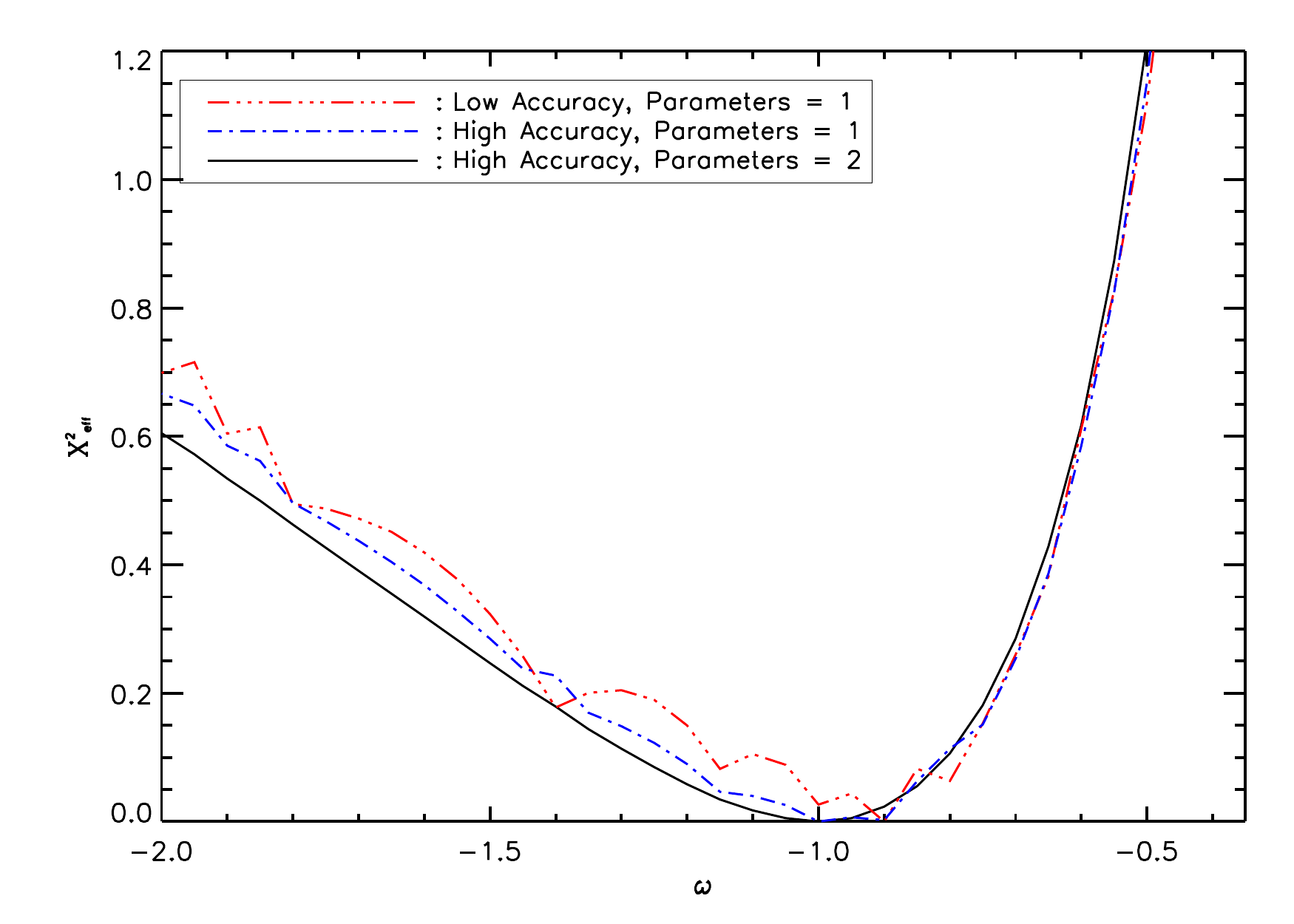}}
 \caption{\emph{Left}: Minimum effective chi-squared values in the $w$-$h$ space for flat
dark energy models, using all $l\geq 2$. The models are computed at low accuracy with accuracy parameters set to 1 (red, triple-dot-dashed), and at high-accuracy with accuracy parameters set to 1 (blue, dot-dashed) and 2 (black, solid). In all cases, the fiducial model has $w=-1$. \emph{Right}: Renormalized chi-squared curves so that the global minimum value is equal to zero.
 \label{FlatDarkEneryChisq}
 }
\end{figure}

We can quantify how much these numerical errors affect the $w$-$h$ degeneracy by looking at the minimum effective chi-squared values obtained in this space for various settings of the accuracy parameters.
Figure~\ref{FlatDarkEneryChisq} plots the minimum effective chi-squared in the $w$-$h$ space for $-2.0 < w < -0.35$ at both standard low and high accuracy settings as well as boosted accuracy.  For standard accuracy there is an offset, in part because of the small numerical `wiggles' shown in Fig.~\ref{FlatDarkEnergyBoosts}. However an offset is usually harmless as it does not affect relative parameter constraints, as demonstrated in the figure by renormalizing the curves to have global minimum values equal to zero.
At low accuracy the curves are not very smooth also due to numerical effects, though only at a low level that goes away when high accuracy is used. The small residual shift in shape compared to boosted accuracy settings can be mostly eliminated as in the previous section by a slight increase of \lSampleBoost\ to $1.1$--$1.2$.
%These results are only using the temperature \(C_{l}\)'s; if polarization information is also included the difference in optical depths between degenerate models can cause a much larger deviation in the chi-squared values for different accuracy settings than that seen here.

\subsubsection{Effect of noise and lensing}

%\begin{figure}[h!]
% \includegraphics[width = 0.5\textwidth]{CAMB_wlam_omega_lambda_likelihood_lsamptest_lensed_planck_new.pdf}
% \caption{A plot showing the effect of increasing \lSampleBoost\ at high accuracy on the minimum chi-squared values for the flat-model dark energy degeneracy when lensing and noise are included.
% \label{FlatDarkEneryChisqlsamp}
% }
%\end{figure}

%From Fig.~\ref{FlatDarkEneryChisqlsamp} it was decided that high accuracy calculations with an \lSampleBoost\ of 1.17, the same as the optimum settings in the previous section, were still accurate enough to allow us to calculate the power spectra and effective chi-squared values reliably, in the presence of Planck-like noise.

\begin{figure}[h!]
 \includegraphics[width = 12cm]{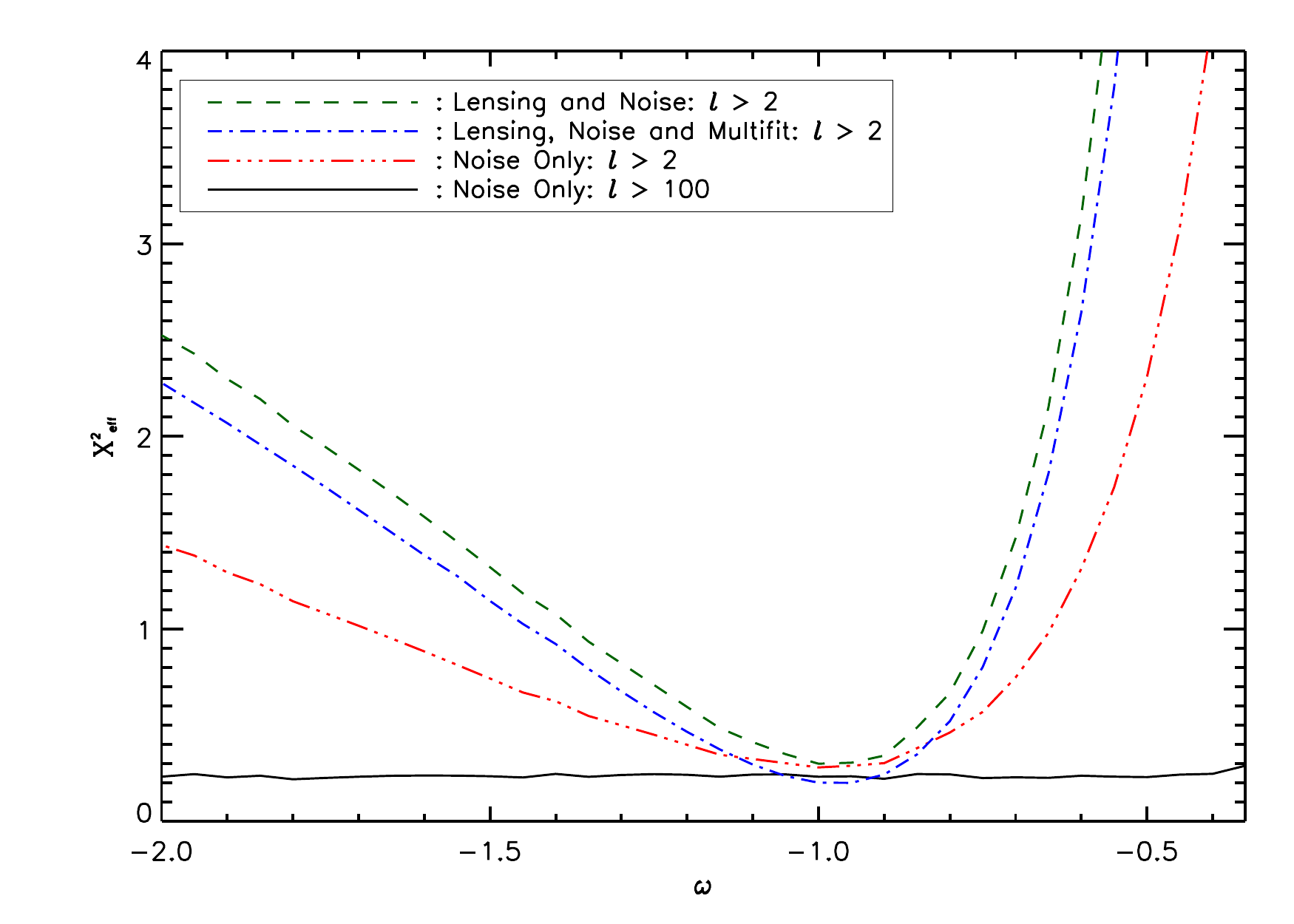}
 \caption{Minimum effective chi-squared values as a function of $w$ in flat
dark energy models.
We consider unlensed spectra with Planck-like noise, using $l\geq 2$ (red triple-dot-dashed) or only $l>100$ (black solid), and lensed spectra with noise using $l\geq 2$.
With lensing included, we consider both a simple minimisation with respect to
$h$ (green dashed) and a standard six-parameter fit (Multifit) minimising with
respect to \(h\), \(\Omega_{b}h^{2}\), \(\Omega_{c}h^{2}\), the spectral index, \(n_{s}\), the optical depth, \(\tau\), and \(A_{s}e^{-2\tau}\) (blue dot-dashed).
%A plot of the minimum effective chi-squared values for each flat $w-H_{0}$ degenerate model, showing the different physical phenomena that can affect the degeneracy. We have plotted versions of the same graph when lensing is used and when it is not, when we use a 6 parameter fit (Multifit) and when we use only high \(l\) as opposed to all \(l\). In all cases Planck-like noise was included.
 \label{FlatDarkEnergyChisqLensedNoise}
 }
\end{figure}

We now use temperature and $E$-mode polarization, as in Eq.~\ref{MatrixChisq}, to explore how lensing, noise and performing a six-parameter fit affect the geometric degeneracy in flat models with general dark energy ($w \neq -1$); see
Fig.~\ref{FlatDarkEnergyChisqLensedNoise}. There is weak degeneracy-breaking at low \(l\) due to the late-time ISW effect, as well as additional breaking from the lensing. In the absence of lensing and low \(l\) information we see that the degeneracy is almost exact.  Finally we can see that fitting \(\Omega_{b}h^{2}, \Omega_{c}h^{2}, A_{s}e^{-2\tau}, n_{s}\) and \(\tau\) as well as \(h\), when using lensing, gives us a slightly better minimization, but the effect is small so that the allowed changes in other parameters cannot effectively mimic the effect of lensing on the power spectrum.

\section{Acoustic-scale degeneracy in a flat {\LCDM} universe}
\label{sec:approxdeg}

\begin{figure}[h!]
 \includegraphics[width = 12cm]{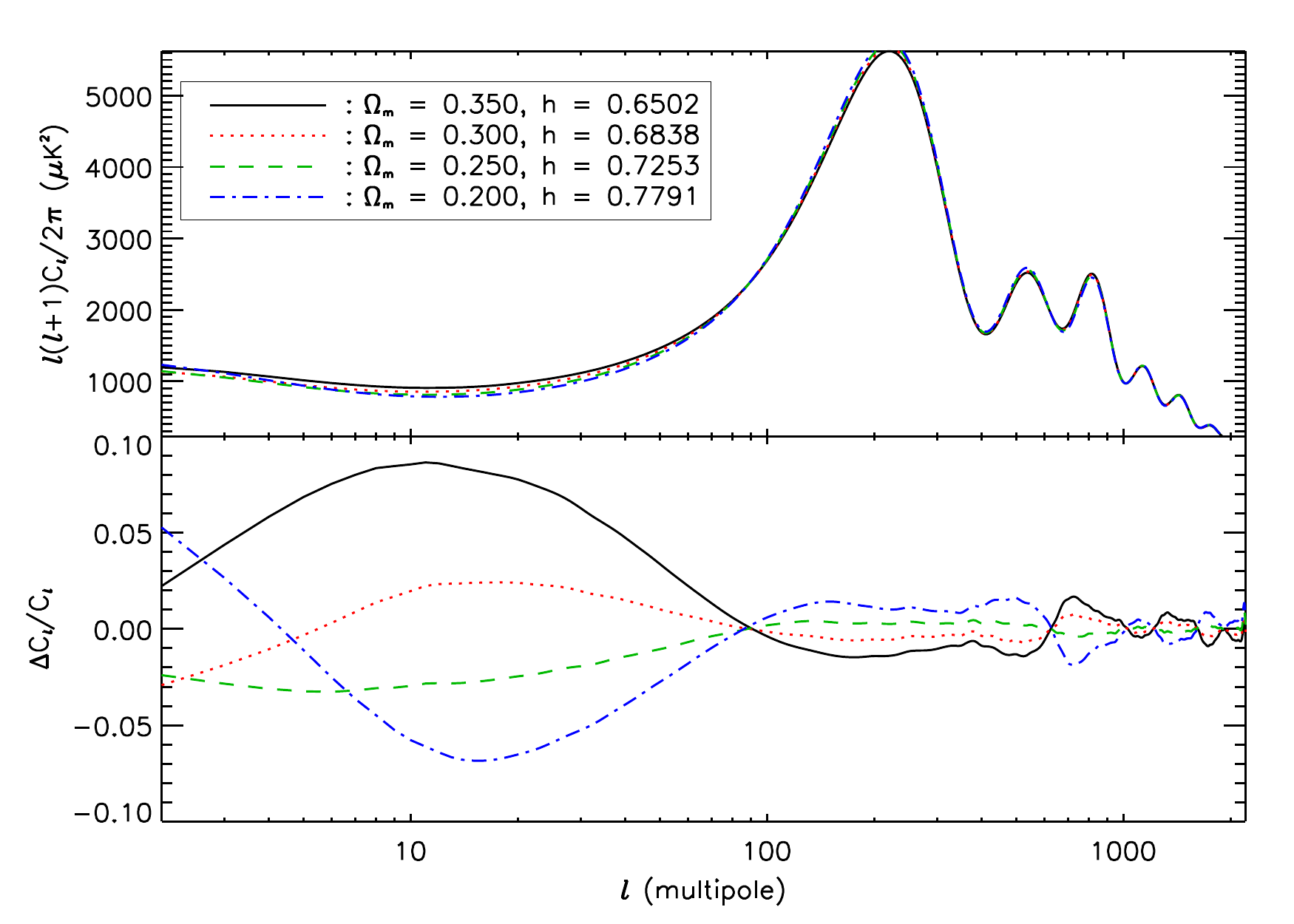}
 \caption{\emph{Top}: Power spectra for four different flat {\LCDM} models that are all nearly degenerate with the fiducial model. \emph{Bottom}: Fractional difference between these degenerate models and the fiducial model.
 %Standard low accuracy settings were used.
 \label{FlatOmegamCl}
 }
\end{figure}

In this section we investigate one further CMB parameter degeneracy, mostly involving \(\Omega_{m}\) and \(H_{0}\), in flat {\LCDM} models. This is not a purely geometric degeneracy --- it involves multiple parameter variables --- but is an important source of uncertainty in individual parameter constraints from the CMB alone~\cite{Zaldarriaga:1997ch,Percival:2002gq}. Depending on the data used, nearly-degenerate models have $\Omega_m h^3$ approximately constant. The exact direction of the degeneracy varies slightly with $\lmax$ and noise under consideration, but degenerate models are such that the observed angular scale of the acoustic peaks is nearly constant. In a flat {\LCDM} model this can only be achieved by changing both the sound horizon at recombination and the angular diameter distance. However changing the sound horizon also involves changes in matter densities, which have other effects on the observed power spectrum: the degeneracy only exists to the extent that changes in other parameters can compensate these effects within the limits of cosmic variance and observational noise.

For all the results given in this section we maximize the likelihood over \(\Omega_{b}h^{2}, \Omega_{c}h^{2}, A_{s}e^{-2\tau}, n_{s}\) and \(\tau\). We use the same fiducial model as before (Table~\ref{ParamsTable}) which has \(\Omega_{m} = 0.267\). For each value of $\Omega_m$ we minimize $\chisqeff$ with respect to the other parameters.

\begin{figure}[h!]
 \vspace{-160pt}
 \includegraphics[width = 18cm]{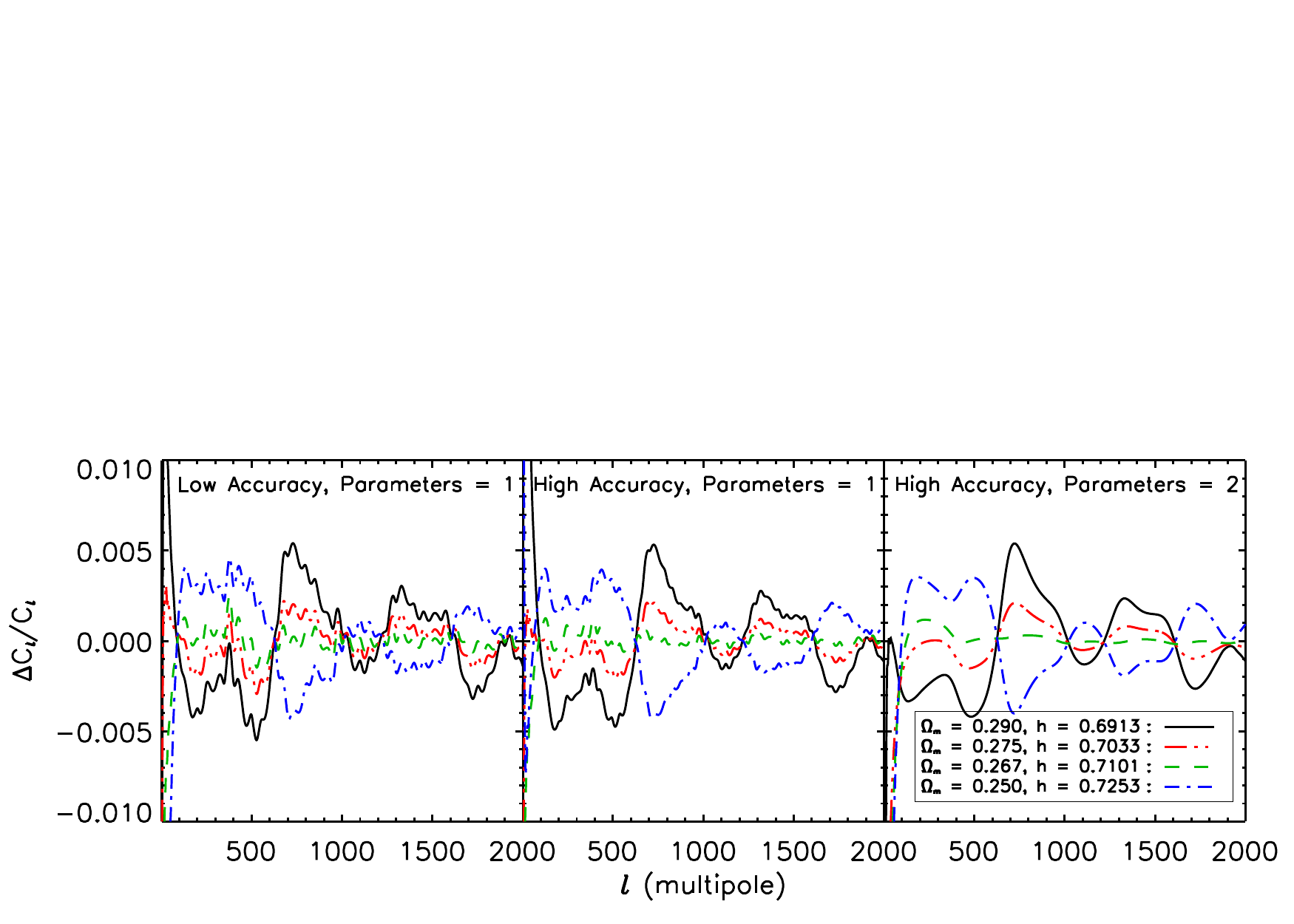}
 \caption{Fractional differences between the CMB power spectra of nearly-degenerate models and the fiducial model. The former are calculated using low accuracy and default accuracy parameters (left), high accuracy and default parameter accuracy parameters (middle) and high accuracy and boosted accuracy parameters (right). The right-hand panel leaves mainly the physical differences between the models.
 \label{FlatOmegamBoosts}
 }
\end{figure}

\begin{figure}[h!]
 \includegraphics[width = 14cm]{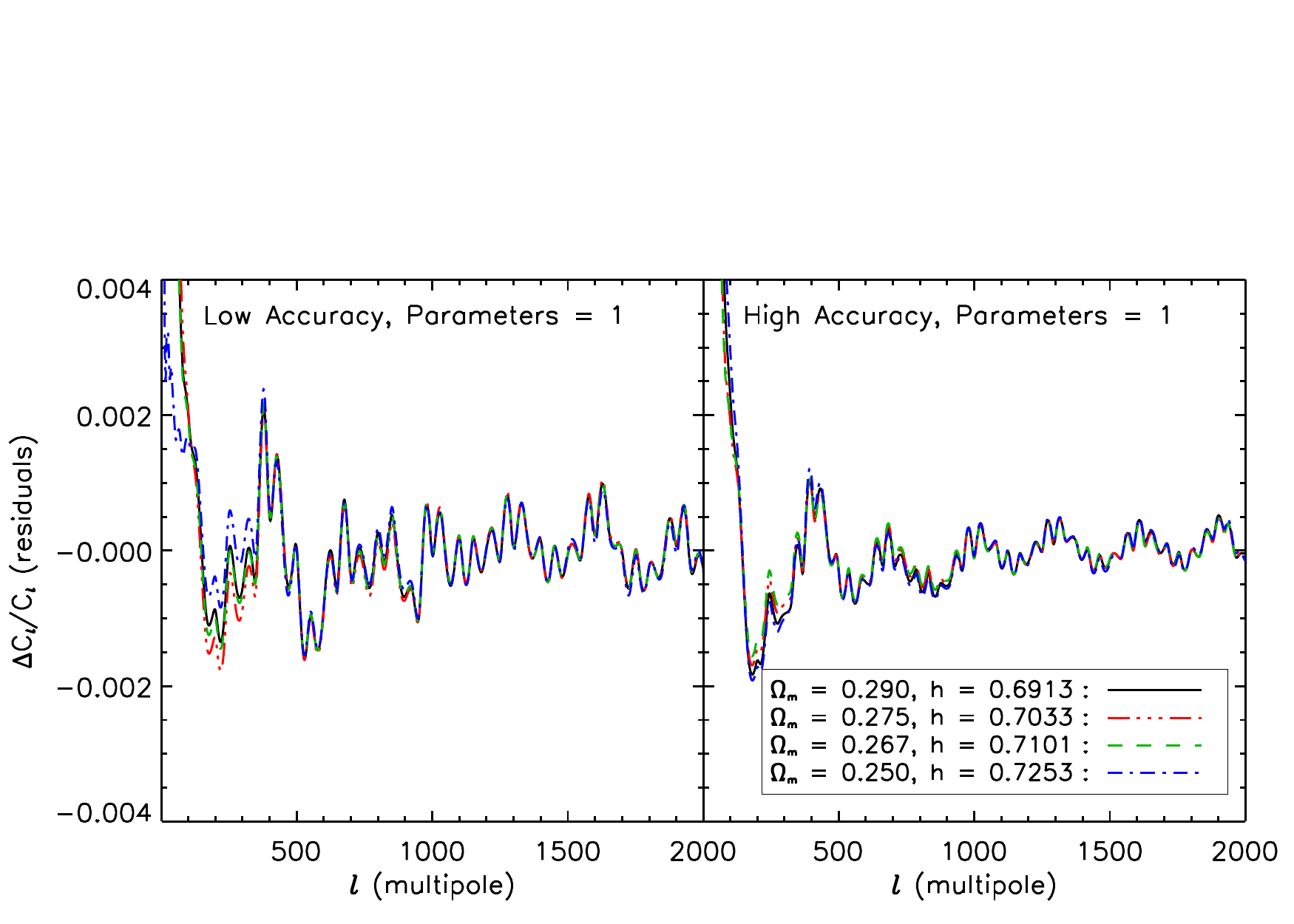}
 \caption{Numerical errors in the computation of flat {\LCDM} models that are nearly degenerate with the fiducial model. For a given model, the numerical errors are estimated by subtracting the spectrum from one calculated at high accuracy with accuracy parameters boosted to 2. Errors are plotted for a low (left) and high (right) accuracy calculation. In all cases, the other accuracy parameters are at their default values.
 \label{FlatOmegamBoostsDoubleDiff}
 }
\end{figure}

\begin{figure}[h!]
 \includegraphics[width = 8.5cm]{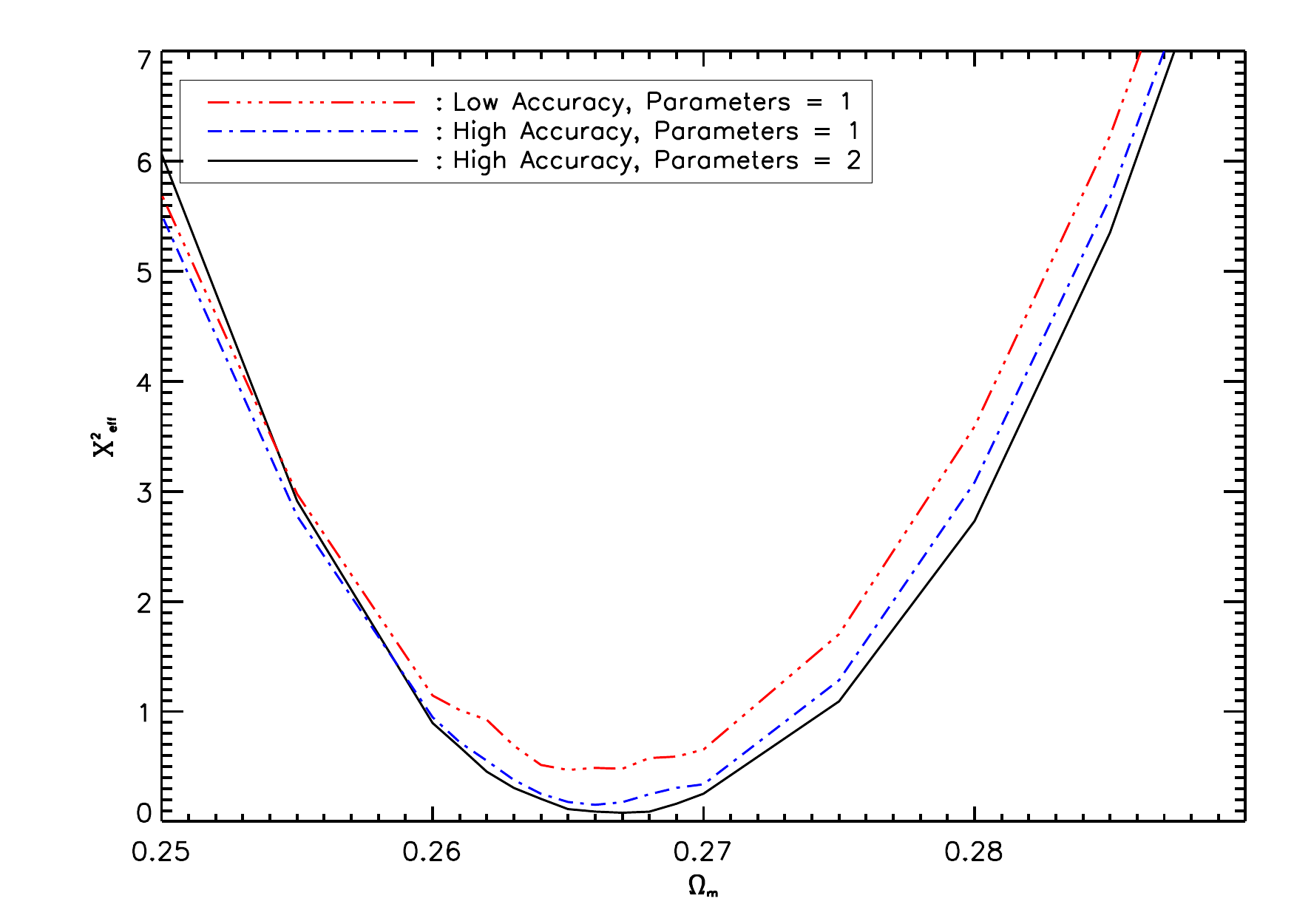}
 \includegraphics[width = 8.5cm]{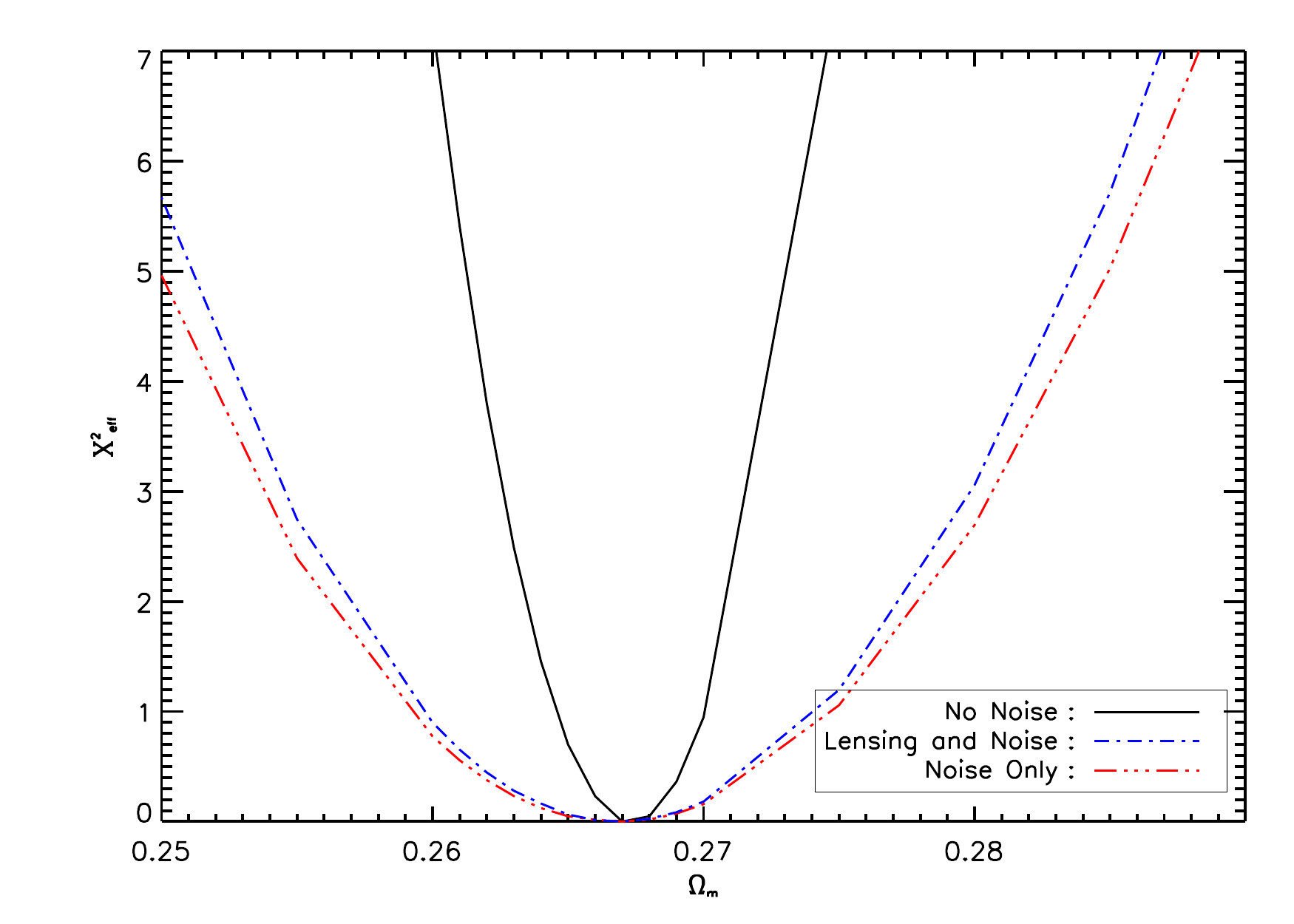}
 %\caption{Likelihood curves for the flat \LCDM\ model looking at the effects of lensing and noise for degenerate models close to the fiducial model, \(\Omega_{m}= 0.267\), in the presence of Planck-like noise. We have normalized the minimum chi-squared value to 0.
 %\label{FlatOmegamChisqLensedNoiseZoom}
 \caption{\emph{Left:} Minimum effective chi-squared values as a function of $\Omega_m$ in flat {\LCDM} models close to the fiducial model. The models are computed at low accuracy with default accuracy parameters (red, dot-dashed), at high accuracy with default accuracy parameters (blue, dot-dashed) and at high accuracy with accuracy parameters boosted to 2 (black, solid). In each case, only temperature data is used and there is no instrument noise. \emph{Right}: Impact of including polarization data with no noise (black, solid), with Planck-like noise (red, triple-dot-dashed) and with lensing and noise (blue, dot-dashed). The models are computed at high accuracy with \lSampleBoost\ increased to $1.17$.
% Results including Planck-like noise are similar.
 %Left is for noise-free temperature, right includes Planck-like noise and polarization. The right figure also shows high accuracy with \lSampleBoost\ increased to 1.17, which removes a good fraction of the small residual shift in likelihood shape.
 \label{FlatOmegamChisq}
 }
\end{figure}

%\subsection{Numerical errors and likelihood shape}

Figure~\ref{FlatOmegamCl} shows the power spectra of four nearly degenerate models, spanning a wide range of \(\Omega_{m}\).
The largest fractional differences are at low \(l\) where the late-time ISW effect breaks the degeneracy due to the different late-time expansion histories. However, there are also significant differences between the spectra on small scales so with precision data these models can be relatively easily distinguished. However the approximate degeneracy will remain over a smaller range of parameter values.

\begin{figure}[h!]
 \includegraphics[width = 10.5cm]{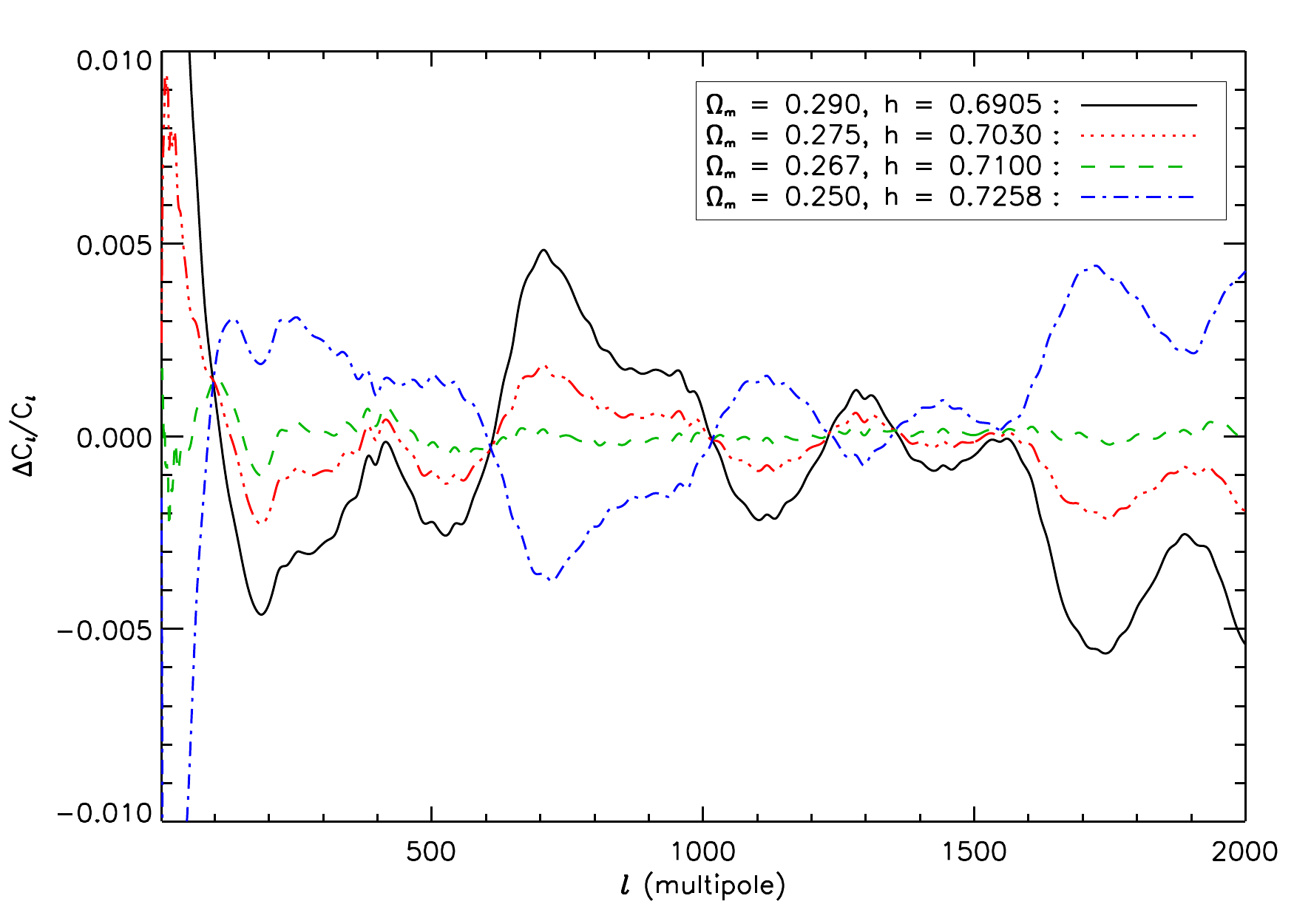}
 \caption{Effect of noise in locating nearly-degenerate flat {\LCDM} models by chi-squared minimisation. Fractional differences from the fiducial model are plotted for four nearly degenerate models; the latter are computed at high accuracy and with \lSampleBoost\ = 1.17 to remove most of the residual interpolation wiggles. Compared to the models in Fig.~\ref{FlatOmegamBoosts}, larger differences in the spectra are allowed at high $l$ where noise dominates.
 \label{FlatOmegamNoiseClDiff}
 }
\end{figure}

%We can see some obvious differences between these degenerate power spectra and the degenerate power spectra in the previous sections.
Figure~\ref{FlatOmegamBoosts} shows the differences between the power spectra and the fiducial model for nearly-degenerate models over a small range of $\Omega_m$
about the fiducial value. The differences seen in high accuracy calculations with boosted accuracy parameters are due to physical effects. Unlike in previous sections, the differences vary in a complicated way across the whole spectrum, stemming from differences in the relative acoustic peak amplitudes and tilt as all the parameters are changed (not quite managing to cancel completely the effect of the changing matter densities at recombination). There are of course also numerical artefacts which are clearly visible at low accuracy, but largely disappear when we use boosted high accuracy settings. Following our earlier treatment of geometric degeneracies, we can isolate the numerical effects by considering the difference between the model spectrum and one calculated at high accuracy with accuracy parameters boosted to 2. Figure~\ref{FlatOmegamBoostsDoubleDiff} shows that the numerical errors at intermediate and high $l$ are below the quoted accuracies of 0.3\% (low accuracy) and
0.1\% (high accuracy).

%As in the previous two sections it is then a relatively simple procedure to isolate the `wiggles' due to the numerical inaccuracies and check that they are below the quoted accuracies of 0.3\% and 0.1\%;  Fig.~\ref{FlatOmegamBoostsDoubleDiff} shows that the numerical effects are at or below the expected level as in the previous cases.

Although these numerical inaccuracies are within the quoted ranges, Fig.~\ref{FlatOmegamChisq} shows that the numerical errors can still give a very small shift in the likelihood. As with the similar results for the non-flat \(\Omega_{\Lambda}\)-\(h\) degeneracy, increasing \lSampleBoost\ slightly removes some of this effect. Including noise-free polarization data helps to break the degeneracy considerably but lensing has little effect. For Planck-like noise, the degeneracy is broadened slightly from the noise-free temperature-only case: noise allows somewhat larger differences in the power spectra at high $l$ where the noise becomes significant, as shown in Fig.~\ref{FlatOmegamNoiseClDiff}.

\begin{figure}[h!]
 \includegraphics[width = 9cm]{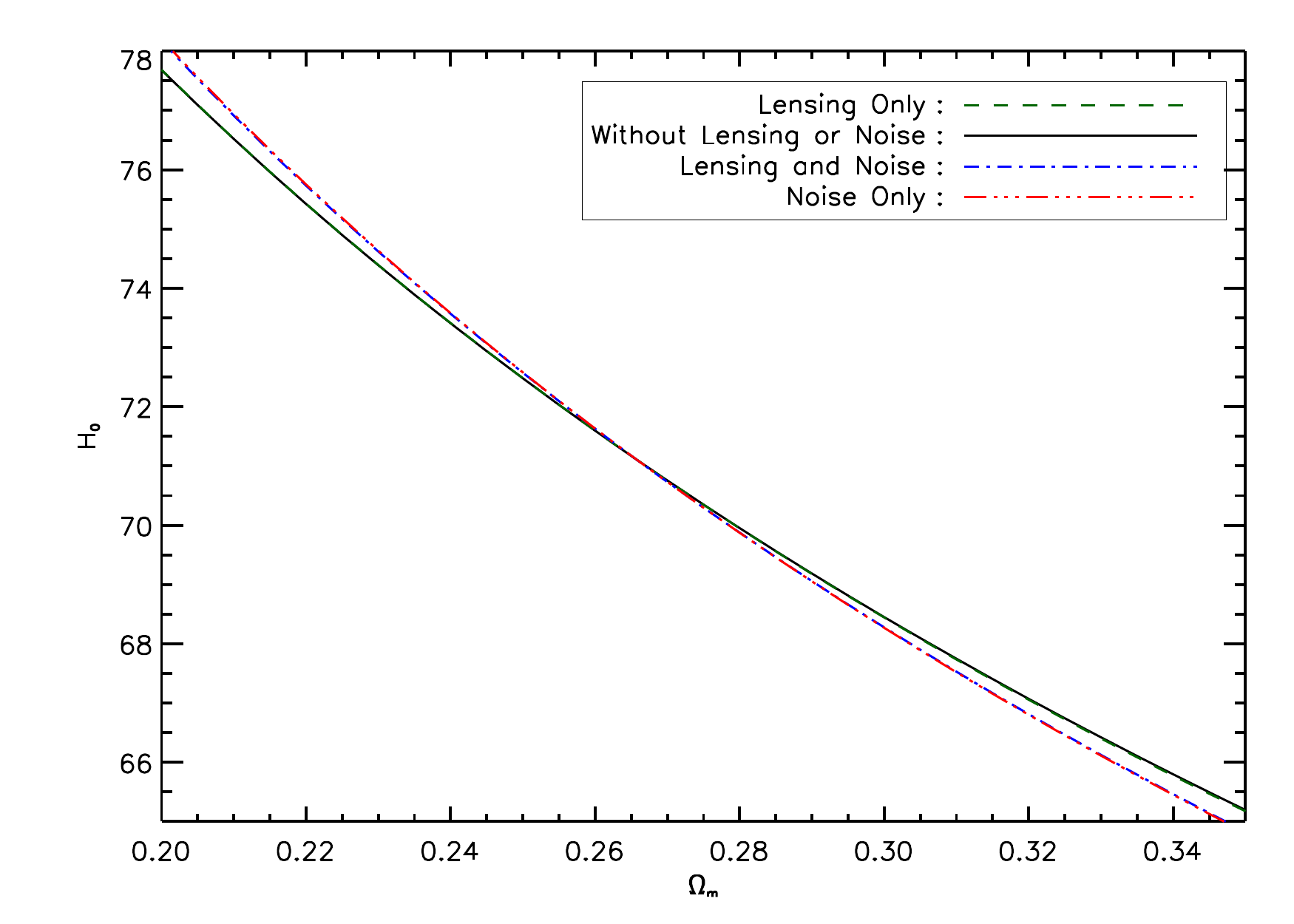}
 \caption{Best-fit values of \(H_{0}\) as a function of $\Omega_m$ without lensing or noise included (black, solid), with only lensing (green, dashed), with only noise (red, triple-dot-dashed), and with lensing and noise (blue, dot-dashed). Optimized accuracy settings were used for all calculations. The degeneracy is close to $\Omega_{m}h^{\alpha} \sim \text{const.}$, with $3\alt\alpha\alt3.1$ having slightly different values in the various cases.
 \label{FlatOmegamH0LensedNoise}
 }
\end{figure}

Although the degeneracy is quite tight, it does not extend over a large range of parameter values; as such including lensing barely has any effect on the chi-squared values returned for each model. In Fig.~\ref{FlatOmegamH0LensedNoise} we show how the best-fit value of \(H_{0}\) varies as a function of \(\Omega_{m}\). We can see that lensing has a negligible effect on the degeneracy direction because the parameters are so strongly constrained without it anyway. However, noise does slightly change the direction; this is expected since the relative importance of different physical effects changes as a function of $l$.
%Fig~\ref{FlatOmegamH0LensedNoisePercent} quantifies the magnitude of this difference, showing the percentage change in \(H_{0}\) as a function of \(\Omega_{\Lambda}\), for lensing, noise and a combination of both. We see that when we include noise, the difference between the calculated parameter set and the real parameter set increases as we move further away from the fiducial model. This is because whilst the degeneracy is still not strong enough to stop us defining the `true' model, the inclusion of noise does make it harder to distinguish between the degenerate models.

%\begin{figure}[h!]
% \includegraphics[width = 8.5cm]{CAMB_omegam_H0_optimum_lensed_6fit_planck_diff.pdf}
% \caption{A similar plot to that previously, but instead showing the percentage difference between the best fit \(H_{0}\) values for lensing, noise (and a combination of both) and the original model, absent of lensing and noise.
% \label{FlatOmegamH0LensedNoisePercent}
% }
%\end{figure}

\section{Geometrical degeneracy in a flat universe with massive neutrinos}
\label{sec:flatnu}

In this section, we consider an approximate geometrical degeneracy that
arises in models with massive neutrinos that are light enough
(masses well below $1~\mathrm{eV}$) to still be relativistic at
recombination. We give a brief description of \CAMB's numerical calculation with massive neutrinos in Appendix.~\ref{app:massive_nu}, including recent changes giving improved performance that are tested for consistency by the analysis here.

We consider a flat fiducial model with the parameters given in Table~\ref{ParamsTable} but with the physical density in massive neutrinos $\Omega_\nu h^2 = 6.0\times 10^{-4}$ (and hence a lower $\Omega_\Lambda$ to preserve flatness). We calculate the fiducial model at high accuracy and with all three accuracy parameters ({\lSampleBoost}, {\lAccuracyBoost} and {\AccuracyBoost}) set to 2 so the numerical errors are very small.
Throughout this section, we assume three neutrino mass eigenstates with squared mass differences given by neutrino oscillation results in
the `normal hierarchy'. In particular, we use the central values from the
2006 update of the global fit in Ref.~\cite{Maltoni:2004ei} (also adopted
in Ref.~\cite{Lesgourgues:2006nd}):
\begin{eqnarray}
m_2^2 - m_1^2 &=& 7.9 \times 10^{-5} \, \mathrm{eV}^{2}, \nonumber \\
m_3^2 - m_1^2 &=& 2.2 \times 10^{-3} \, \mathrm{eV}^2 ,
%m_2^2 - m_1^2 &=& 7.59 \times 10^{-5}\, \mathrm{eV}^{2}, \nonumber \\
%m_3^2 - m_1^2 &=& +2.46 \times 10^{-3} \, \mathrm{eV}^2 ,
\label{eq:masssq}
\end{eqnarray}
with $m_1$ the lightest neutrino mass, and $m_3$ the heaviest.
While these central values are only consistent with more recent fits,
e.g.\ Ref.~\cite{GonzalezGarcia:2010er}, at around the $2\sigma$ (of the
marginal errors) this should not impact our conclusions on
numerical robustness and physical breaking of degeneracies.
The energy density of neutrinos in our fiducial model corresponds to the
minimal-mass ($m_1 = 0 \, \mathrm{eV}$) normal hierarchy.

We consider the two-parameter approximate degeneracy between
$\Omega_\nu h^2$ and $h$ in flat models. All other parameters are fixed
to their fiducial values throughout this analysis (including $\Omega_c h^2$; only $\Omega_\Lambda$ changes to preserve flatness).
Physically this degeneracy arises because, on small scales, the only
significant change to the power spectrum is through a change in the angular
diameter distance to last scattering, $d_A(z_*)$. For the sub-eV neutrino
masses considered here, the
neutrinos are relativistic at recombination, so the difference they make
on the dynamics of the pre-recombination universe compared to massless
neutrinos is very small. However, at late times at least two of the mass
eigenstates become non-relativistic increasing the energy density relative to
a massless model with otherwise the same parameters. This decreases
$d_A(z_*)$ and causes the CMB acoustic peaks to shift to larger angular scales,
since the sound horizon at recombination has only a weak dependence on (light)
neutrino masses. The value of $d_A(z_*)$, and hence the peak locations,
can be restored by reducing the
Hubble constant (and hence decreasing the dark energy density which is a
derived parameter in our analysis of flat models).

\begin{figure}
\centering
\includegraphics[width=\textwidth]{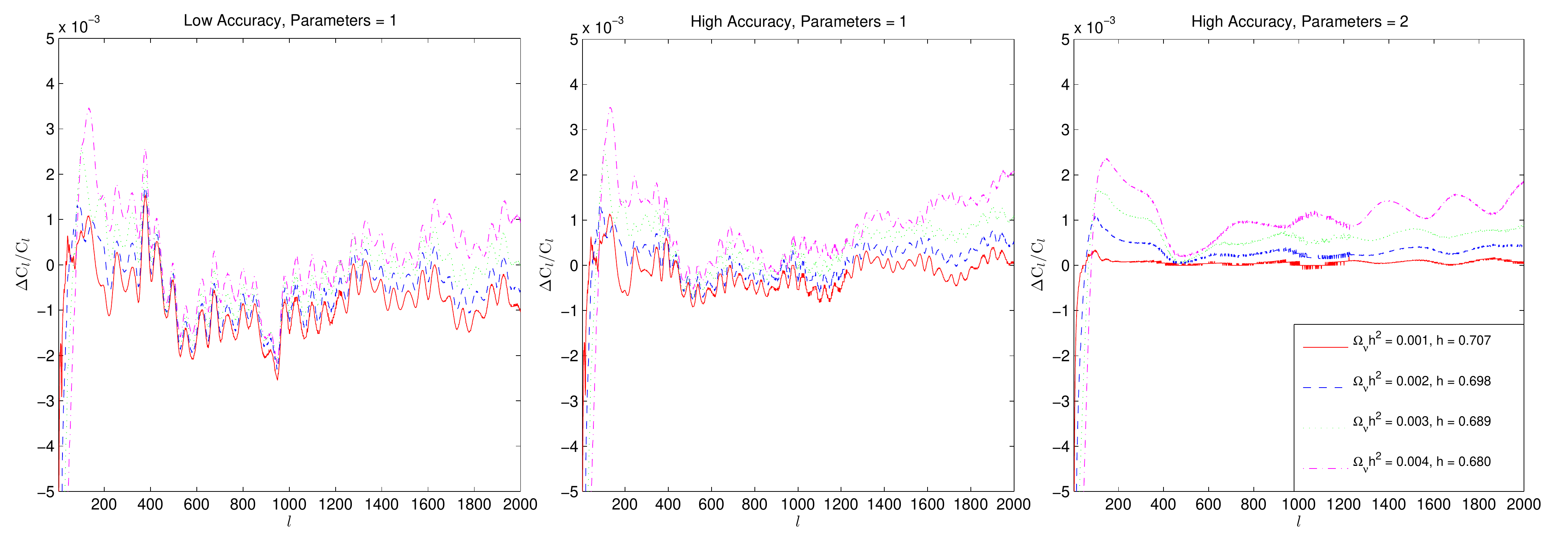}
\caption{Fractional differences between four degenerate models and the fiducial model. The degenerate models are computed at low accuracy with no boost in accuracy parameters (left), high accuracy with no other accuracy boost (middle) and at high accuracy with the accuracy parameters boosted to 2 (right). The right-hand panel reveals small physical degeneracy-breaking effects between the models.
}
\label{fig:flatneutrinos_diff}
\end{figure}

We construct several nearly degenerate models by locating parameters in the $\Omega_\nu h^2$-$h$ plane all with the same ratio of the sound horizon at last-scattering to $d_A(z_*)$ as in the fiducial model. Examples of the fractional differences between the unlensed power spectra of these models and the fiducial model are shown in Fig.~\ref{fig:flatneutrinos_diff}. The most massive degenerate model has summed masses of $0.38\,\mathrm{eV}$
corresponding to almost degenerate masses
$m_1 \approx m_2 = 0.122\,\mathrm{eV}$ and $m_3 = 0.131\,
\mathrm{eV}$. In the right-hand panel of Fig.~\ref{fig:flatneutrinos_diff}, the degenerate models are calculated at high accuracy and with all accuracy parameters boosted to 2. The degeneracy is nearly exact on small scales with physical differences in the power spectra below $0.2\%$ for $100 \leq l \leq 2000$; the difference would be only marginally detectable (assuming a known template shape for $\Delta C_l$ and all other parameters fixed)
for the our largest mass case with perfect temperature data.

%%%%%%%%%%%%%%%%%%%%%%%%%%%%%%%%%%%%%%%%%%%%%%%%
%%%%%%%%%%%%%%%%%%%%%%%%%%%%%%%%%%%%%%%%%%%%%%%%

\subsection{Degeneracy breaking effects}

\subsubsection{Numerical accuracy}

We see the same general trends in Fig.~\ref{fig:flatneutrinos_diff} to variation of the accuracy settings as in the models discussed earlier. At low accuracy, there is high frequency numerical noise and slowly-varying numerical drifts in the spectra. However, these numerical effects are comfortably smaller than the quoted accuracy of 0.3\%. With {\highaccuracydefault} set true, most of the drifts are removed and the amplitude of the high-frequency noise is significantly reduced though not eliminated entirely. The high-frequency noise can be further reduced by boosting {\lSampleBoost} or, better, with the improved interpolation method introduced in Sec.~\ref{newinterpolation}.

%%%%%%%%%%%%%%%%%%%%%%%%%%%%%%%%%%%%%%%%%%%%%%%%
\begin{figure}
\centering
\includegraphics[width=0.6\textwidth]{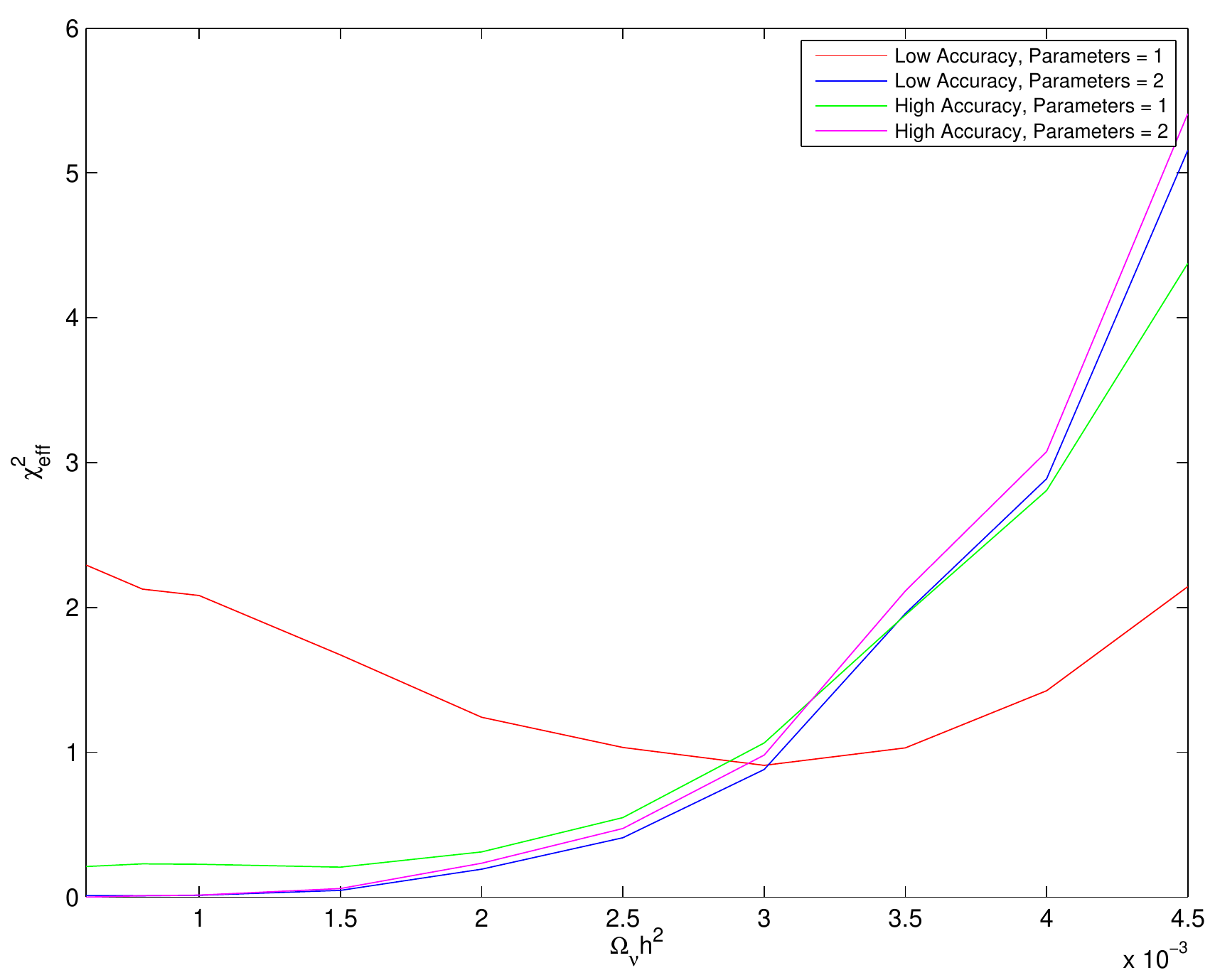}
\caption{Effective mean $\chi^2$ about the fiducial model
for a range of degenerate models and accuracy settings: low accuracy with default values (red) or boosted values of 2 (blue) for {\lSampleBoost}, {\lAccuracyBoost} and {\AccuracyBoost}; and high accuracy with default (green) and boosted (to the value 2;
magenta) parameters.
In all cases, the fiducial model in $\chisqeff$
is computed at high accuracy and all parameters boosted to 2. Only unlensed temperature spectra are included and there is no noise.
}
\label{fig:chisq}
\end{figure}
%%%%%%%%%%%%%%%%%%%%%%%%%%%%%%%%%%%%%%%%%%%%%%%%

We now compute $\chisqeff$ from Eq.~(\ref{eq:chisq})  up to $l_{\mathrm{max}}=2000$ along the degeneracy line in the $\Omega_\nu h^2$-$h$ plane. Results are plotted in Fig.~\ref{fig:chisq} for a range of accuracy settings.
The magenta curve is computed with the same high accuracy as the fiducial model used as input to $\chisqeff$ and therefore represents our best estimate
of the truth. In this case, the minimum $\chi^2_{\mathrm{eff}}$ is achieved at
the fiducial value $\Omega_\nu h^2 = 0.6 \times 10^{-3}$, and
the slow variation of $\chi^2_{\mathrm{eff}}$ arises from the degeneracy not
being physically exact (see later). Comparing to the low accuracy calculations and default parameter settings, we see that in this case the shape of the likelihood is significantly altered
with the minimum of $\Omega_\nu h^2$ displaced from the true value by
$\sim 1\sigma$ along the degeneracy direction, and the width of the
distribution artificially tightened. These errors likely arise from the low frequency
numerical errors visible in Fig.~\ref{fig:flatneutrinos_diff}.

We further see in Fig.~\ref{fig:chisq} that default high accuracy setting get close to the `true' likelihood shape (with a slight offset). The numerical
noise that remains in this case (see Fig.~\ref{fig:flatneutrinos_diff}) has only a small
effect on the likelihood, and is removed by a further boost in accuracy.

\subsubsection{Physical effects}
\label{sec:degs}

The degeneracy is not exact as illustrated in the right-hand panel of Fig.~\ref{fig:flatneutrinos_diff}. On the large scales relevant for the late-ISW effect, the neutrinos are able to cluster and their additional contribution to the
expansion rate at late time does not impede the growth of structure. However,
simultaneously reducing the dark energy density to keep $d_A(z_*)$ fixed
lessens its tendency to halt structure formation and so there is less
decay of the gravitational potential and a smaller late-ISW effect.
On intermediate scales near the first peak, the early-ISW effect is
significant. This arises from the decay of the gravitational potential as the equation of state changes during the matter-radiation transition. Compared to the massless case, neutrinos of mass $m_\nu$ that are still relativistic at temperature $T_\nu(z)$ have their pressure reduced by a fractional amount $\clo(m_\nu / k_B T_\nu)^2$ and the energy density increased by a similar fraction. This leads to additional decay of the gravitational potential shortly after recombination (i.e.\ when the neutrinos are no longer ultra-relativistic but still make a significant contribution to the energy density) and an increase in the early-ISW effect.
Further effects of the enhanced energy density are small reductions in the physical sound horizon and damping length but the former is compensated by adjusting $d_A(z_*)$. Moreover,
the dynamics of the small-scale neutrino perturbations themselves acquire
$\clo(1-v_\nu)$ corrections, where the typical neutrino thermal speed
$v_\nu(z) = 1 - \clo(m_\nu / k_B T_\nu)^2$ changes the free-streaming scale from
the (light) horizon which feeds back into the evolution of other perturbations.
The expected fractional size of these effects on the CMB power spectrum is $\sim \clo(m_\nu / k_B T_\nu)^2 \rho_\nu / \rho_{\mathrm{tot}}$, where $\rho_\nu$ is the neutrino energy density and $\rho_{\mathrm{tot}}$ the total energy density. In our most massive model, this suggests effects at the $\clo(10^{-3})$ level.
The right-hand panel of Fig.~\ref{fig:flatneutrinos_diff} shows the
relative physical differences of our degenerate model
power spectra to the fiducial model. The expected changes in the late- and early-ISW effects can clearly be seen. In addition, on smaller scales we see effects of $\clo(10^{-3})$ for our most massive model, consistent with the estimate above.

Note that our analysis here does not include the effect of weak gravitational lensing
of the CMB. If this effect were included, the degeneracy would be further
broken since massive neutrinos reduce the power spectrum of the lensing
deflections on small scales compared to a model with massless neutrinos.

%The behaviour on small scales is due to the degeneracy-breaking effects discussed in Section \ref{sec:degs}. We see that as the neutrino mass is increased, the spectra become more distinct from the fiducial model. We can discern two effects at high masses: an overall increase in amplitude, and the appearance of low amplitude oscillations. Both effects are small in the mass range considered, the combination being at most $0.1\%$.

%We may understand the low amplitude oscillations as arising from physical effects, as opposed to finite accuracy effects. Perturbations in the neutrino distribution function prior to recombination have a small effect on the gravitational potentials, the dependence being upon $\sum m_{\nu}^2$ in the limit of light masses (Lewis and Challinor 2002). This arises because the relevant terms in the perturbation equations depend on the mass only through the comoving energy, $\sqrt{q^2+a^2m^2}$ where $q$ is the comoving momentum, and $a$ is the scale factor. We thus expect this effect to become greater as neutrino mass increases.

\section{Three-parameter geometric degeneracy with massive neutrinos and curvature}
\label{sec:nonflatnu}

We now extend the parameter space of the previous section to include spatial curvature. We expect a three-parameter geometric degeneracy involving $h$, $\Omega_\nu h^2$ and $\Omega_K$ at fixed $\theta$ since neither curvature nor sub-eV neutrino masses are dynamically significant before last-scattering. We retain the fiducial model used in the previous section, and continue to probe only sub-$\mathrm{eV}$ neutrino masses. We do not include lensing.

We find degenerate models by fixing the angular size $\theta$ of the sound horizon to its value in the fiducial model. The likelihood is not exactly constant over the two-dimensional degeneracy surface due to the small physical degeneracy-breaking effects (and numerical effects). We sample the degeneracy surface at discrete points over a regular $\Omega_\nu h^2$-$h$ coordinate grid with parameter ranges chosen to give a good spread of $\chisqeff$.

In the absence of numerical effects, a Taylor expansion of $C_l$ in $h$ and $\Omega_\nu h^2$ at fixed $\theta$ should be accurate close to the fiducial model. The derivatives with respect to these parameters at the fiducial model can be already be inferred from the results in Sec.~\ref{sec:geo} and Sec.~\ref{sec:flatnu}. We thus expect physical differences in the spectra that are linear combinations of those described earlier. Note that our previous calculations have used either a flat code with massive neutrinos or a curved code with massless neutrinos. Our main aim here is to check the numerical robustness of the curved code with massive neutrinos.

%%%%%%%%%%%%%%%%%%%%%%%%%%%%%%%%%%%%%%%%%%%%%%%%
\begin{figure}
\centering
\includegraphics[width=\textwidth]{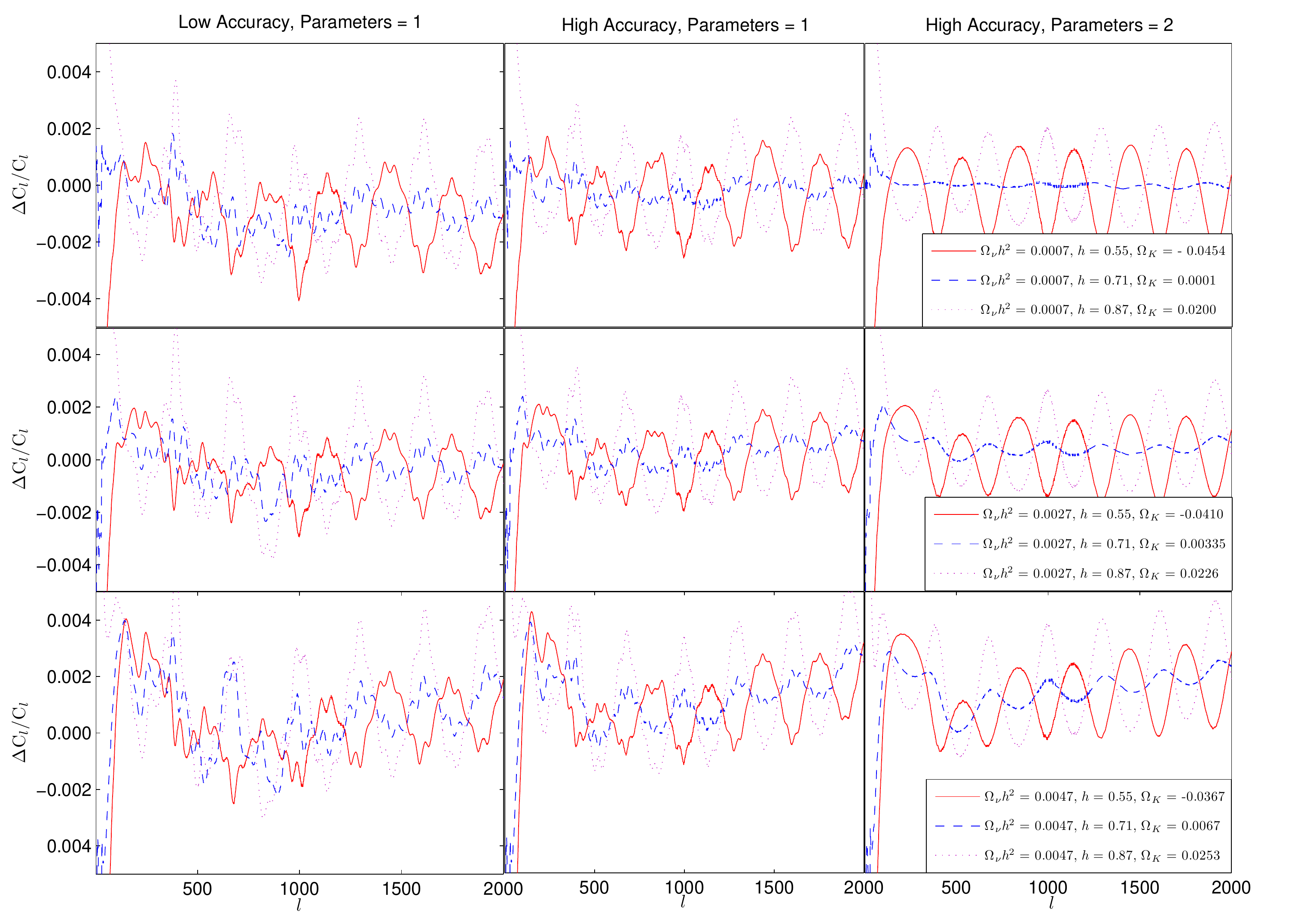}
\caption{Fractional differences between nine degenerate models (three in each
row) and the fiducial model exploring the $\Omega_\nu h^2$-$\Omega_K$-$h$ degeneracy. The degenerate models are computed at low accuracy with no boost in accuracy parameters (left column), high accuracy with no other accuracy boost (middle column) and at high accuracy with the accuracy parameters boosted to 2 (right column). The right-hand panels reveal small physical degeneracy-breaking effects between the models. Moving from the top row to the bottom at fixed line style (colour), the neutrino mass is increasing at fixed $h$ (and $\theta$). Within a panel, the mass is constant.}
\label{fig:curvedneutrinos_diff}
\end{figure}
%%%%%%%%%%%%%%%%%%%%%%%%%%%%%%%%%%%%%%%%%%%%%%%%

In Fig.~\ref{fig:curvedneutrinos_diff}, we plot the fractional differences between our degenerate models and the fiducial model with the former calculated at various accuracy settings. In all cases, the numerical errors are under control and for high accuracy with the other accuracy parameters at their default values the numerical errors are below the quoted $0.1\%$ level.

The physical differences revealed in the right-hand panels of Fig.~\ref{fig:curvedneutrinos_diff} are consistent with our earlier findings.
Increasing the neutrino mass scale at fixed $h$, the changes in the curvature are rather small and the physical effects of the neutrino masses dominate the evolution seen in the power spectra. At fixed neutrino mass, the evolution in the spectra is due to the change in curvature, i.e.\ the late-ISW and projection effect discussed in Sec.~\ref{sec:geo}.

%An additional degeneracy-breaking effect is the large scale ISW effect. We saw that in the flat scenario, increasing the neutrino mass reduces the power from the ISW effect. The scaling is reversed for $\Omega_K$. At fixed total mass, we see that increasing $\Omega_K$ \emph{increases} the large scale power from the ISW effect. With fixed total matter density, increasing $\Omega_K$ and $h$ increases the expansion rate at all redshifts, leading to enhanced ISW effect. \adc{Discontinuous change of slope of power spectra at $l=31$ for these models relative to flat universe, irrespective of sign of curvature or magnitude of neutrino mass.}

%%%%%%%%%%%%%%%%%%%%%%%%%%%%%%%%%%%%%%%%%%%%%%%%
\begin{figure}
\centering
\includegraphics[width=0.32\textwidth]{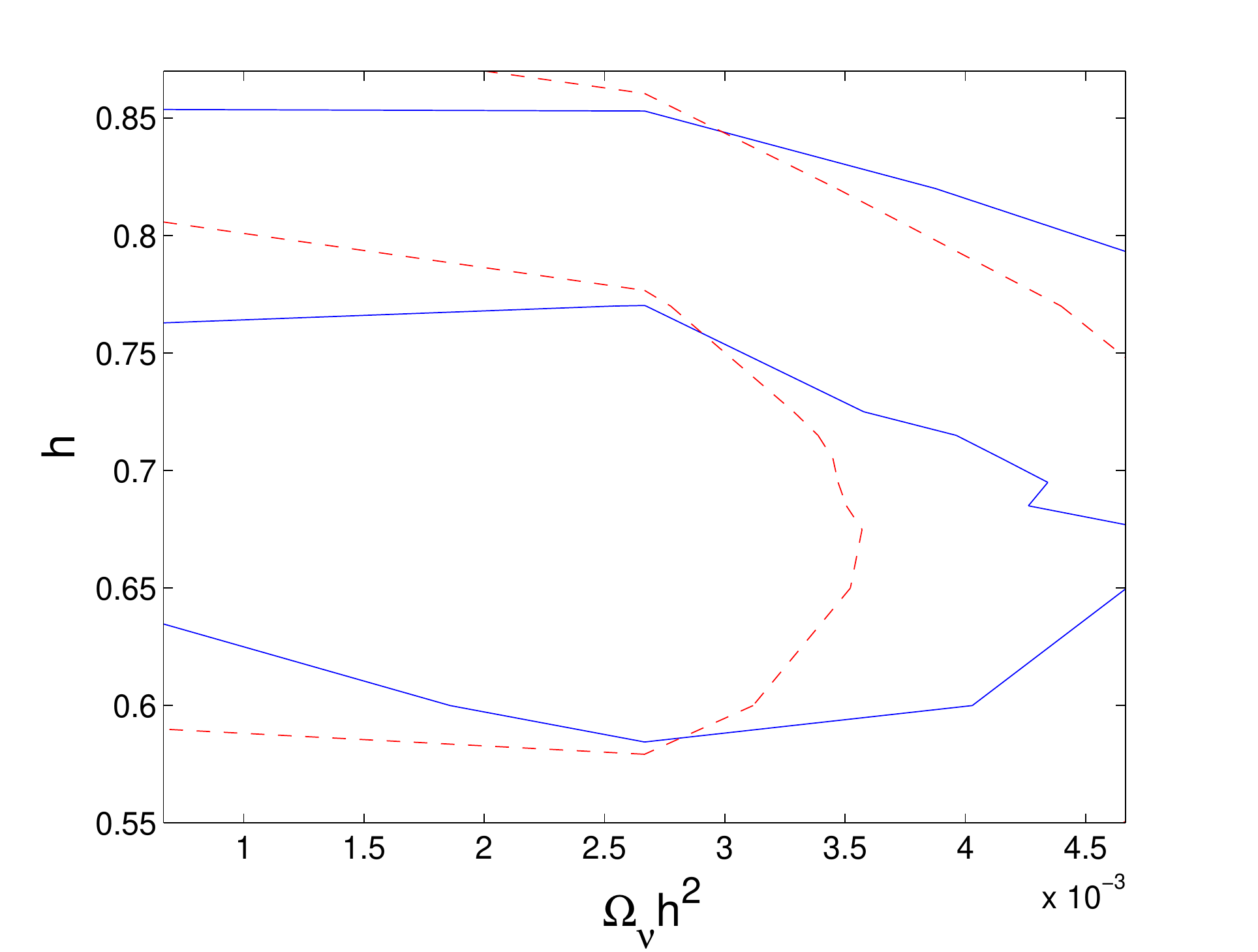}
\includegraphics[width=0.32\textwidth]{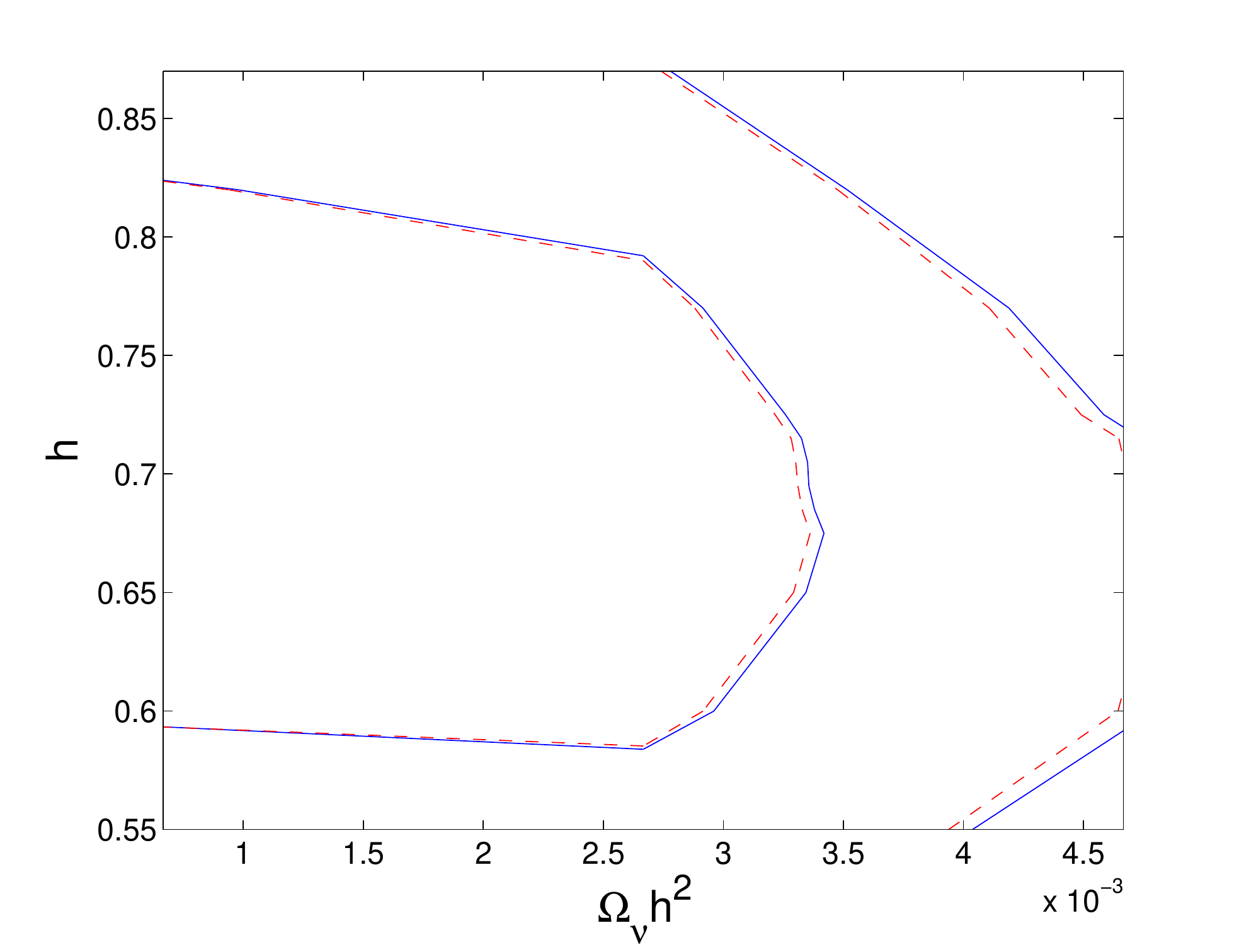}
\includegraphics[width=0.32\textwidth]{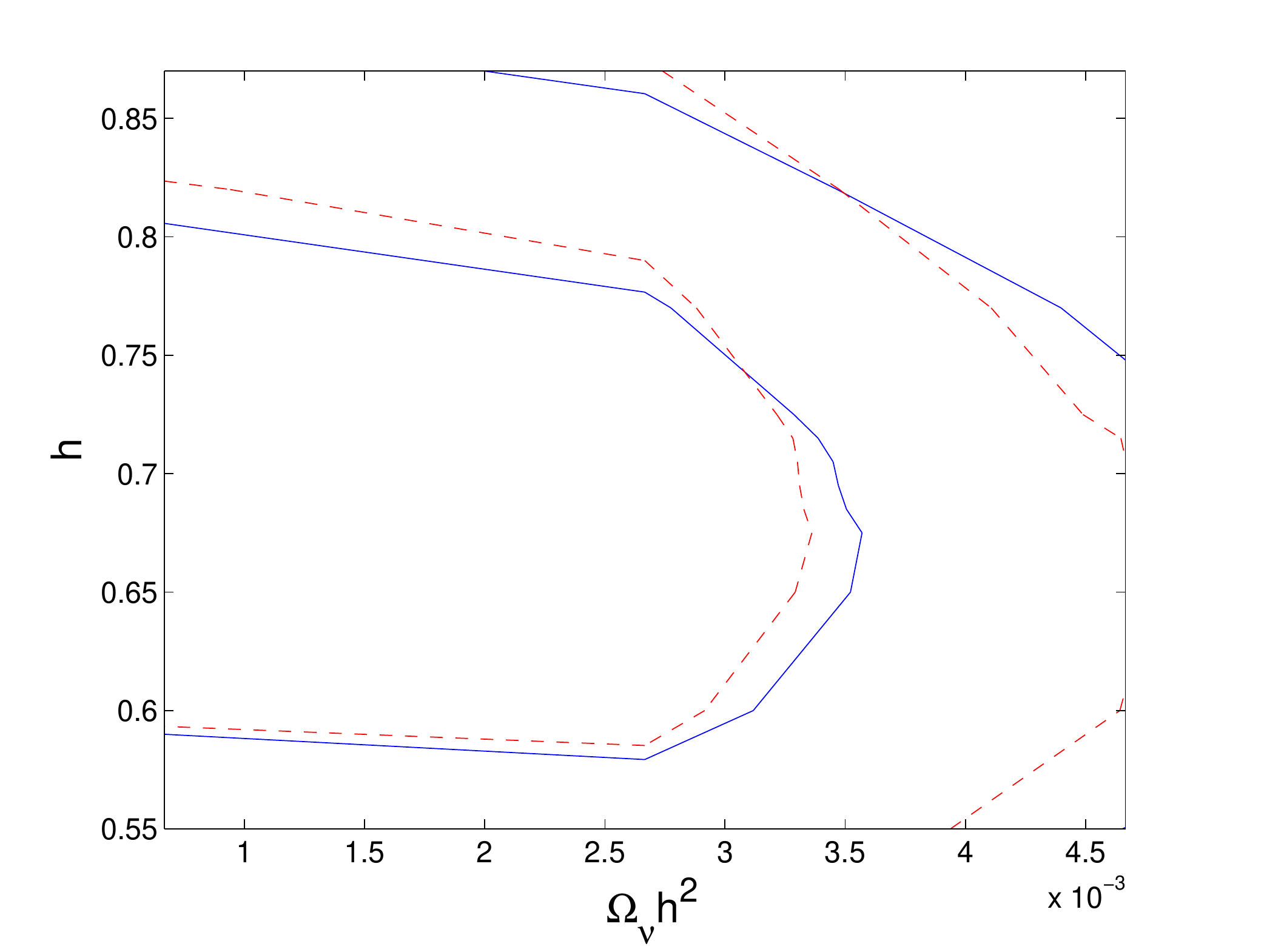}
\caption{Contours of $\chisqeff$ over the two-dimensional degeneracy surface,
$\theta=\mathrm{const.}$, in the $\Omega_\nu h^2$-$h$-$\Omega_K$ space. We
parametrize the surface by $h$ and $\Omega_\nu h^2$ and sample over a regular grid in these parameters. The contours enclose 68\% and 95\% confidence regions. \emph{Left}: default values of {\lSampleBoost}, {\lAccuracyBoost} and {\AccuracyBoost} with high accuracy calculations (red dashed) and low accuracy (blue solid).
\emph{Middle}: accuracy parameters boosted to 2 and high accuracy (red dashed) and low accuracy (blue solid). \emph{Right}: high accuracy and default parameters (blue solid) and boosted to 2 (red dashed). The dashed (red) contours in the middle and right plots represent the true situation (i.e.\ with minimal numerical errors). Lensing and instrumental noise are not included.
}
\label{fig:chisq_nuhk}
\end{figure}
%%%%%%%%%%%%%%%%%%%%%%%%%%%%%%%%%%%%%%%%%%%%%%%%

In Fig.~\ref{fig:chisq_nuhk}, we show contours of $\chisqeff$ over the two-dimensional degeneracy surface of constant $\theta$ parametrized by $h$ and $\Omega_\nu h^2$. The models are computed at various accuracy settings. The dashed (red) contours in the middle and right-hand plots have minimal numerical errors. The conclusions are similar to previous sections. Numerical noise in low accuracy calculations with default accuracy parameters distorts the likelihood significantly, shifting the best-fit point and artificially broadening the distribution in the $\Omega_\nu h^2$ direction. High accuracy calculations with default parameters perform better with only minor shifts in the contours. In this case, the high-frequency
noise apparent in the middle column of Fig.~\ref{fig:curvedneutrinos_diff} has little
effect on parameter inferences. The physical degeneracy-breaking effects of curvature and neutrino masses have rather different spectral shapes (see Fig.~\ref{fig:curvedneutrinos_diff}) and this is reflected by the lack of correlation seen
between parameters in Fig.~\ref{fig:chisq_nuhk}.

\section{MCMC sampling analysis}
\label{sec:MCMC}

\begin{figure}[h!]
 \includegraphics[width = 12cm]{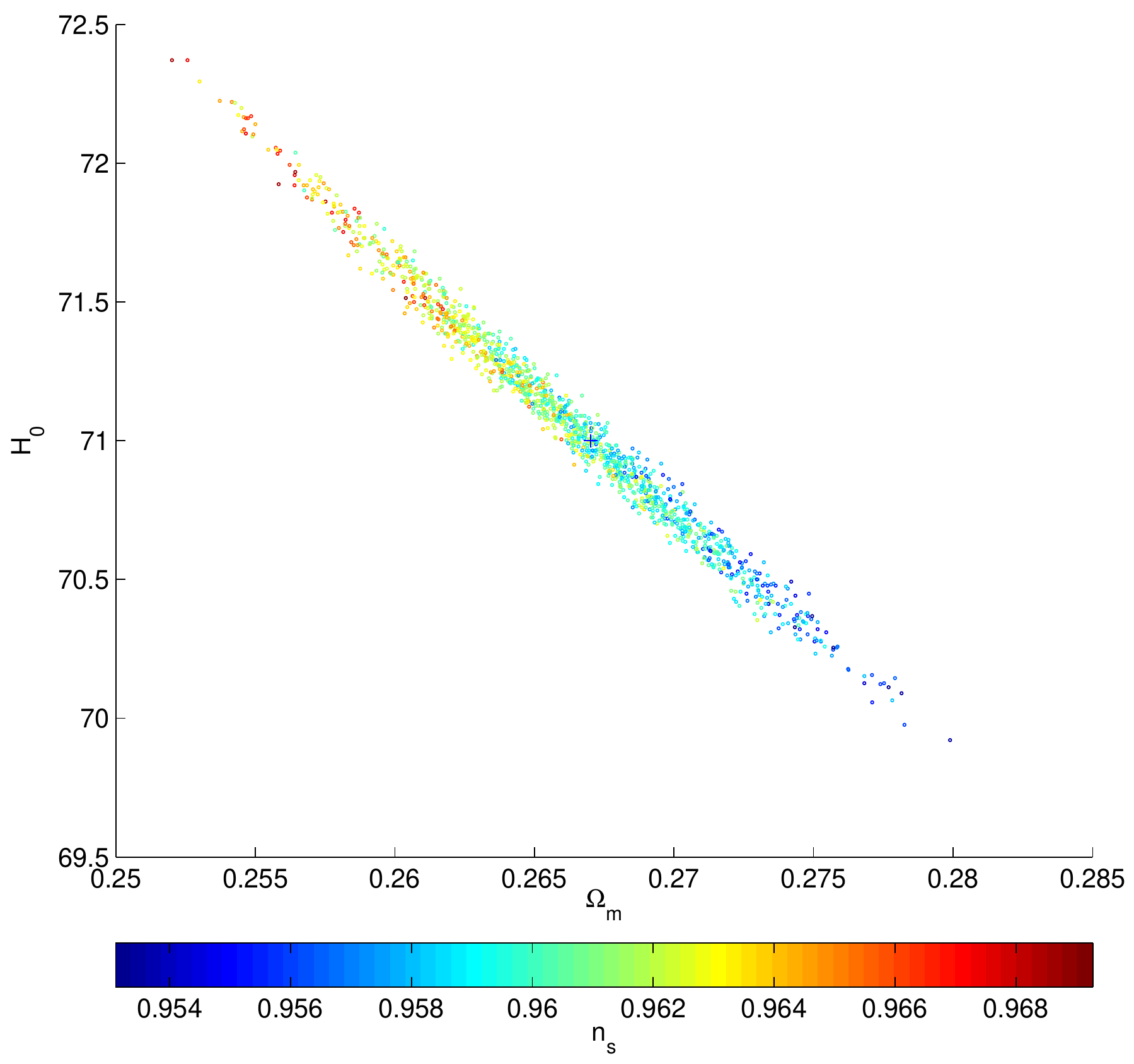}
 \caption{MCMC samples from the posterior of Eq.~(\ref{MatrixChisq}) with
$\hat{\mC_{l}}$ given by the fiducial {\LCDM} model described in the text
(marked with a cross) and Planck-like noise, including CMB lensing and polarization.
The constraint on $\Omega_m h^3$ is nearly twenty times tighter than on the orthogonal direction.
 %$\Omega_m h^{-3}$.
 The degeneracy involves other parameters, so the accuracy, for example, of $n_s$ constraints from CMB alone is limited by this near degeneracy (see Fig.~\ref{FlatMCMCOtherParams}).
%\adc{The true orthogonal family of curves have $3 \Omega_m^2 - h^2 = \mathrm{const}.$ while the obvious orthogonal direction has (close to the fiducial point) $\Omega_m h^{-1/3} = \mathrm{const}.$.}
 \label{FlatMCMC}
 }
\end{figure}

\begin{figure}[h!]
 \includegraphics[width = 12cm]{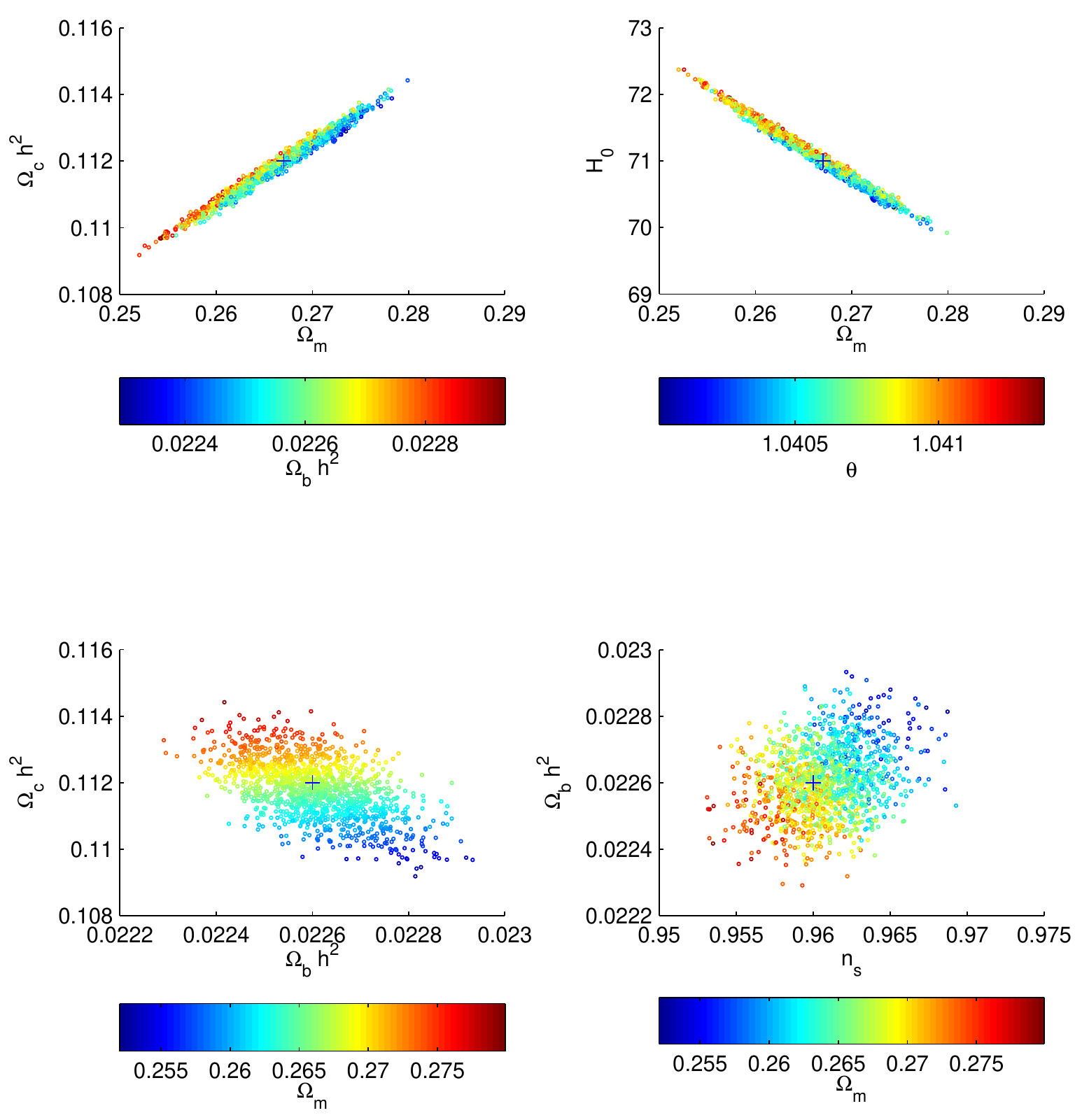}
 \caption{As Fig.~\ref{FlatMCMC} but exploring correlations between the other parameters. The parameter $\theta$ is approximately 100 times the ratio of the sound horizon to the angular diameter distance, and is very well constrained by the observed acoustic peak positions and nearly constant along the degeneracy~\cite{Kosowsky:2002zt}. Changes in $\Omega_m$ shift the angular scale of the CMB peak locations, though not very strongly, which is compensated by a shift in sound horizon at recombination due to a change in the physical matter density $\Omega_m h^2$~\cite{Percival:2002gq}. However a change in physical matter densities changes the amplitudes of the acoustic peaks, so $\Omega_b h^2$ and $n_s$ also have to change for partial compensation.
 \label{FlatMCMCOtherParams}
 }
\end{figure}

\begin{figure}[h!]
\begin{minipage}[l]{0.5\textwidth}
 \includegraphics[width = 10cm]{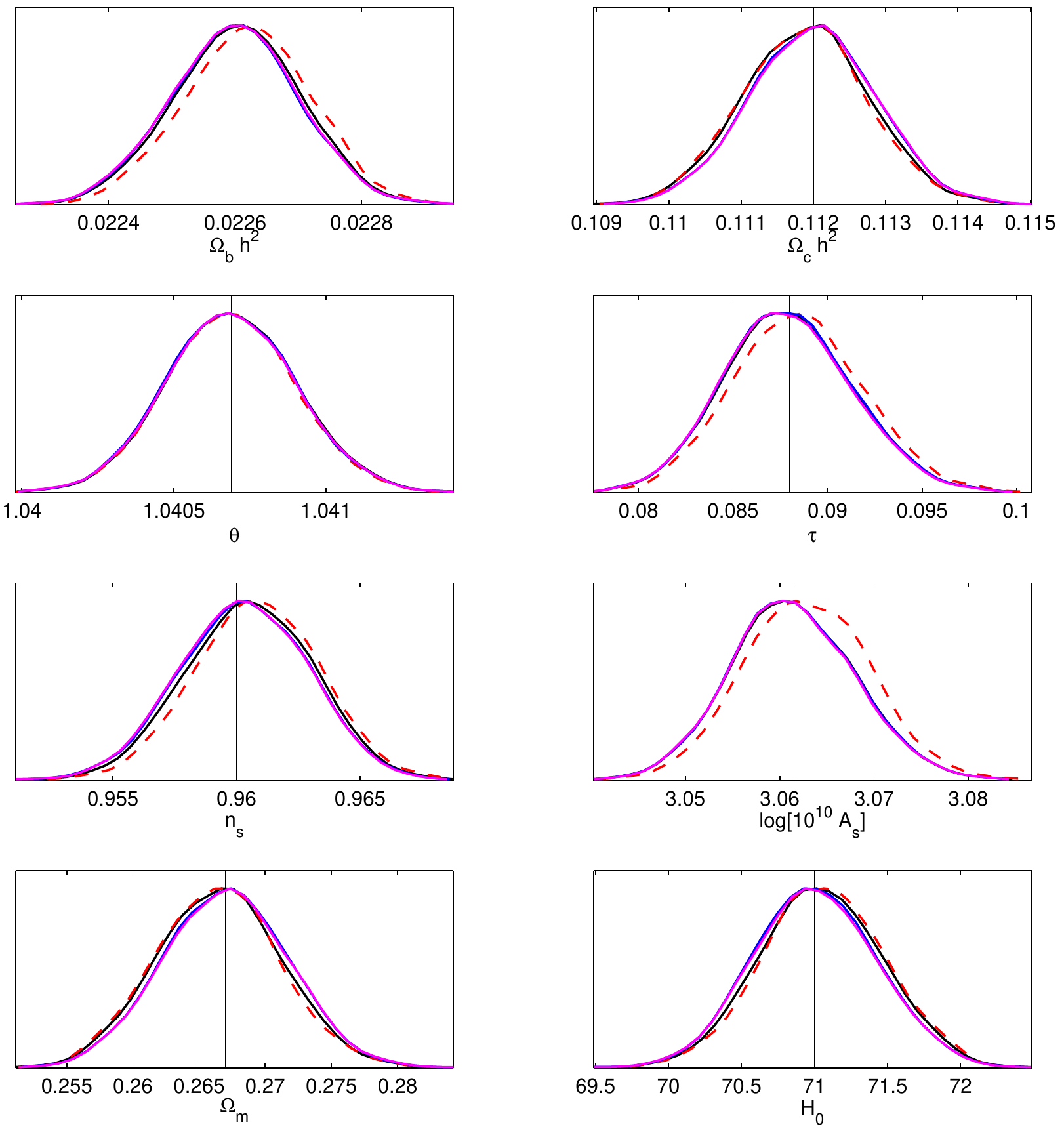}
\end{minipage}
\begin{minipage}[r]{0.49\textwidth}
\hspace{0.5cm} \includegraphics[width = 5.8cm]{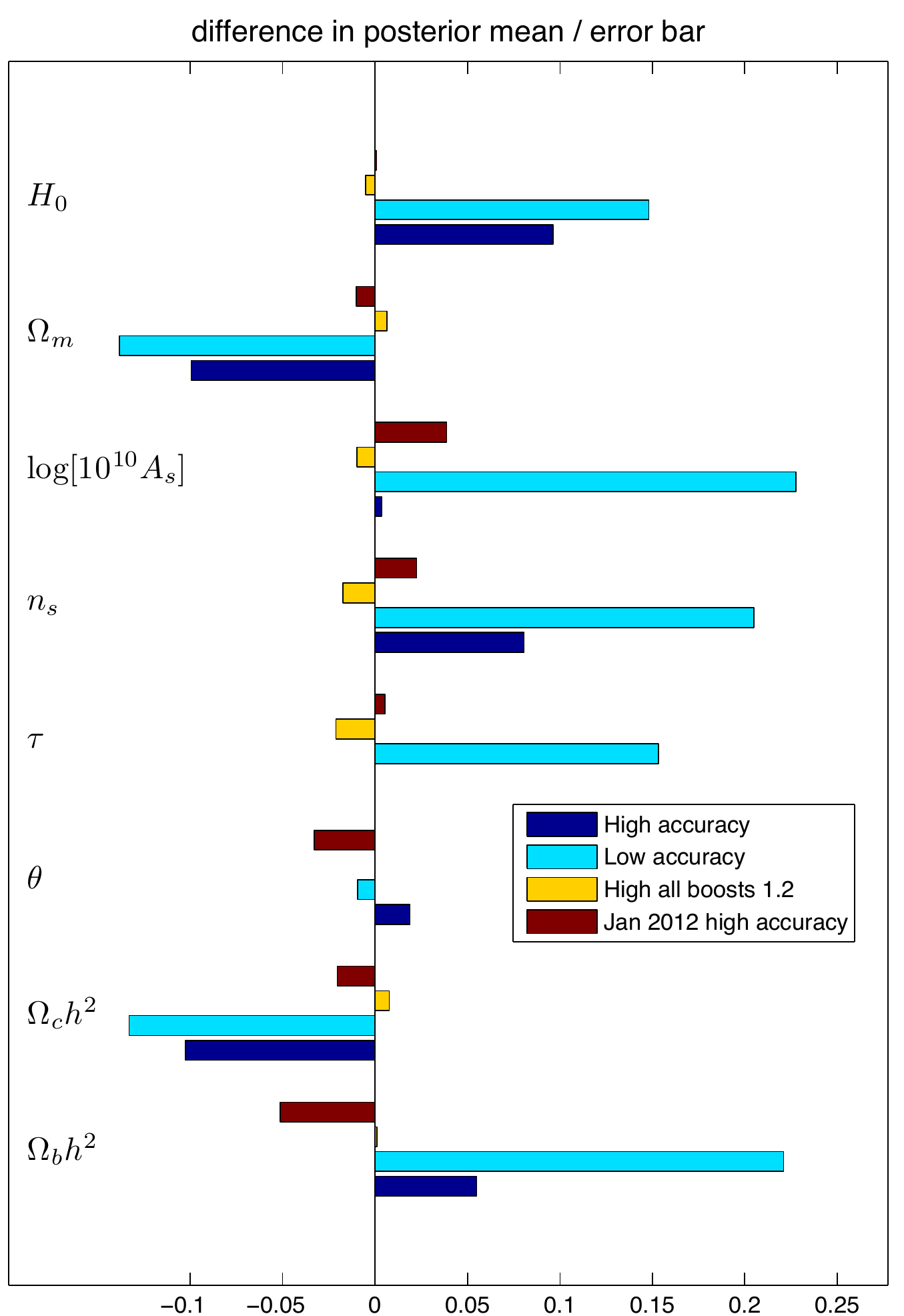}
\end{minipage}
 \caption{\emph{Left}: Marginalized parameter constraints for the fiducial flat {\LCDM} model with Planck-like noise, generated by using importance sampling from a well-converged MCMC run with default high accuracy settings (so all curves are generated using samples at the same parameter values). Solid black lines show the result using {\CAMB} with default high accuracy settings; red dashed lines show the slightly biased result from using low accuracy settings. Magenta and blue curves use boosted high accuracy settings of $2$ and $1.2$ respectively, which are almost on top of each other.
  \emph{Right}: Fractional difference in posterior means for different accuracy settings compared to the result from high accuracy with boosts set at 2 (as in the fiducial power spectrum).
  Default high accuracy settings are sufficient for biases to be $\alt 10\%$ of the error bar, but a small increase in \lSampleBoost\ to $\sim 1.2$ is sufficient to remove most of the small residual bias (yellow bars show boosted accuracy). Red bars show the result comparing to the new January 2012 version of {\CAMB} at default high accuracy settings: this interpolates the difference between the $C_l$ and a fiducial spectrum, significantly reducing interpolation errors for higher speed, so residual differences are then $\alt 5\%$ of the error bar.   Less accuracy is likely to be required with more realistic data.
 \label{importance1DMCMC}
 }
\end{figure}

\begin{figure}[h!]
\includegraphics[width = 5.5cm]{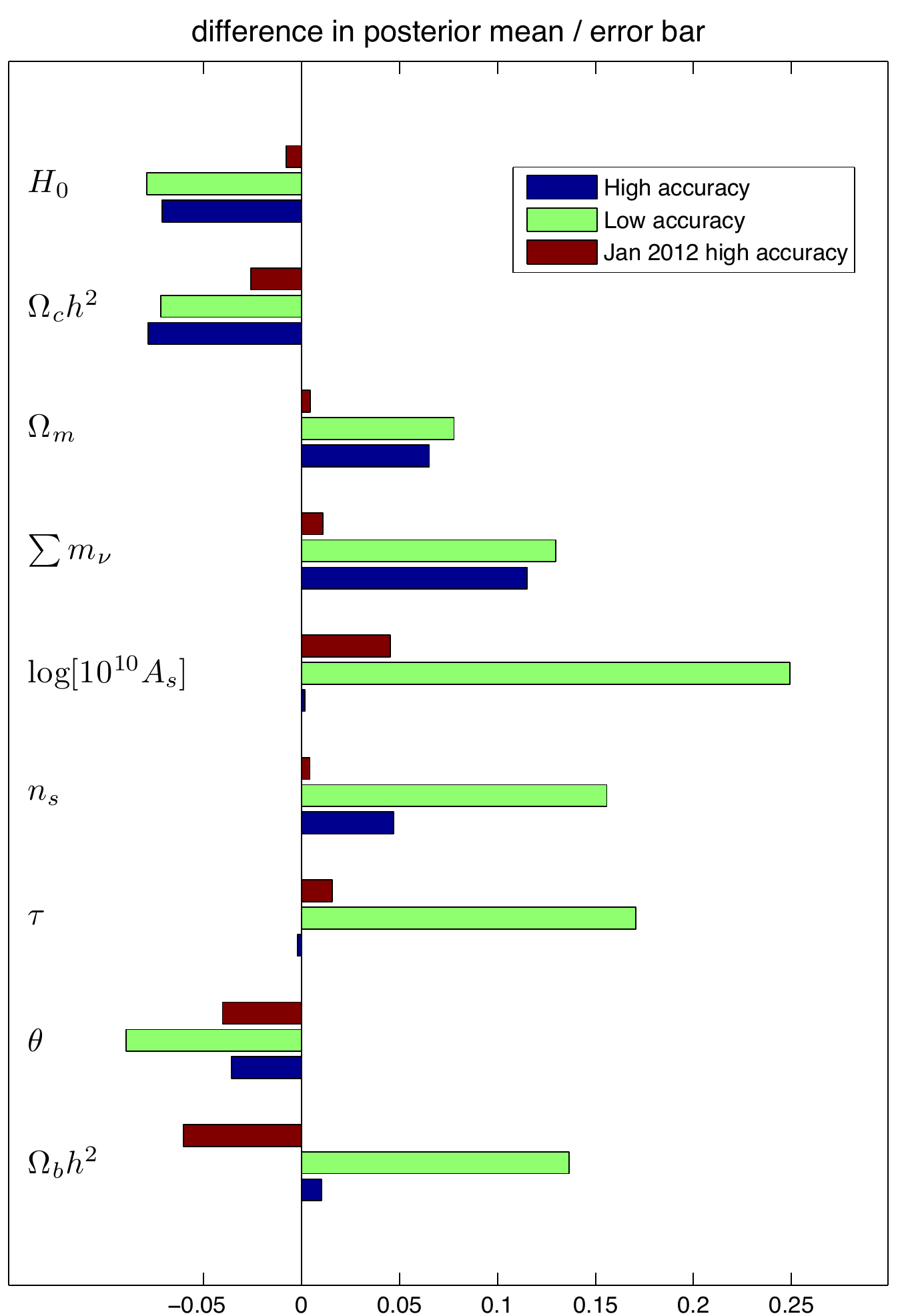}
\includegraphics[width = 5.5cm]{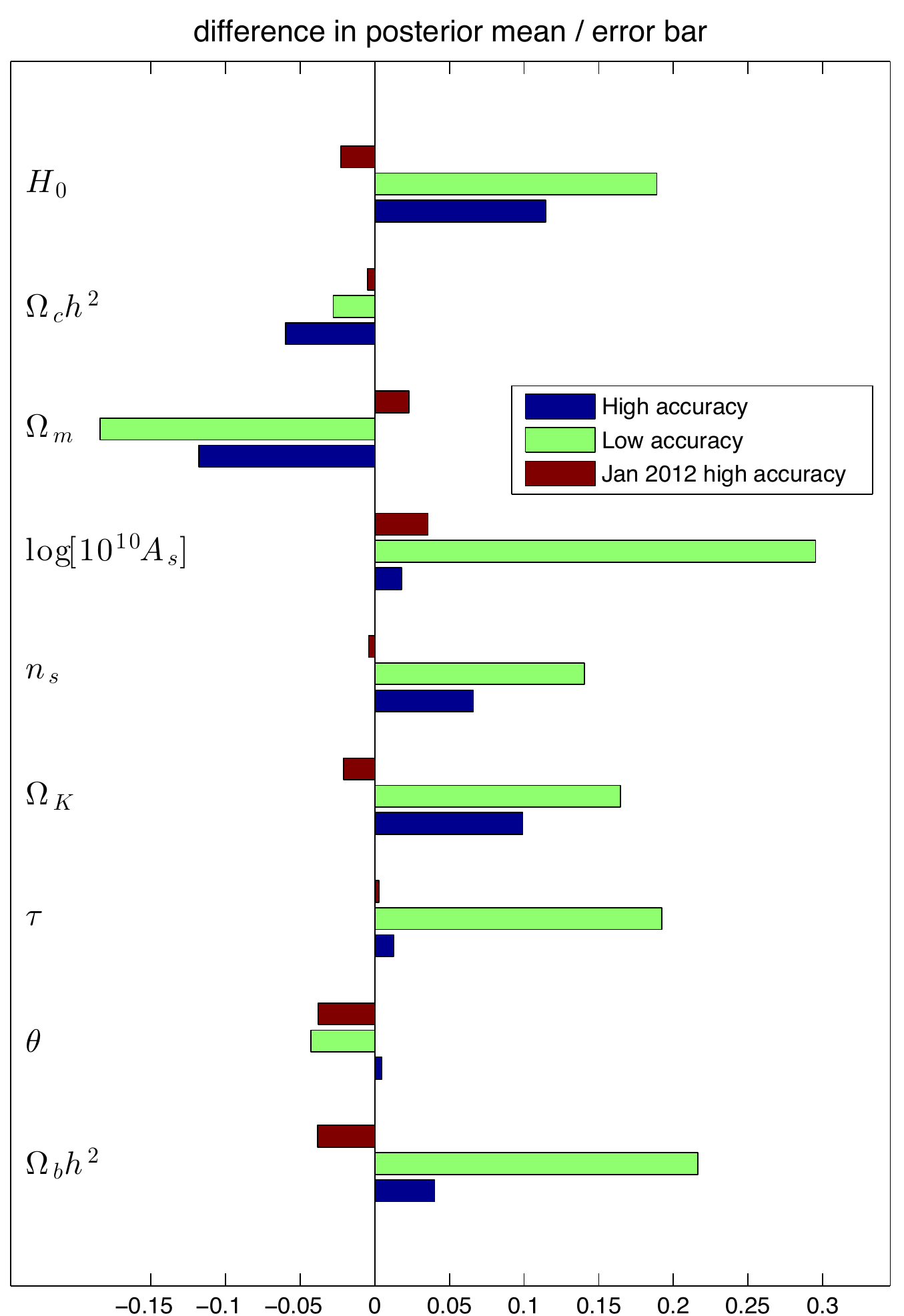}
 \includegraphics[width =5.5cm]{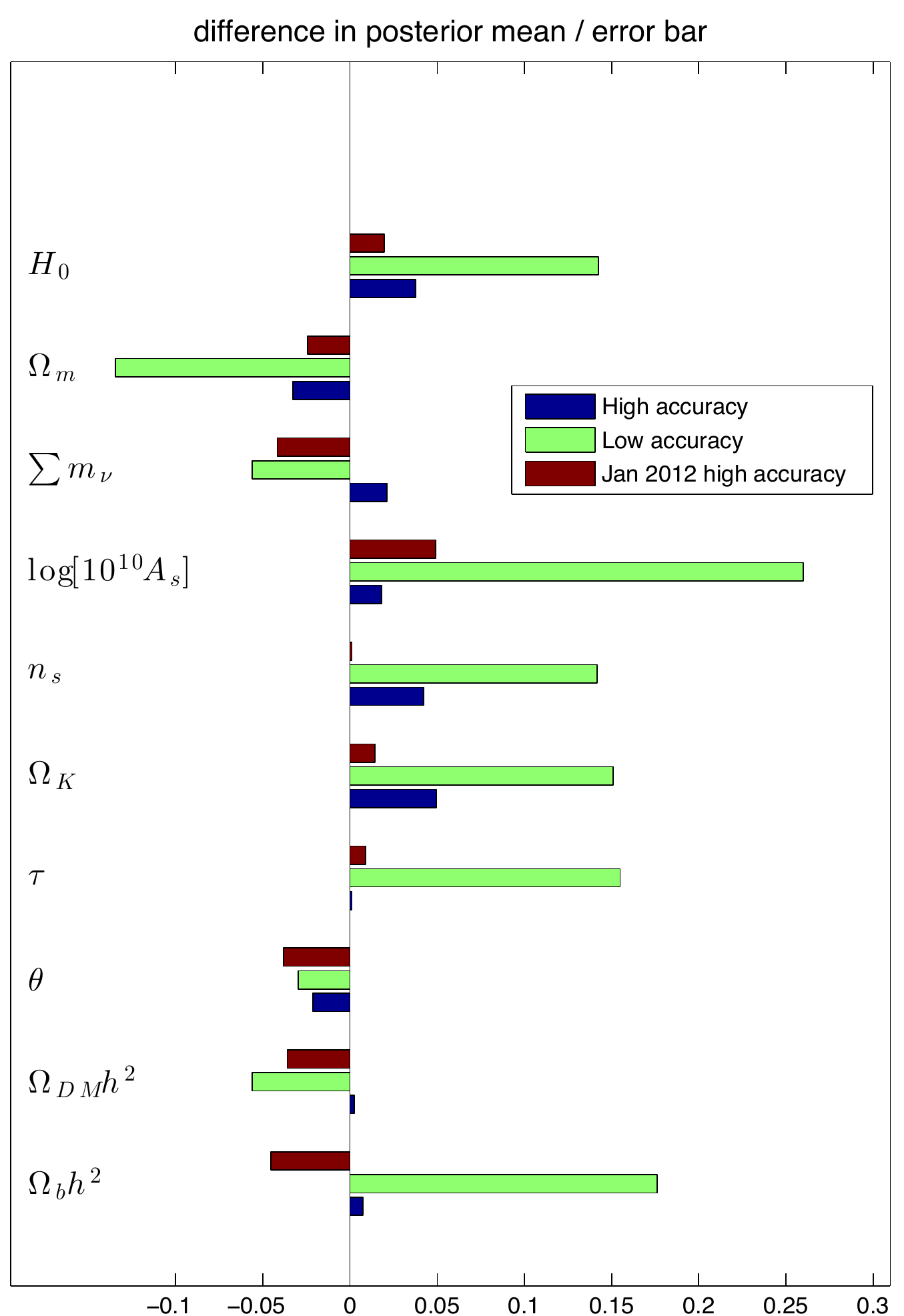}

 \caption{
 Effect of different accuracy settings on posterior parameter constraints for Planck-like noise compared to using high accuracy with accuracy parameters boosted to 2. \emph{Left:} flat \LCDM\ model with massive neutrinos (assumed degenerate); \emph{Centre:}  non-flat
\LCDM\ model with massless neutrinos; \emph{Right:} non-flat \LCDM\ model with massive neutrinos. As in the vanilla flat case, default high accuracy setting generally give results correct to within $10\%$ of the error bar, and using the new January 2012 interpolation scheme gives differences $\alt 5\%$ of the error bar.
 \label{MCMCAccuracyTests}
 }
\end{figure}

We have focused in this paper on maximizing the likelihood for different parameter values. This procedure has the advantage of being prior-independent, and is a rather stringent test of numerical accuracy because for convergence to a maximum of the likelihood the basin of attraction all has to be smooth and not give false numerical minima. However in practice most parameter analyses use sampling methods, which are rather less sensitive to small numerical errors. Figure~\ref{FlatMCMC} shows a set of MCMC samples from {\COSMOMC}~\cite{Lewis:2003an} assuming a flat \LCDM\ model, using the likelihood of Eq.~(\ref{MatrixChisq}) with
$\hat{\mC_{l}}$ given by our fiducial {\LCDM} model and Planck-like noise.
The degeneracy that we explored in Sec.~\ref{sec:approxdeg} is clearly apparent, and by showing the value of other parameters for each posterior sample we can also clearly see the importance of other parameter variations in addition to $\Omega_m$ and $H_0$; see Fig.~\ref{FlatMCMCOtherParams}.

We used the \highaccuracydefault\ setting for the MCMC runs, and obtained results that are fully consistent with the input parameters (as shown in Figs.~\ref{FlatMCMC} and~\ref{FlatMCMCOtherParams}).  The numerical cost for high accuracy runs is quite manageable: assuming a good covariance matrix is available from previous forecasting runs, a \LCDM\ six-parameter analysis using high accuracy settings only takes a few hours using two CPUs per chain (assuming the likelihood function is fast). By contrast, just increasing the accuracy parameters by hand to large values can easily waste very large amounts of computer time, and is unnecessary at Planck precision.

It is worth noting that the parameter $\theta$, which measures fairly directly the observed angular scale of the acoustic peaks~\cite{Kosowsky:2002zt}, is very well measured (at the $0.2\%$ level); as such it is very sensitive to small changes in model, for example a change in the CMB temperature from $2.725$~K to $2.726$~K, or in the effective number of neutrinos from $3.04$ to $3.046$, will shift the calculated value of $\theta$ by order $1\sigma$ for fixed data. However the values of the other parameters are virtually unaffected compared to their error bars, since their relative precision is much lower because of the degeneracies.

Small corrections due to changes in accuracy settings can easily be assessed from posterior samples by use of importance sampling, as explained in Ref.~\cite{Lewis:2002ah}. This only requires re-calculation of likelihoods at a subset of semi-independent samples from the original chains, and has the advantage of being almost independent of sampling noise (since the original samples and the importance-sampled samples are at the same points in parameter space).
In Fig.~\ref{importance1DMCMC} we show fully-marginalized 1D parameter constraints using various accuracy settings for the simplest six-parameter model; Fig.~\ref{MCMCAccuracyTests} shows equivalent results for non-flat and massive neutrino models.
The default high accuracy setting produces significantly more accurate results than low accuracy, though even low accuracy only has a mean bias of about $20\%$ of the error bar, and may be sufficient (at least as a base for importance sampling). The default high accuracy settings are still slightly biased compared to boosted accuracy results, though at a level that is $<10\%$ of the error bar. The residual bias can partly be removed using \lSampleBoost$\sim 1.2$.

\subsection{More accurate multipole interpolation}
\label{newinterpolation}
\begin{figure}[h!]
\includegraphics[width = 16cm]{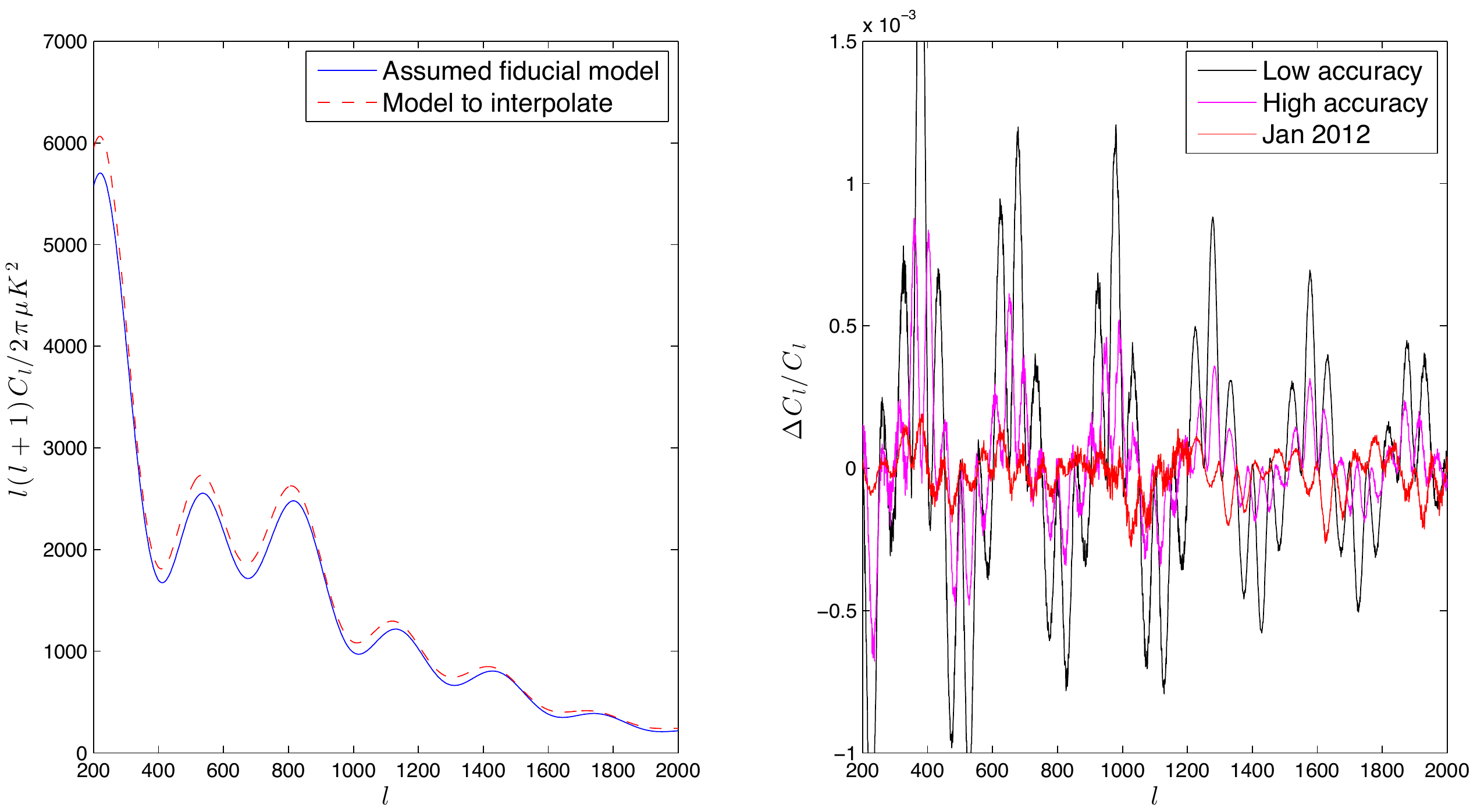}
 \caption{
  Power spectrum errors due only to cubic spline interpolation in $l$. The red and black lines on the right-hand side show the fractional error using a fixed $l$ sampling with $\Delta l=50$ and $\Delta l=42$ respectively. The red line shows the error using the new January 2012 interpolation method based on interpolating (with sampling $\Delta l=50$) the difference to the fiducial model shown on the left (which is many sigma away from the model we are trying to interpolate). The new method is faster, and also significantly more accurate ($\Delta C_l/C_l \ll 10^{-3}$) even for models not very close to the assumed fiducial template; as the model gets closer to the template the interpolation error goes to zero. Similar results hold for the polarization, and using lensed/unlensed power spectra.
 \label{spline_error}
 }
\end{figure}

The main results of this paper have been calculated using the October 2011 version of {\CAMB}, which interpolates the $C_l$ in $l$ using cubic-spline interpolation. The effect of the high accuracy setting (of Oct 2011) on this sampling is to increase the $l$ sampling slightly (by $\sim 20\%$).

However the interpolation errors are in fact very well known, because we now have measurements of the CMB power spectrum that fix the location of the acoustic peaks very well. Differences in the $C_l$ between probable models only move the power spectrum by a small fractional amount compared to the best fit, so the interpolation errors evaluated for some fixed fiducial model near the best-fit model will also be very close to those in any nearby model. We can therefore simply subtract a fixed model of the interpolation errors; this makes the interpolation exact for the fiducial model, very accurate for close models, and only increases the error for models that are very different (by an amount that goes to zero as the interpolation accuracy parameter is increased). This is equivalent to interpolating the difference between the $C_l$ and a fiducial $C_l^f$, and adding the interpolated result to $C_l^f$ calculated with dense $l$ sampling\footnote{The interpolation is actually done on $l(l+1)C_l/2\pi$ as this is more nearly constant on small scales.}. This interpolation scheme has been implemented in the January 2012 version of {\CAMB}, and numerical interpolation errors are compared in Fig.~\ref{spline_error}. The accuracy of the MCMC sampling results using this scheme is shown in Figs.~\ref{importance1DMCMC} and~\ref{MCMCAccuracyTests}, and is generally at a level $\alt 5\%$ of the error bar size with default high accuracy settings for Planck. This interpolation scheme is significantly more accurate for near-fiducial models, and as such high accuracy settings no longer need to increase the $l$ sampling density; it is therefore faster than the high accuracy settings of October 2011 used in the bulk of this paper (and higher accuracy; the sampling density could be further decreased slightly to gain additional speed advantage).

We conclude that the default high accuracy settings should be adequate for analysis of the standard {\LCDM} model at Planck sensitivity, and the January 2012 interpolation scheme achieves this at somewhat higher speed than using an interpolation scheme without error correction. Data analysis and physical model uncertainties are likely to be much larger than numerical issues.

\section{Conclusions}

We have shown how parameter degeneracies remain very important even with high-precision CMB data. However, small degeneracy-breaking effects become quantitatively more important, especially CMB lensing on small-scales. The approximate acoustic-scale degeneracy in flat {\LCDM} models will shrink significantly with future data, however it will remain a significant limitation on the ultimate precision of CMB-alone parameter inferences.

We have shown that {\CAMB} has sufficient numerical stability at high accuracy settings for degeneracies to be explored reliably, and quantified the size of residual numerical artefacts. Our tests conclude that {\CAMB} is performing at or below its quoted numerical accuracy, and that the default high accuracy calculation is likely to be sufficient for interpretation of Planck data. Small residual biases due to multipole interpolation can be eliminated by subtracting the spline errors at a fiducial model, as implemented in the January 2012 version of {\CAMB}; typical parameter biases are then $\alt 5\%$ of the random error expected with Planck.

\section{Acknowledgements}
CH acknowledges support from the University of Sussex Research Placement (RP) scheme.
AL acknowledges support from the Science and Technology Facilities Council [grant number ST/I000976/1].
Some of the calculations for paper  were performed on the
 COSMOS Consortium supercomputer within the DiRAC Facility jointly
 funded by STFC, the Large Facilities Capital Fund of BIS and the
 University of Sussex.
AH is supported by an Isaac Newton Trust European Research Studentship
and the Isle of Man Government.

\appendix

\section{Numerical evolution of massive neutrino perturbations}
\label{app:massive_nu}
The evolution of linear perturbations in the massive neutrinos is well understood~\cite{Ma95}, but is more complicated and hence numerically slower than for massless species because neutrinos moving with different velocities evolve differently. Here we briefly describe {\CAMB}'s implementation as from July 2011.

Scalar mode perturbations with comoving momentum $q$, multipole moment $l$, and wavenumber $k$ evolve with conformal time as
$$
F_l' + \frac{k v}{2l+1}\left[ (l+1)\kf_{l+1}F_{l+1}-l F_{l-1}\right] + \left[ \delta_{2l} \frac{2}{15}k\sigma -\delta_{0l}h'\right]
\frac{\ud \ln F}{\ud \ln q}=0,
$$
where the time-dependent velocity is $v\equiv q/\epsilon$ and $\epsilon$ is the comoving energy, and we follow the conventions of~\cite{camb_notes,Challinor:1999xz,Lewis:2002nc}. Here $F$ without a subscript is the background distribution function, which is assumed to be Fermi-Dirac when the neutrinos are fully relativistic. For convenience we can divide through by defining $\nu_l \equiv -4 F_l/(\frac{\ud \ln F}{\ud \ln q})$, giving
$$
\nu_l' = \frac{kv}{2l+1}\left[
l  \nu_{l-1}-  \nonflat{\kf_{l+1}}(l+1) \nu_{l+1} \right]
 + \frac{8}{15} k \sigma \delta_{l2} -
\frac{4}{3}k \clz\delta_{l0}.
$$
When the neutrinos are relativistic so that $v=1$, this is identical to the massless neutrino equation, and $\nu_l$ is then independent of $q$. The only difference with massive neutrinos is that they travel at a different (time dependent) speed once the mass becomes important. The hierarchies are truncated at $l_{\rm max}$ using
$$
\nu_l' = kv \nu_{l-1} - (l+1)\cot_K \nu_l.
$$

Evolution is started when neutrinos are highly relativistic, when $\nu_l=J_l$, where $J_l$ are the massless neutrino multipoles. To get the leading correction from the mass we can write
$$
\nu_l = J_l +  \frac{m^2}{2q^2} \Delta J_l
$$
and use the series expansion $v\approx 1- a^2 m^2/(2q^2)$ so that
$$
\Delta J_l' = \frac{k}{2l+1}\left[
l  \Delta J_{l-1}-  \nonflat{\kf_{l+1}}(l+1) \Delta J_{l+1} \right] - \frac{ka^2}{2l+1}\left[
l  J_{l-1}-  \nonflat{\kf_{l+1}}(l+1) J_{l+1} \right].
$$
This lets us calculate the evolution of any $q$ momentum mode while that mode has $|a m/q| \ll 1$, which for light neutrinos and larger $q$ can be a while: only one additional hierarchy has to be evolved (that for $\Delta J_l$) until  modes start to become significantly non-relativistic, at which point the full mode equation must be integrated for a sample of momenta. The time saving from this approximation is not large in itself, but it does allow lower $l_{\text{max}}$ to be used when switching to integrating the momentum modes separately - without it the momentum modes need to be integrated from the beginning with roughly the same $l_{\rm max}$ as the massless neutrinos. It also makes clear the leading $q$ dependence of the perturbed distribution that we use below.

{\CMBFAST}, \COSMICS\ and {\CAMB} previously used a rather brute force integration sampling, using a fix grid of many $q$ samples, which leads to numerically slow evolution: we would like to evolve as few different $q$-mode hierarchies as possible. Ref.~\cite{Lesgourgues:2011rh} pointed out that one can do much better by using a more intelligent sparse sampling of $q$ samples, e.g using Gauss-Laguerre quadrature. Here we describe a sampling scheme specifically optimized for the problem in hand. We need integrals of the form
$$
\frac{1}{4}\int_0^\infty dq \frac{q^4 e^q}{(1+e^q)^2} v^w \nu_l
$$
in order to calculate the massive neutrino density, heat flux, and other perturbations enter the equations for the evolution of other species. In the perturbatively relativistic regime, so that we can do an expansion in $a^2 m^2/(2q^2)$ as before, the integrals are sums of terms involving integrals of the form
$$
\int_0^\infty dq  \frac{q^4 e^q}{(1+e^q)^2} q^n.
$$
At late times we also expect $v\sim q$, also giving terms roughly of this form, though the distribution has evolved away from anything simple.
We do not attempt to integrate the distribution accurately in the intermediate sub-Hubble regime where there can be oscillations in $q$: ignoring these seems to be harmless at required precision, presumably because averaged over time or $k$ they are smoothed out. So the idea is to choose a sampling in $q$ so that integrals with $n=-4,-2..2$ are evaluated exactly, which gives a set of constraint equations for the points and weights that can be solved, and if more points are desired the solution can be made unique by adding other constraints or making a choice of a few points (e.g. at high $q$ would expect Gauss-Laguerre point sampling to be nearly optimal~\cite{Lesgourgues:2011rh}). For 3 points we find\footnote{Mathematica: \url{http://camb.info/maple/NeutrinoIntegrationKernels.nb}} the remarkably sparse sampling $q=(0.913201, 3.37517, 7.79184)$ produces results accurate at the $2\times 10^{-4}$ level with
$$
\frac{1}{4}\int_0^\infty dq \frac{q^4 e^q}{(1+e^q)^2} v^w \nu_l \approx \sum_i K_i v^w \nu_l
$$
and kernel weights $K=(0.0687359, 3.31435, 2.29911)$. A four-point sampling is accurate at the $< 10^{-4}$ level, e.g. with $q=(0.7, 2.62814, 5.90428, 12)$, $K=(0.0200251, 1.84539, 3.52736, 0.289427)$.

At late times further speedups are possible. For highly sub-horizon perturbation $k\tau \gg 1$ (as for massless neutrinos), and once significantly non-relativistic, $l_{\rm{max}}$ can be reduced down to 2 or 3~\cite{Blas:2011rf} Once the neutrinos become very non-relativistic we can evolve velocity-integrated equations (i.e. a truncated fluid hierarchy); this is described in detail in Ref.~\cite{Lewis:2002nc} and was previously implemented in {\CAMB}.

Further speed ups may be possible using the fluid approximations of Ref.~\cite{Lesgourgues:2011rh}, but the scheme described above is sufficient to dramatically reduce the computing compared to a naive evolution of a large sampling of $q$ modes, so that calculations involving massive neutrinos only take $\clo(1)$ times longer than with massless neutrinos.

%\bibliography{../antony,../cosmomc}

\providecommand{\aj}{Astron. J. }\providecommand{\apj}{Astrophys. J.
  }\providecommand{\apjl}{Astrophys. J.
  }\providecommand{\mnras}{MNRAS}\providecommand{\aap}{Astron. Astrophys.}

\end{document}